# Microstructure-Aware Deep Learning Bridges Atomistics to Macroscale for Shock-to-Detonation Prediction

Simon Gonzalez-Zapata[1]†, Aidan Pantoya[2]†, Chunyu Li[2], Marisol Koslowski[1], and Alejandro Strachan[2]*

[1]School of Mechanical Engineering, Purdue University; West Lafayette, IN, 47907, USA.
[2]School of Materials Engineering, Purdue University; West Lafayette, IN, 47907, USA.

†Equal Contributions

*Corresponding author. Email: strachan@ purdue.edu

The shock-to-detonation transition in energetic materials is governed by coupled processes spanning Ångstroms to millimeters and femtoseconds to microseconds, where traditional multiscale models fail due to the lack of scale separation. We address this grand challenge by directly bridging large-scale molecular dynamics (MD) simulations with continuum finite-element (FE) models using MISTnetX, a convolutional deep neural network. Trained on MD simulations of shock propagation through complex microstructures, MISTnetX captures shock–microstructure interactions, hotspot formation, and the transition to deflagration, supplying critical sub-grid information to FE simulations of mechanics, shocks, thermal transport, and chemistry. Applied to a synthetic but realistic nanostructured plastic-bonded RDX composite, MISTnetX enables parameter-free prediction of the full run-to-detonation transition.

## Introduction

Understanding the shock initiation of explosives is important for their safe handling, use, and decommissioning. The experimental determination of this property, e.g. via pop plots that relate run to detonation to shock pressure (*1*), requires time-consuming and costly experimental campaigns. Thus, predictive models are highly valuable and significant efforts have been devoted to their development over decades (*2*). Despite experimental (*3–6*) and theoretical (*7–10*) progress in the understanding of the underlying mechanisms, current models cannot predict detonation initiation or failure for plastic bonded explosives (PBXs) require significant experimental calibration. The shock-to-detonation transition (STDT) remains a modeling grand challenge at the intersection of shock physics, materials science, and chemistry, that involves thermal, mechanical, and chemical processes of materials at extreme conditions. The complexity of the physics involved is compounded by the disparate range of scales that need to be captured and the complex microstructure and defects in PBXs. Run to detonation distances are in the millimeter scale, grain sizes range from tens to hundreds of microns, internal voids whose collapse leads to hotspots range from tens to hundreds of nanometers, shock-induced jetting, friction, and chemistry require molecular and atomic resolution. Unlike other materials phenomena where multiscale modeling or coarse graining has resulted in predictive capabilities (*11–13*) these phenomena are tightly coupled, and traditional separation of scales is not applicable. To address this challenge, we use a microstructure-aware machine learning (ML) model trained from large-scale MD simulations to capture sub-micron phenomena associated with the formation of hotspots due to the interaction between the shock and the materials microstructure and the subsequent thermo-chemical evolution. The ML model, denoted MISTnetX, provides a speedup of up to $10^8$ with respect to MD and is coupled, on the fly, to a finite elements (FE) framework to provide sub-grid resolution in run to detonation simulations. Thus, the challenge of lack of scale separation is captured by *learning* from explicit MD simulations and not modeling unit processes. Direct coupling to molecular simulations through surrogate modeling has recently been used to describe boundary lubrication (*14*).

The chemical reactions that lead to detonation initiate at hot spots that form due to the interaction of the shock wave with microstructural features and defects. Models can, thus, be classified into those that explicitly capture microstructure and the associated hotspots and those that model their effect as sub-grid phenomena. Reactive burn models (*15–17*) belong to the latter group and describe the process via two distinct stages. The ignition phase estimates the density and characteristics of reacting hotspots based on the shock strength. The growth phase describes the expansion and coalescence of reaction zones which can lead to detonation. While these models can reproduce experimental detonation thresholds and pop plots, they inherently rely on empirical calibration. To remove some empiricism, recent efforts describe ignition starting from the distribution of voids in the material and include a pressure-dependent critical void surface above which activation occurs (*18*). Within the former category of models, direct numerical simulations of shock initiation capture the effect of selected microstructural features. Ref. (*19*) combines isotropic plasticity and a four-stage chemical decomposition model to predict the thermal, chemical, and mechanical response of pressed HMX (1,3,5,7-Tetranitro-1,3,5,7-tetrazocane) samples under shock loading. These simulations provide valuable insight but necessarily approximate or ignore molecular-level processes like localized plasticity, pore collapse, jetting and other sub-grid phenomena. Mesoscale modeling of processes like pore collapse remain challenging (*20*) and solutions are resolution-dependent (*21*). A recent development is the use of ML to accelerate direct numerical simulations. PARC (Physics-aware Recurrent Convolutional Neural Network) (*22*) (*23*) combines a recurrent convolutional neural network with a time-stepping, partial differential equation (PDE) like update structure. PARC can predict both initiation and run to detonation distances around 9,000 times faster than direct simulations (*22*). Increasingly, these continuum simulations are being informed by MD simulations (*24*).

We introduce a new multiscale model capable of modeling STDT with microstructure and composition as the only inputs. To address the lack of scale separation and the challenge in describing phenomena like pore collapse and localized plasticity with mesoscale models, we use large-scale MD simulations to provide sub-grid information to a continuum model using FE. The MD simulations capture the interaction of the shockwave with the materials microstructure and the formation of hotspots with molecular resolution, making no approximation beyond the use of classical mechanics and those associated with the interatomic potential used. Key to our approach is a microstructure-aware ML model trained from MD simulations and invoked, on-the-fly, in FE simulations, which enables millimeter scale simulations.

**Results**

***Multiscale model with microstructure-aware ML sub-grid enhancement.***
We describe the STDT in a PBX consisting of nanoscale 1,3,5-Trinitro-1,3,5-triazinane (RDX) grains and a polystyrene binder. The microstructure, with grain sizes between 5 and 20 nm, is synthetic but has realistic features (*25*); additional details can be found in the SM and Refs. (*26*) (*27*). We use FE to model the full scale of the STDT, bottom of Fig. 1. The model incorporates shock physics via JWL equations of state (*28*), plasticity, thermal transport, and chemistry. All these continuum models have been parameterized from MD, see SI. Each element (approximately 1x1x2 mm$^3$) is assigned a sub-grid microstructure at random, and its response is described using MISTnetX, middle panel of Fig. 1. The direct bridging of explicit MD simulations of the shock interaction with the materials microstructure (top panel of Fig. 1) and the FE simulations via MISTnetX is key to our ability to predict the STDT with chemistry and microstructure as only inputs and with minimal approximations.

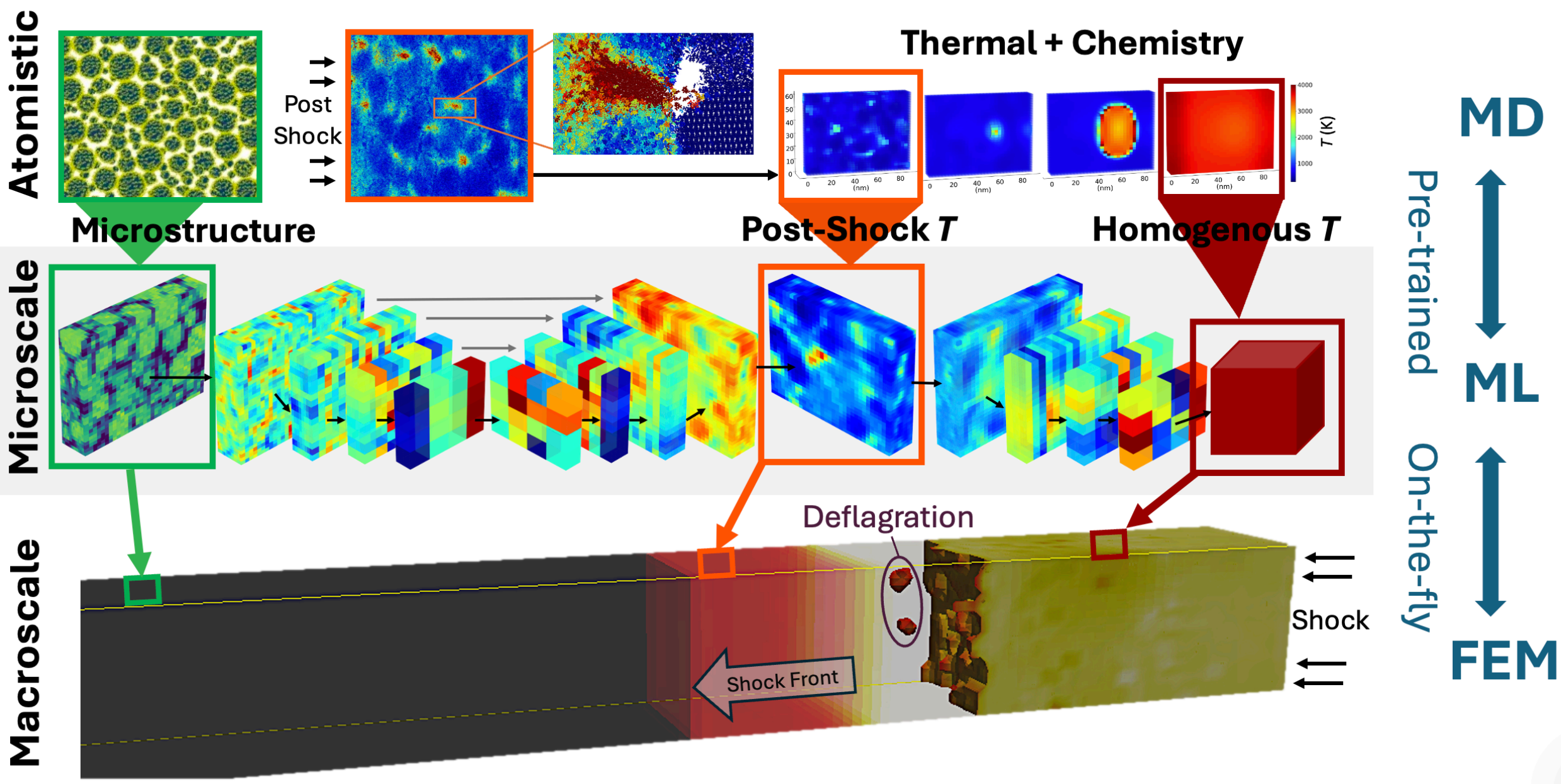

**Fig. 1. MISTnetX, trained from MD simulations, predicts the shock response at the nanoscale and couples with FE to predict the deflagration to detonation transition at the continuum scale.**

MISTnetX builds on and extends recent work on a conditional convolutional neural network to predict the temperature field resulting from the shock loading of energetic composites (*27*). Here we add the capability to describe the post-shock chemical and thermal evolution of the system until local thermal and chemical equilibrium is attained. MISTnetX predicts the temperature field right after the passage of the shock and the final homogeneous temperature, after the system either quenches or deflagrates. The model was trained on MD simulations of various microstructures shocked with particle velocities ranging from 1 to 2.5 km/s to predict the post-shock temperature fields. The simulations include PBX systems and RDX crystals with single and multiple pores, see supplementary materials (SM) and Refs. (*25*, *26*). The final homogeneous temperatures were obtained by evolving the shock temperature fields from the MD simulations using thermal transport and a reduced chemical kinetics model derived from reactive MD simulations (*24*). Importantly, the MD simulations used to train MISTnetX make no approximations regarding the mechanisms of energy localization. Processes like localized plasticity, interfacial friction, pore collapse and jetting, including their shock-strength dependence, are explicitly modeled. Additional details including architectural details of MISTnetX are provided in the SM Sections S1.1 through S1.3.

At the macroscopic level, the continuum problem is solved using an element size similar to the MISTnetX domain, which is assumed to be representative of the entire element's sub-grid resolution behavior. The FE simulations involve 3D domains with a cross section of $66\ \times 66\ \mu\mathrm{m}^2$ and a length along the shock direction between $380\ \mu m$ to $570\ \mu m$. The FE simulation enforces conservation of linear momentum, energy, and mass, through a set of coupled equations to integrate the shock thermo-chemo-mechanical problem, see Methods and Sections S2.1 through S2.5 in SM for details. When the arrival of the shockwave is detected in an element and the local particle velocity ($u_p$) is established, MISTnetX is invoked using the local $u_p$ and the underlying microstructure as inputs. At this point, the FE solver temporarily delegates chemical species evolution, chemical heat release, and sub-grid thermal transport to MISTnetX and the *takeover* stage begins, see Section 2.6 in the SM and Figure S2. During the first period of this stage, a heat source term is added to take the element system from the initial temperature ($T_0 = 300$ K in our case) to the average post-shock temperature predicted by MISTnetX in a timescale $\tau_{shock}$, the time required for the shock to travel across the element, see SM Section S2.7. This heat source accounts for sub-grid and microstructure-dependent heating processes captured by our MD simulations. During the second stage heat source takes the system to the homogeneous deflagration or quench state within a timescale $\tau_{hom}$, as described in the SM. Once the takeover stage ends, the continuum constitutive equations regain full control and evolve the temperature, chemical compositions, and mechanical response (*10*).

Figure 2 shows snapshots of the key events during the STDT in our nanocomposite system shocked with an impact velocity $u_p = 1.75$ km/s. Pressure is shown as a transparent field and temperature as solid surface for volumes where final products exceed 95% of completion. Random variation in microstructure results in the formation of localized critical hotspots followed by their ignition and growth, see snapshots between $t = 5$ ns and $t = 6$ ns. We observe ignition and growth of hotspots with sufficient mass and temperature, while smaller and cooler hotspots remain relatively unchanged. At $t \sim 7$ ns we observe the coalescent of hotspots due to their fast deflagration velocities. The large overall rate of reaction ultimately leads to the release of a secondary wave that catches up with the leading shock; this process leads to the deflagration to detonation transition at approximately $t = 12$ ns. The see-through pressure contour enables the visualization of the leading shock front and the secondary wave, followed by reactions dominated by the local microstructure. The frame at $t = 16\ ns$ shows the complex reaction patterns seen during steady state detonation at the reaction front that will be further discussed below. We find identical processes and steps for a slightly stronger shock with $u_p$ = 1.85 km/s, see Section S.3 in the SM.

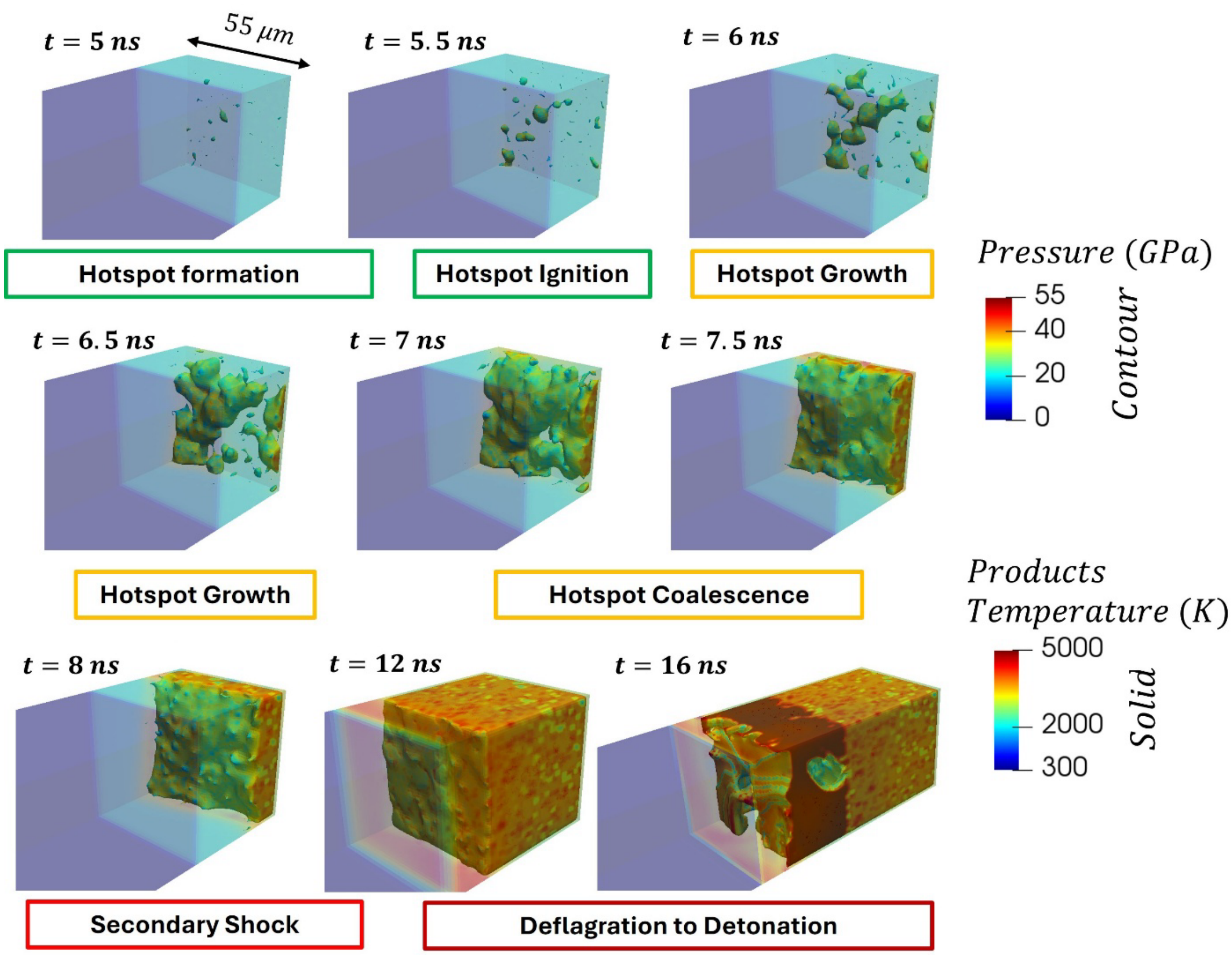

**Fig. 2. Evolution of a STDT for an initial impact of $u_p = 1.75$ km/s.** Key steps of ignition, growth and coalescence of deflagration waves, and the launch of a secondary shock are clearly visible before the transition to detonation.

The time evolution of pressure profiles, see Fig. 3a for $u_p$ =1.75 km/s results, provides a clear picture of the STDT. The solid black lines represent the average pressure of the cross section, and the shaded regions show the four quartiles of the pressure distribution. We clearly see the strengthening of the initial shock into a detonation and an overshoot in pressure before a steady-state detonation. This transient occurs as the pressure and particle velocity buildup from the deflagrating regions accelerate the leading shock and has been observed both in experiments (*29*, 30), (30) and simulations (15, 31). Our simulations predict a von Neuman spike at the shock front and a reduction of pressure within the reaction zone to the Chapman-Jouet (CJ) state (32). This overall behavior and pressure values match what is observed experimentally (33) and our results highlight the significant variability in local pressure, especially during the STDT and between the von Neuman spike and the CJ plane where microstructural effects are dominant. The temperature profiles in Fig. 3b show the initial shock induced temperature increase at early times followed by additional heating behind the shock front due to chemistry. During the transition to detonation the rapid increase in shock pressure, driven by the chemical reactions in the deflagration wave, leads to significant heating right behind the shock front. We note the correspondence between the overshoot in the pressure spike prior to a steady detonation, Fig. 3a, and the significant increase in temperature right behind the shock front, Fig. 3b.

Before diving into additional details of the STDT, we compare our predicted run to detonation distances as a function of shock pressure to experimental results for similar materials, Figure 3b. To determine the run to detonation, we track the position of the shock front over time and find the intersection between two linear fits for the pre-detonation and the steady detonation portions of the x-t plot, see additional information in Section S2.8 of the SM. We compare our results with well-characterized RDX explosives (PBX-9407, RDX/2.5 WAX/2.5 ELVAX) and HMX ones (PBX9501, Pressed HMX). PBX-9407 uses the vinyl-chlorine copolymer Exon 461 as a binder and has a grain size between 10 and 50 mm while PBX9501 has a bimodal particle size distributions with peaks at ~50 mm and ~200 mm (34). Typical voids and cracks in the experimental samples are in the 10 to 100 nm size range (35) but larger voids are often present (36). Our model captures the expected trend of decreasing run to detonation distance with increasing shock strength and good agreement with the slopes and run distances of PBX 9407 and ELVAX, which are the closest systems in terms of RDX/Binder ratio. The larger microstructural features in the experimental PBXs can activate hotspots at lower velocity regimes (17, 36, 37) explaining the higher pressures in our simulations.

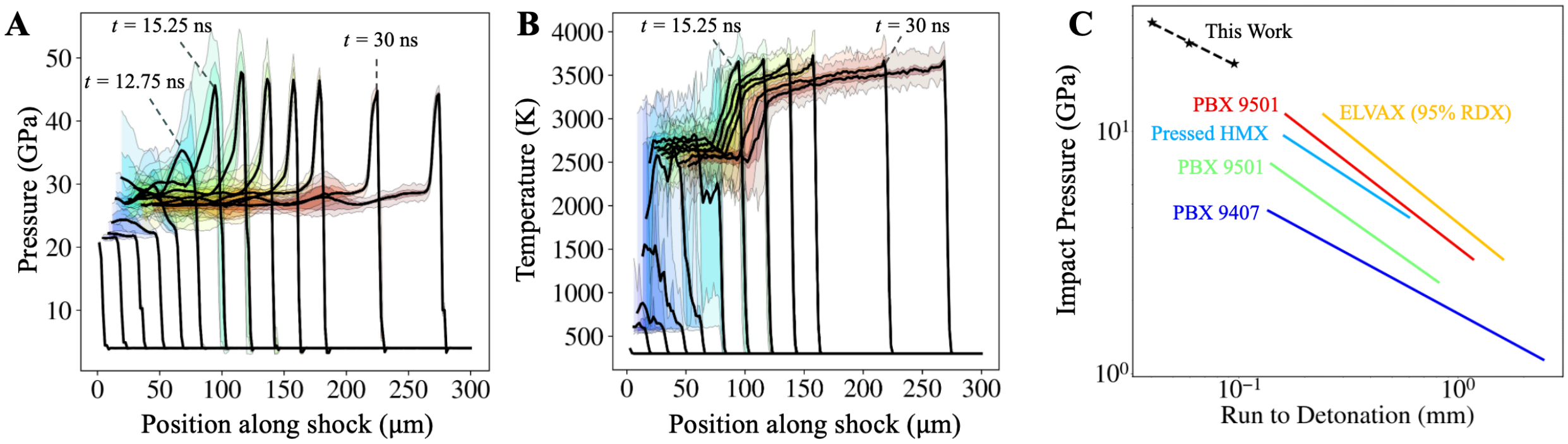

**Fig. 3. Shock to detonation profile.** (**A**) Pressure-position profiles at different times for an initial impact velocity of 1.75 km/s. Subsequent profiles are spaced by 2.5 ns, with the initial one corresponding to t = 0.25 ns after the initial impact. The last two lines, t = 35 and 30 ns, show the stead detonation pressure profile. (**B**) Temperature-position profiles at the same time as (**A**). The temperature profiles show a distinct transition between hotspot burn behind the shock front and concave down temperature profiles during detonation. (**C**) Pop plot comparing this work and experimental results from (1).

A detailed analysis of the simulations indicates that the morphology and evolution of ignited hotspots and the reaction front are governed not only by the initial distribution of microstructures but also by complex dynamic phenomena that arise during the shock-induced reaction process. Secondary shock interactions within the reaction zone, shock wave reflections from the evolving microstructures, and the local coupling between thermal softening and mechanical deformation collectively influence how energy localizes and propagates, see Figs. 4A and B. These coupled effects can modify the pressure and temperature fields, alter the growth and coalescence of the hotspots, and ultimately affect the continuity and stability of the overall reaction front. The correlation and mutual influence between temperature, pressure, and reaction product fields during the STDT is apparent in Fig. 4A. We observe the ignition of multiple hotspots and note the importance of their spatial correlation: the dominant burning region, see fields between 10 and 14 ns, originates from the interaction between three neighboring ignited hotspots. We also note a strong coupling with the pressure field as these deflagration sites grow and coalesce. During the initial stages of the shock to deflagration transition, reactions proceed rather isotopically, consistent with reactive MD simulations (*10*). Eventually, the entire trailing region forms a deflagrating front that travels from the beginning of the reaction zone up to the shock front. This, in turn, increases the temperature and density of ignited hotspots due to an increase in shock velocity driven by higher pressures. This acceleration eventually results in the STD transition, see snapshots at 15 ns onward in Figure 4B. We note that even after steady state detonation is achieved, the variation in local microstructure results in slight spatial variations in the temperature field, see 21 ns snapshot in Figure 4B. The evolution of local pressure-volume relations over time provides additional insight and helps to visualize the different stages of the STDT as the material transitions from reactants to combustion products. Figure 4C shows datapoints in pressure-volume space, averaged across each cross-section, at different simulation times. Early times show the Rayleigh line from the initial state to the initial shock (approximately 20 GPa). The increase in pressure during the STDT is also apparent and the later times (red) show the material jumping to the von Neuman state and relaxing back to the CJ point. The equations of state of the reagent and products are included as references; we note the CJ point is below the products EOS since our PBXs contain 12% polymer. See Appendix of the SM for details of the microstructures.

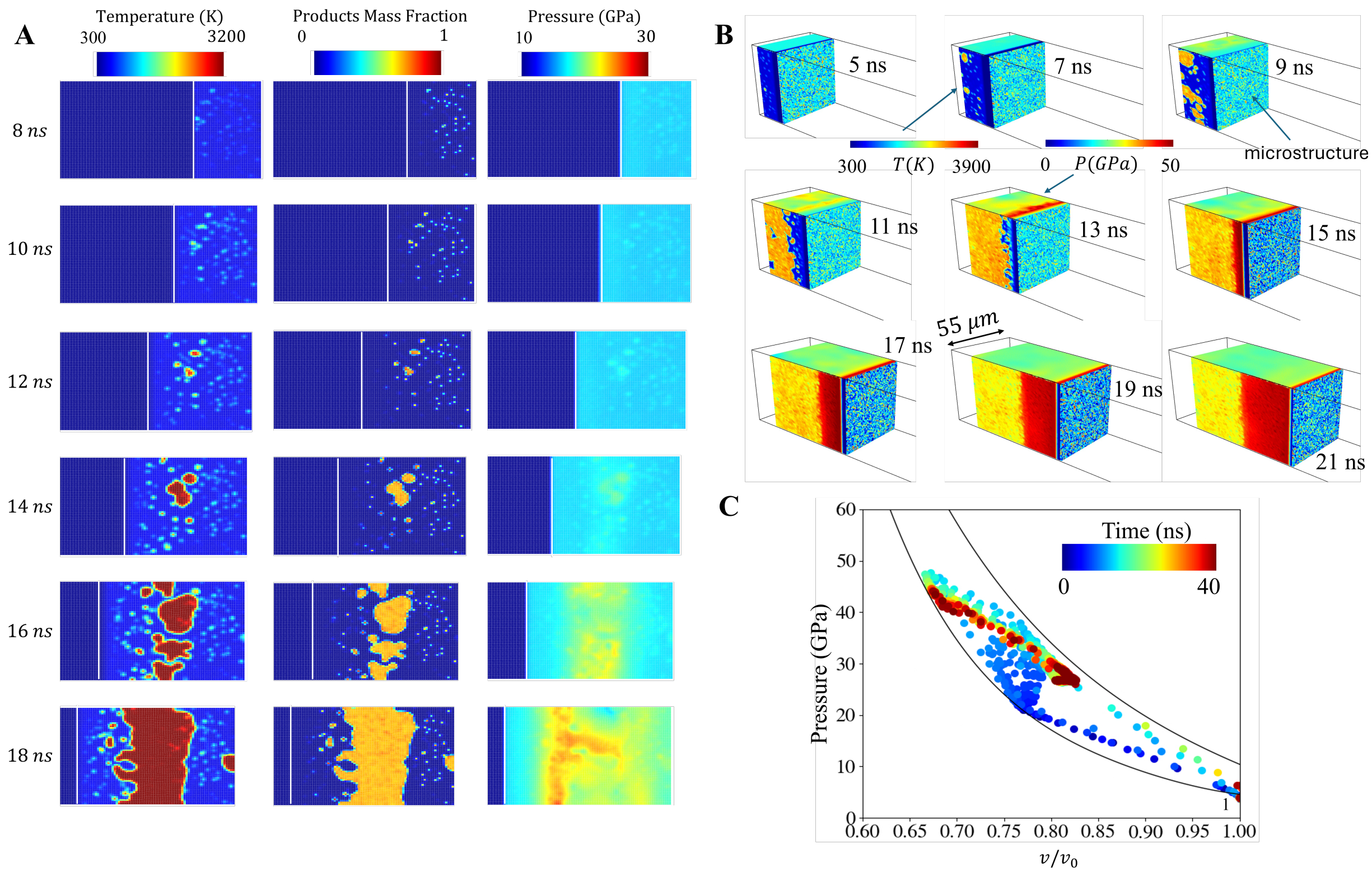

**Fig. 4. 2D and 3D visualization of the shock to detonation transition.** (**A**) Temperature, reaction extent, and pressure field during the early stages of the STDT showing hotspot ignition and growth. (**B**) Temperature, pressure, and microstructure 3D profiles showing DDT at later stages of STD transition. (**C**) Reconstructed Hugoniot from a 1.75 km/s shock.

## Discussion

A novel approach to learn microstructural effects from large-scale MD simulations using deep learning enables multiscale modeling for a problem where separation of scales is not possible. Simulations of nanoscale PBX systems reveal the STDT with unprecedented resolution. MISTnetX *learns* the relative important of microstructural features by example, using convolutions and pooling operations that can extract general features and high-order correlations. Thus, we map an explicit description of the microstructure (as opposed to descriptors like average quantities, two-point correlations, or distributions) to properties, this is important since initiation is dominated by microstructural extremes. In addition, by directly connecting MD simulations with continuum simulations, we avoid mesoscale modeling of processes like localized plasticity, interfacial shear, pore collapse and jetting that are notoriously hard.

While MISTnetX provides sub-grid resolution, the FE model can achieve macroscopic scales and capture microstructure with larger scales. To demonstrate this, we performed preliminary calculations of a more complex microstructure consisting of RDX grains with internal porosity and a binder consisting of our nano PBX, see Figure 5. The preliminary simulations show how larger voids dominate initiation, especially for relatively weak shocks. Further studies on such microstructures are ongoing and will be the focus of a separate publication.

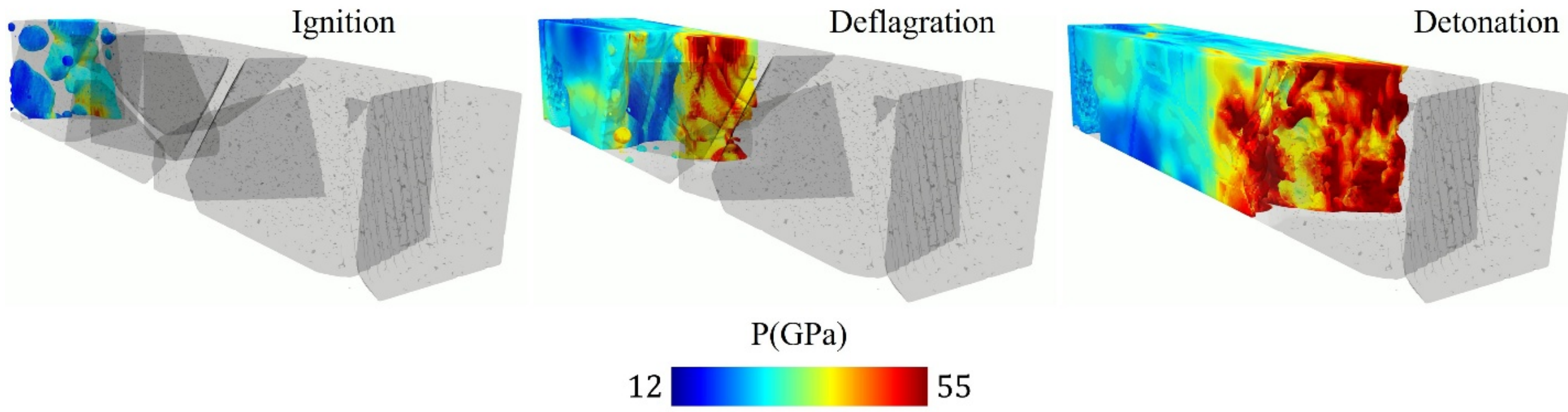

**Fig. 5. Pressure over different SDT stages for a shocked multi-fidelity PBX microstructure.** Three stages are depicted, starting with ignition, then deflagration, then detonation.

Our approach can be generalized to other problems in multiscale modeling. Examples include the mechanical response of advanced alloys (*38*) and ultrasonic initiation of chemistry (*39*). Finally, we note that the underlying model does not necessarily need to be MD, but it should capture all the possible underlying mechanisms. In the case of mechanical response, these models could be for example dislocation dynamics (*40*).

**Acknowledgments:**

**Funding**: This research was sponsored by the United States Air Force Office of Scientific Research Award Number FA9550-23-1-0674 (program managers Maj. Derek Barbee and Dr. Chiping Li). Computational resources from Purdue University's Rosen Center for Advanced Computing are gratefully acknowledged.

**Author contributions:** Conceptualization, supervision, and funding acquisition: AS and MK. Investigation, data analysis, and visualization: SG, AP, CL. Methodology and writing: all authors.

**Competing interests:** Authors declare that they have no competing interests.

**Data, code, and materials availability:** All data are available in the main text or the supplementary materials.

# Supplementary Materials for

## Microstructure-Aware Deep Learning Bridges Atomistics to Macroscale for Shock-to-Detonation Prediction

Simon Gonzalez-Zapata, Aidan Pantoya, Chunyu Li, Marisol Koslowski, and Alejandro Strachan

Corresponding author: strachan@purdue.edu

**The SM includes:**

- Methods
- Supplementary Text
- Figs. S1 to S11
- Tables S1 to S5
- Appendix

# Supplementary Materials

## S1. MISTnetX Additional Details

### S1.1 MISTnetX Architecture

MISTnetX uses a U-Net architecture conditioned on shock-strength and voxel dimensionality to map an initial atomistic level microstructure to the immediate post shock temperature distribution and final equilibrium temperature following thermochemical evolution, see Figure S1. This is achieved in two stages, first we predict the shock temperature immediately following the passage of the shock (implemented in MISTnet2 (*27*)), then this prediction is passed to the exothermic branch, hence the 'X' in MISTnetX, to predict the evolved equilibrium temperature. Depending on shock strength and microstructure, the final equilibrium state could be a quenched sample with negligible chemistry or a reacted deflagrated sample.

The first branch is a reproduction of the MISTnet2 model, where a convolutional U-Net is used to map an initial microstructure description [$\rho_{\text{total}}$, $\rho_{\text{RDX}}$, $\rho_{\text{RDX•PS}}$], voxel dimensionality [x (nm), y (nm), z (nm)], shock particle velocity [$u_p$ (km/s)], and simulation type [0 = all-atom, 1 = dissipative particle dynamics with energy conservation] to the resulting post-shock temperature map. The second branch is a convolutional encoder network. The predicted post-shock temperature field and the voxel dimensionality is passed through the encoder, where the probability of the temperature field resulting in a deflagration or a diffused state is predicted [*P*(reaction)]. If the state is most likely a deflagration, the predicted homogeneous high temperature (*T* if reactive) is returned. If the state is most likely diffusive, the predicted homogeneous low temperature (*T* if non-reactive) is returned. The result is two outputs: the immediate post-shock temperature field as a three-dimensional volume and the homogeneous temperature as a scalar.

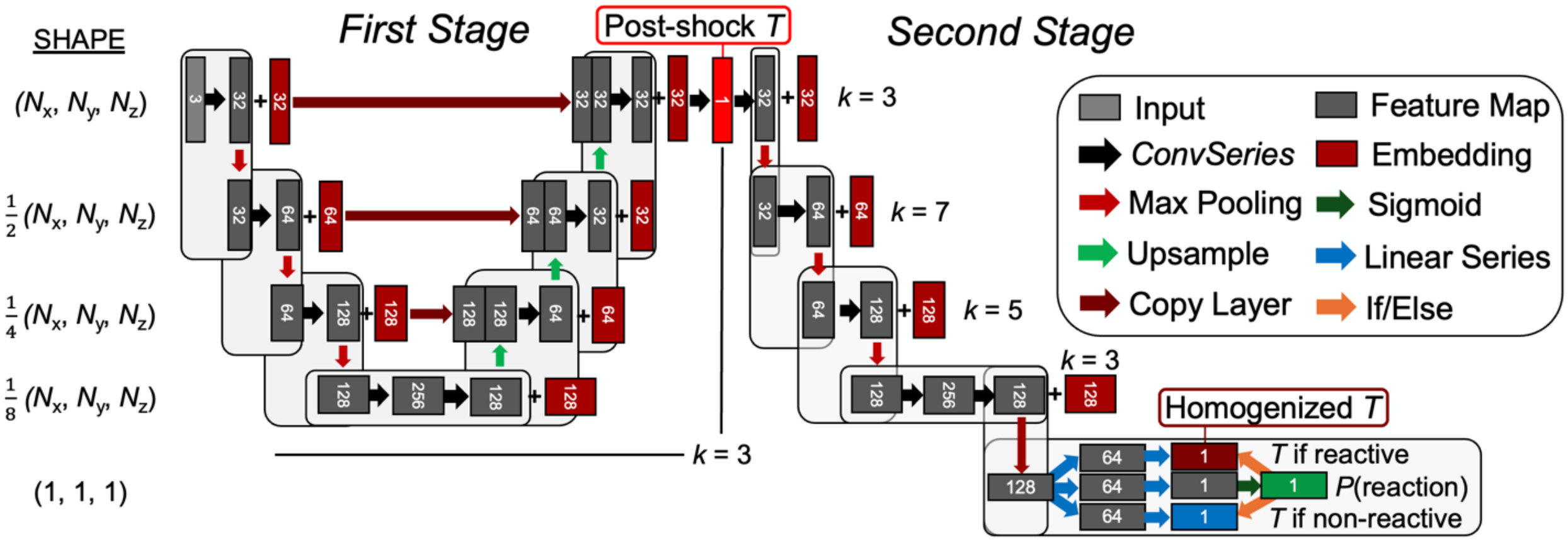

**Fig. S1. MISTnetX Architecture.** A multi-stage convolutional U-Net + encoder network mirroring physical shock progression. Stage 1 predicts the immediate post-shock temperature field from input microstructure and shock strength. This result is passed to the subsequent encoder layers to predict the final evolved temperature. This intermediate temperature field ensures physical interpretability through the prediction pipeline.

### S1.2 MISTnetX Training and Validation

The nonreactive MD derived datasets utilized to train, validate, and test MISTnetX are from atomistic MD and dissipative particle dynamics with energy conservation (DPDE) simulation, shocked at a $u_\text{p}$ of 1.0, 1.5, 2.0, or 2.5 km/s (*27*). The dataset includes MD simulations results from four large (~30 million atoms; ~84 x 18 x 220 nm$^3$) and seven smaller (~10 million atoms; ~60 x 18 x 100 $nm^3$) PBX nanostructures, generated by coating RDX crystals with polystyrene and compacting via atomistic MD to preserve realistic interfaces and porosity. Single-pore systems (diameters: 25, 50, 125 nm) and multi-pore systems (diameters: 20 – 100 nm) are also included, constructed by inserting a circular void into pure RDX crystals. Including augmentations, 443 total temperature fields are derived, 218 resulting from the atomistically simulated PBX systems used in this application. To provide a room-temperature foundation to the model, each training set is enhanced with the results corresponding to zero $u_p$, where the post-shock and equilibrium ground-truth temperatures are a uniform 298 K. These datasets are further detailed in (*24,28,41*), along with the augmentation descriptions. The systems used for training (61.6%, n = 273), validating (21.7%, n = 96), and testing (16.7%, n = 74) MISTnetX is described in Table S1.

| **Small nano-PBX** | | | **Large nano-PBX** | | | **Multi-Pore** | | |
|---|---|---|---|---|---|---|---|---|
| Train | Validate | Test | Train | Validate | Test | Train | Validate | Test |
| 3, 4, 5, 6 | 1, 7 | 2 | 2, 4 | 1 | 3 | 2, 3, 6, 7, 8, 9 | 4, 5 | 1, 10 |

**Table S1. Train, validation, and test sets.** Data split used for training and assessing MISTnetX.

As discussed above, MISTnetX extends MISTnet to include prediction of the homogenous temperature resulting from the evolution of shock-induced temperature field. This evolution is simulated using a reduced order chemical kinetics model derived from reactive MD simulations and thermal transport, with parameters also from MD (*24*). We refer to Lee et al. (*27*) for our implementation of the thermal-chemical kinetics model. The accuracy of MISTnetX is described in (*27*). The extension of MISTnetX to predict the final homogeneous temperature is enabled through a two-stage architecture. The first stage of the network (a U-Net) predicts the post-shock temperature field given a microstructure description and shock strength, $u_p$. The second stage (a convolutional encoder) utilizes the predicted post-shock temperature field as a physically grounded intermediate state to predict the resulting thermal and chemically evolved homogenous temperature. The homogenous temperatures reside between ~300 K and ~700 K, reflecting a diffused state, or between ~3300 K and ~3900 K, reflecting deflagration.

MISTnetX is trained in two stages. First, MISTnetX is trained to predict the post-shock temperature field. To do this, the homogeneous temperature prediction branch's parameters are frozen, and the remaining parameters are free to train. The objective function for the post-shock temperature field prediction aligns with the previous MISTnet2 implementation, consisting of a weighted mean squared error (MSE) loss where temperature regions below 300 K are weighted by one, regions from 300 K to 600 K are weighted by 5, and regions equal to or greater than 600 K are weighted by 10. We train for 500 epochs, using a weight decay of 0.2 and start with a learning rate of 1e-4. We reduce the learning rate by 50% after 100 consecutive epochs without an increase in validation accuracy. We save the model with the best performance on the validation dataset.

Following training of the post-shock temperature prediction layers, we load in the saved MISTnetX model with the best validation performance. The post-shock temperature prediction layers are then frozen, and the homogeneous temperature prediction branch is unfrozen. Using the AdamW optimizer, we train the equilibrium temperature layers for 500 epochs, with an initial weight decay of 0.2 and learning rate of 3e-5. We reduce the learning rate by 50% after 100 consecutive epochs without an increase in validation accuracy. We save out the model with the best validation performance for final testing.

The loss function used for the evolved homogeneous temperature prediction combines a classification, regression, and physical constraint term to ensure both predictive accuracy and thermodynamic realism. A binary cross-entropy (BCE) loss, scaled by a factor of 11, is applied to the model logits to classify each system as either *quenched* or *deflagrated*, based on whether the ground-truth homogeneous temperature exceeds 2000 K. Conditional regression is then performed using a MSE term, where the low-temperature branch is penalized only for non-reactive samples and the high-temperature branch only for reactive ones. To maintain physically realistic equilibrium behavior, an additional penalty term weighted by 4 is applied via a ReLU operation that discourages predicted temperatures falling inside the prescribed range of 900 K to 3300 K (unrealistic equilibrium temperature region). The total loss is thus defined as the sum of the scaled classification loss, the conditional regression error, and the equilibrium-bound penalty.

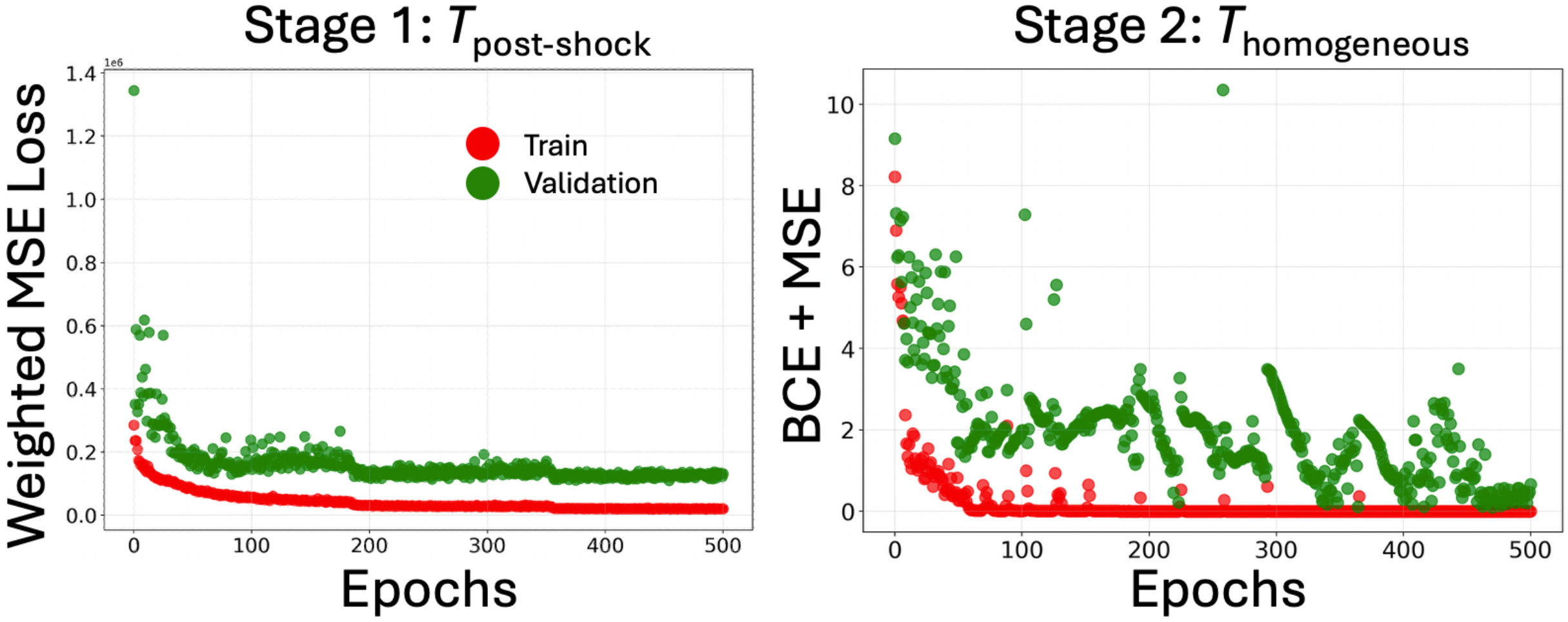

**Fig. S2. Training and validation of MISTnetX.** Stage 1 of MISTnetX [Left] learns to predict the post-shock temperature field ($T_{\text{post-shock}}$) over 500 epochs. Stage 2 of MISTnetX [Right] learns to predict the homogeneous temperature ($T_{\text{homogeneous}}$). During stage 1 the $T_{\text{homogeneous}}$ prediction layers are frozen, and the unfrozen layers are penalized by a weighted MSE loss function. During stage 2, the $T_{\text{post-shock}}$ layers are frozen, and the remaining layers are trained with a BCE + MSE objective.

S1.3 MISTnetX accuracy evaluation

MISTnetX predicts the immediate post-shock temperature field following shock loading, then uses this prediction as the input to predict the evolved homogeneous temperature. This second stage of this process relies on the accuracy of the first stage. To assess the ability of MISTnetX to predict these two outputs, we quantify the root mean squared error (RMSE) and mean absolute percentage error (MAPE) of the post-shock temperature fields as well as the $R^2$ score, classification accuracy (correct 'go' or 'no-go' prediction), mean absolute error (MAE), and mean absolute percentage error (MAPE) of the predicted homogeneous temperatures. These metrics are described across the training, validation, testing, and total dataset in Table S2.

| | $T_{\text{post-shock}}$ Metrics | | $T_{\text{homogeneous}}$ Metrics | | | |
|---|---|---|---|---|---|---|
| Type | RMSE (K) | MAPE (%) | $R^2$ | Classification % | MAE (K) | MAPE (%) |
| Training | 67.43 ± 25.62 | 6.70 ± 1.64 | 0.999720363 | 100 | 18.87 ± 13.25 | 1.47 ± 2.07 |
| Validation | 96.53 ± 47.90 | 8.96 ± 1.98 | 0.999517901 | 100 | 27.06 ± 20.70 | 2.50 ± 2.76 |
| Test | 105.12 ± 47.68 | 9.16 ± 2.33 | 0.999406382 | 100 | 25.69 ± 18.23 | 1.82 ± 2.44 |
| Total | 80.03 ± 39.18 | 7.60 ± 2.17 | 0.999628358 | 100 | 21.78 ± 16.40 | 1.75 ± 2.33 |

**Table S2. MISTnetX error assessment.** MISTnetX's post-shock and homogeneous temperature predictions are evaluated.

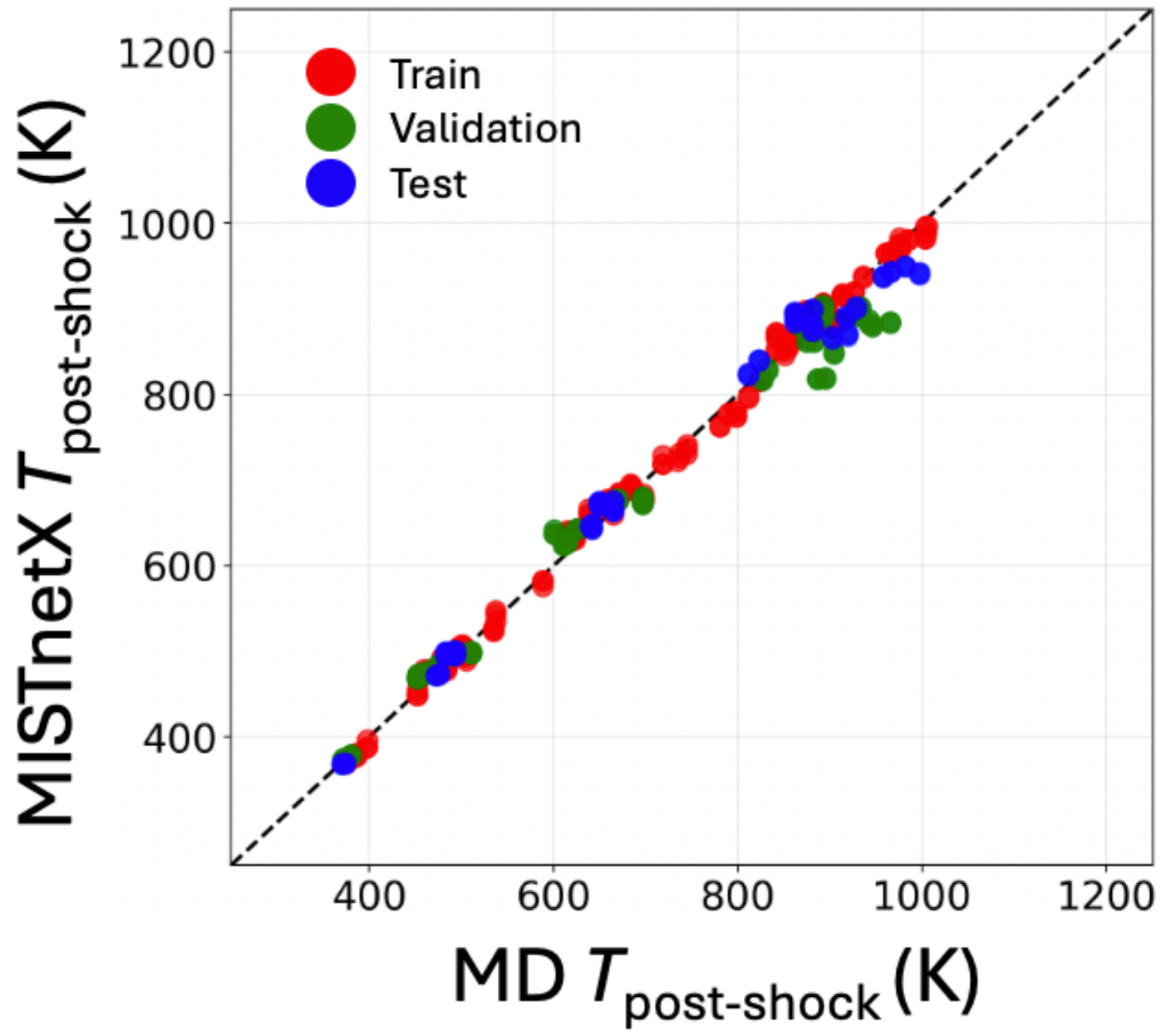

**Fig. S3. Parity of averaged MD and MISTnetX predicted post shock temperature fields.** The averaged MD post-shock temperature fields (x-axis) are compared to the averaged MISTnetX post-shock temperature fields across the train (red), validation (green), and test (blue) dataset.

The predicted homogeneous temperatures are further compared in Figure S4.

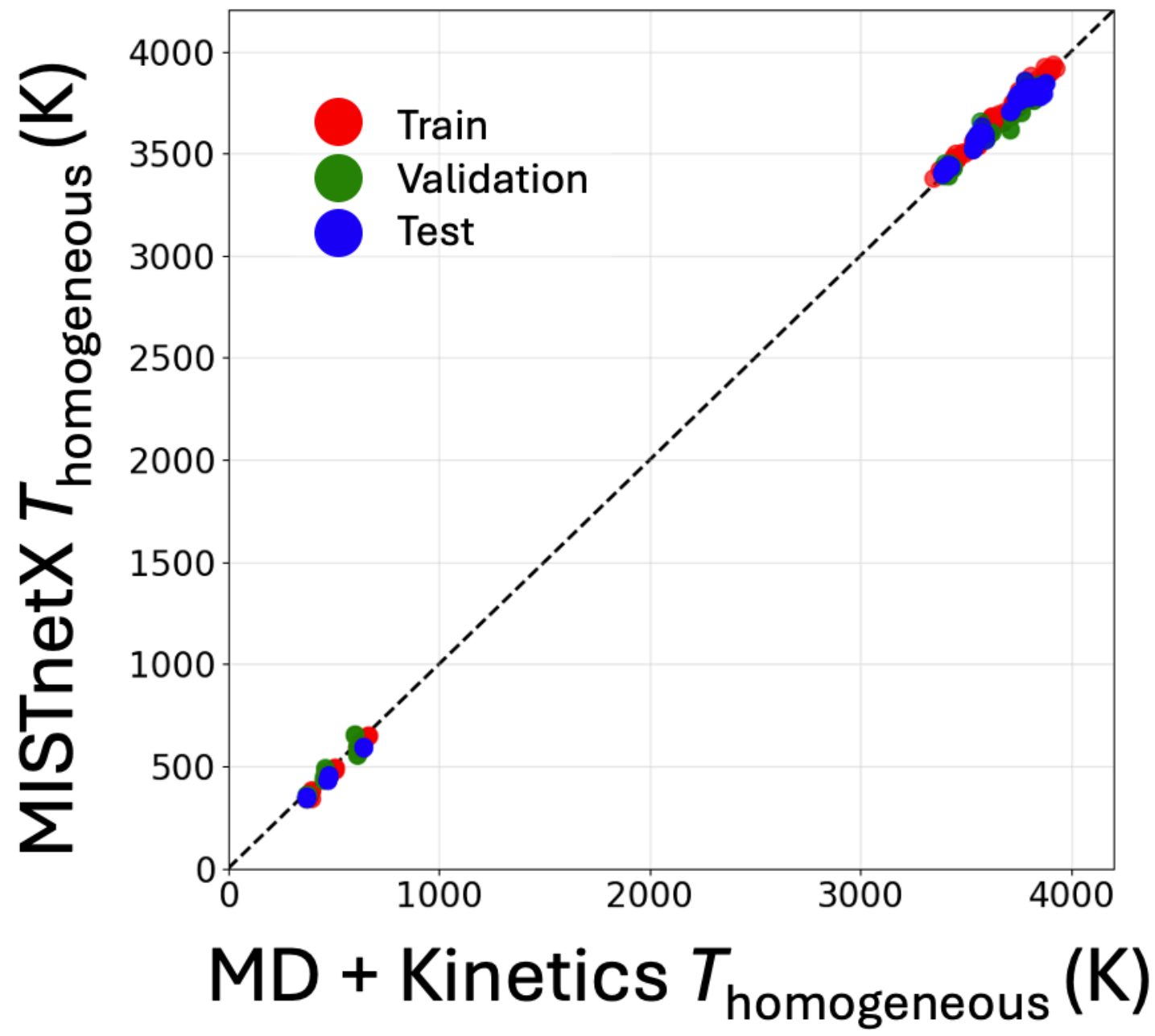

**Fig. S4. Parity plot of predicted and MD evolved homogeneous temperatures.** Captured MD post-shock temperature fields thermal and chemically evolved to a homogeneous state (x-axis) compared to MISTnetX homogeneous temperature predictions (y-axis). The training set is denoted by red circles, validation by green, and test set by blue.

## S2. FE Model

S2.1 Continuum Model Formulation

We model the shock response of the PBX using the Jones-Willkins-Lee (JWL) equation of state (EOS), parameterized from reactive MD simulations for unreacted and the final products of RDX (*28*). The model describes plasticity in the PBX system with a $J_2$ yield function and a Johnson-Cook model for the yield stress that displays kinematic hardening, rate dependency, and a modified thermal softening rule given by a pressure dependent melt model (*44*). The thermal decomposition kinetics are described using the reduced order chemical model from Sakano et al. (*24*) used for MISTnetX. The two-reaction, three-component model was obtained from reactive MD simulations using non-negative matrix factorization for dimensionality reduction. The evolution of temperature plays a crucial role in the coupling of the sub-grid MISTnetX model, not only because it predicts a temperature field, but also because it provides closure to the system of equations during the takeover stage, when MISTnetX describes sub-grid phenomena, accounting for the thermal contribution of plastic dissipation, work of pressure, and chemical reactions (*41*), see equation (S.16).

To account for the contribution of the mechanical properties of the binder to the overall response, the binder is modeled as a hyper-elastic material and mechanical equilibrium of each element is enforced by a rule of mixture that uses the initial fraction of RDX as an interpolation rule to obtain the bulk averaged response of each element.

S2.2 Continuum Model Implementation

We describe the coupled system of equations, and the constitutive models assigned to the condensed and gas phases for their mechanical, thermal, and chemical representation. The chemical reactions, conservation of mass, momentum, and energy equations are solved using the finite element MOOSE framework (*43*).

S2.3 Kinetics

The kinetics model used for the decomposition of RDX assumes a two-step, first order reaction. This model assumes the existence of reactants ($Y_1$), intermediates ($Y_2$), and final products ($Y_3$) with a temperature-only Arrhenius reaction rate for each sequential transition

$$\begin{cases} \dot{Y}_1 = -r_1 Y_1 \\ \dot{Y}_2 = r_1 Y_1 - r_2 Y_2 \\ \dot{Y}_3 = r_2 Y_2 \end{cases} \tag{S.1}$$

where each reaction rate is modeled by $r_i = Z_i e^{-E_i/RT}$, $i = 1,2,3$ for each component. The heat of reaction for each stage of the decomposition is temperature dependent and fitted to MD data.

S2.4 Mechanical Response

We decompose the deformation gradient, $\boldsymbol{F}$, into elastic, $\boldsymbol{F}^e$, and plastic parts, $\boldsymbol{F}^p$ (*44*)

$$\boldsymbol{F} = F^e \boldsymbol{F}^p \tag{S.2}$$

and enforce the conservation of linear momentum through the conservative equation

$$\rho_0 \frac{\partial^2 \boldsymbol{u}}{\partial t^2} = \boldsymbol{\nabla} \cdot \boldsymbol{P} \tag{S.3}$$

where $\boldsymbol{u}$ is the displacement vector, $\mathbf{P}$ is the First Piola-Kirchhoff stress tensor, and $\rho_0$ is the mass density in the reference configuration. The reference mass density is estimated with a rule of mixture based on the volume fractions of the reactants and voids, see System 1. The First Piola-Kirchhoff stress tensor is obtained with the relation

$$P = J \sigma F^{-T} \tag{S.4}$$

where $\boldsymbol{\sigma}$ is the Cauchy stress and $J = \det \boldsymbol{F}$. The Cauchy stress is split into volumetric and deviatoric parts

$$\sigma_{ij} = -p\delta_{ij} + s_{ij} \tag{S.5}$$

The volumetric part of the stress follows a Jones-Willkins-Lee (JWL) EOS of the form

$$p = Ae^{-R_1 J} + Be^{-R_2 J} + \frac{\omega \rho_0 c_v T}{J} \tag{S.6}$$

where $A$, $B$, $R_1$ and $R_2$ are parameters fitted from reactive MD data for the reactants and products, $\omega$ is the Grüneisen parameter, $c_v$ is the specific heat capacity at constant volume, and $T$ is the local temperature (*45*). The values of the parameters for reactants and products are listed in Table S3.

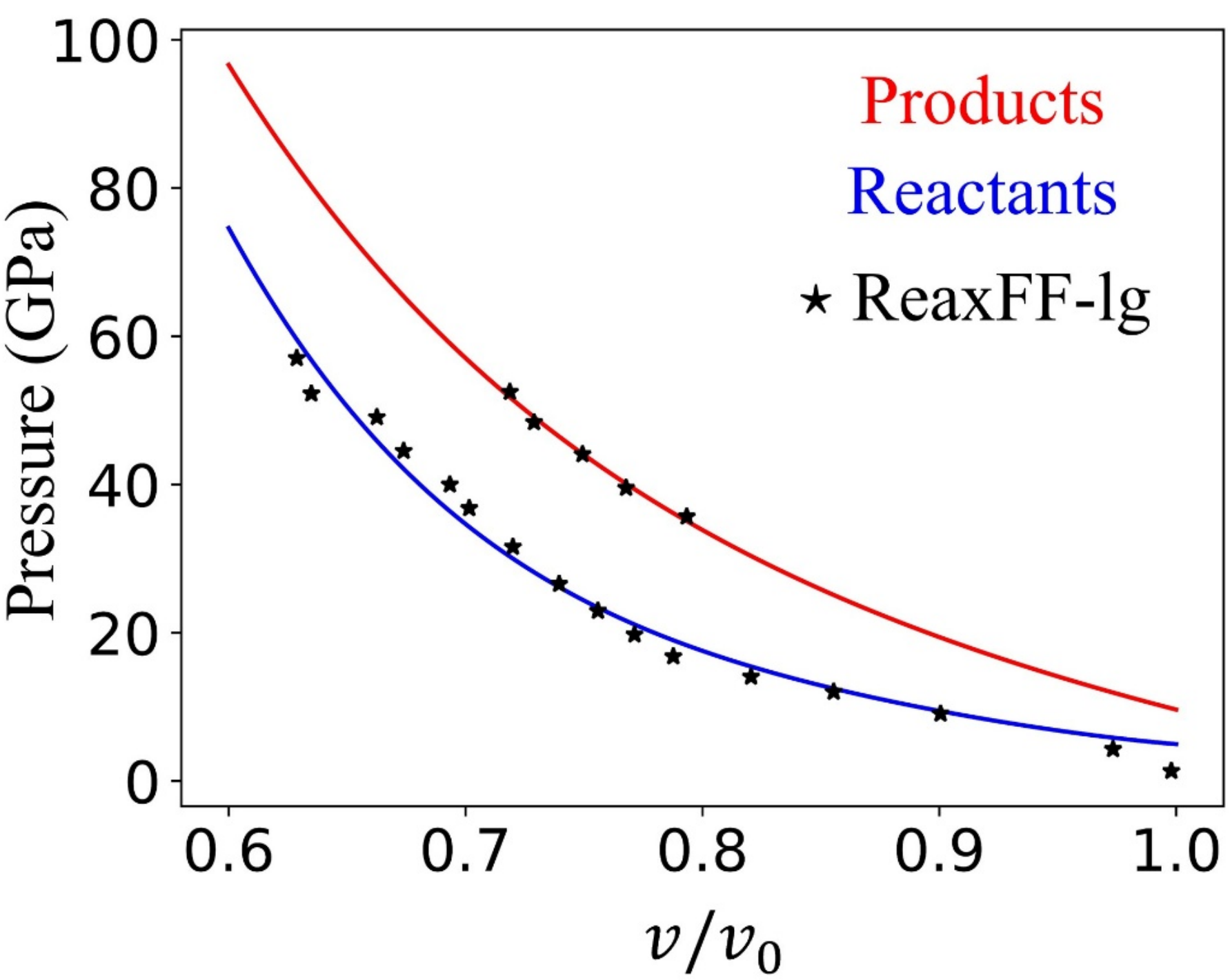

**Fig. S5. Fitted JWL EOS from reactive MD data.** The green line represents Rayleigh path from the reference state, through the Von Neumann spike, and finally back to the CJ state during detonation.

| JWL Parameter | Reactants EOS | Products EOS |
|---|---|---|
| $A$ [GPa] | 13398.42 | 1000 |
| $B$ [GPa] | 3810.37 | 2738.02 |
| $R_1$ [-] | 68.12 | 60 |
| $R_2$ [-] | 6.81 | 6 |
| $\omega$ [-] | 0.37 | 0.77 |

**Table S3. EOS Parameters.** Parameters obtained are for the JWL functional form using data from Figure S5.

Additionally, we introduce a second order artificial viscosity pseudo-pressure to enforce a finite shock width for numerical stability (*46*)

$$p_{av} = C_0 \rho_0 \frac{\dot{J}|\dot{J}|}{j^2} h^2 + C_1 \rho_0 c_0 \frac{\dot{J}}{J} h \tag{S.7}$$

where $C_0 = 0.1$ and $C_1 = 1.0$ are the von-Neumann and Landshoff parameters that control the numerical width of the shock front, $\rho_0$ is the mass density of the material, $\dot{J}$ is the time derivative of the Jacobian, $h$ is the element size, and $c_0 = \sqrt{K_0/\rho_0}$ is the longitudinal sound speed of the unshocked material along the shock propagation direction.

We append this pressure term as a volumetric contribution to the stress tensor, getting the final expression for the pressure as the interpolation between the reactants and products EOS, with the addition of the artificial viscosity terms as:

$$p = Y_1 p_r + (1 - Y_1) p_p + p_{av} \tag{S.8}$$

where $p_r$ is the EOS pressure contribution from reactants with a mass fraction $Y_1$, and $p_p$ is the pressure contribution from the products with a mass fraction $1 - Y_1 = Y_2 + Y_3$ via conservation of mass.

For the plastic part of kinematics, we consider a $J_2$ plasticity model with a Johnson-Cool yield stress formulation of the form (*42*)

$$\sigma^Y = (A_Y + B_Y(\epsilon^p)^n)\left(1 + C_Y \log\frac{\dot{\epsilon}^p}{\dot{\epsilon}_0^p}\right)(1 - \theta^m) \tag{S.9}$$

where $A_Y$ represents the initial yield stress, $B_Y$ is a hardening modulus, $\epsilon^p$ is the effective plastic strain, $n$ is a hardening exponent, $C_Y$ controls the contribution of plastic strain into hardening, $\dot{\epsilon}^p$ is the effective plastic strain rate, $\dot{\epsilon}_0^p$ is a reference parameter for scaling of the plastic strain rate, $\theta$ is a normalized temperature measure to account for thermal softening, and $m$ is the sensitivity exponent to thermal softening of the material.

To compute the normalized temperature parameter $\theta$, we use a Simon-Glatzel melting function of the form (*47*)

$$T_{melt} = T_{0,m}\left(1 + \frac{P - P_0}{a}\right)^{1/c} \tag{S.10}$$

which is then employed in

$$\theta = \max_{T > T_0}\left(\frac{T - T_0}{T_{melt} - T_0}, 0.9\right) \tag{S.11}$$

where the upper limit in $\theta$ is enforced so that the local yield stress is strictly positive, required for numerical consistency.

The deviatoric component of the stress tensor is computed using a linear relation of the form

$$\boldsymbol{s} = Y_1 G_{mix} \boldsymbol{b}^{e'} \tag{S.12}$$

where $G_{mix}$ is the mixture shear modulus interpolating mechanical properties of RDX and Polystyrene using a rule of mixtures of the form $G_{mix} = Y_1 G_{RDX} + (1 - Y_1) G_{PS}$, and with $\boldsymbol{b}^{e'}$ defined as the deviatoric component of the left Cauchy-Green deformation tensor. This model assumes that only the condensed phase material ($Y_1$) is allowed to withstand deviatoric stresses. Thus, the multiplication by the mass fraction of reactants allows to capture the loss of shear stiffness as the material decomposes during deflagration.

| **Parameter** | **Value** |
|---|---|

| $A_Y$ [GPa] | 0.3 |
|---|---|
| $B_Y$ [GPa] | 0.1 |
| $n$ [-] | 0.1 |
| $C_Y$ [-] | 1 |
| $\dot{\epsilon}_0^p$ [1/ns] | $10^3$ |
| $m$ [-] | 3 |
| $K_0$ [GPa] | 20 |
| $\rho_0$ [g/cm$^3$] | 1.45-1.50 |

**Table S4. Material parameters used in the numerical model.** The mass density values are within the shown range as different mass fractions of RDX and polystyrene are present at the reference configuration.

The yield stress is used to form the yield surface function

$$f = s^{eff} - G_{mix}\epsilon^p tr(\boldsymbol{b}^{e'}) - \sigma^Y < 0 \tag{S.13}$$

where $s^{eff} = \sqrt{(3/2)}\sqrt{(\boldsymbol{s}:\boldsymbol{s})}$ is the effective scalar deviatoric stress, and $\epsilon^p$ is the scalar accumulated plastic strain. The accumulated plastic strain is solved implicitly from this equation so that it satisfies the $f < 0$ condition for the stress state to be located at the yield surface. The increment in accumulated plastic strain at the current configuration is therefore obtained implicitly from equation S12. The plastic deformation gradient is evolved incrementally over time using the relations

$$\boldsymbol{b}^e = \boldsymbol{b}^e_{trial} - \frac{2}{3}\Delta\epsilon^p tr(\boldsymbol{b}^e_{trial})\sqrt{\frac{3}{2}}\frac{\boldsymbol{s}}{|\boldsymbol{s}|} \tag{S.14}$$

$$\boldsymbol{C}^{p^{-1}} = \overline{\boldsymbol{F}}^{-1}\boldsymbol{b}^e\overline{\boldsymbol{F}}^{-T} \tag{S.15}$$

where $\boldsymbol{b}^e_{trial}$ is the assumed-elastic left Cauchy-Green deformation tensor, $\boldsymbol{C}^p = \boldsymbol{F}^{p^T}\boldsymbol{F}^p$ is the plastic right Cauchy-Green deformation tensor, and $\overline{\boldsymbol{F}} = J^{-1/3}\boldsymbol{F}$ is the volume-preserving deformation gradient satisfying $\det\overline{\boldsymbol{F}} = 1$. The plastic deformation gradient is then obtained from equation S14 by applying a spectral decomposition of $\boldsymbol{C}^p = \boldsymbol{PDP}^T$, taking advantage of the symmetry of $\boldsymbol{C}^p$. Therefore $\boldsymbol{F}^p = \boldsymbol{P}\sqrt{\boldsymbol{D}}\boldsymbol{P}^T$, where the columns of $\boldsymbol{P}$ contain the eigenvectors of $\boldsymbol{C}^p$ and $\boldsymbol{D}$ contains the respective eigenvalues along the principal diagonal. Finally, the elastic deformation gradient is obtained from the direct expression $\boldsymbol{F}^e = \boldsymbol{F}\boldsymbol{F}^{p^{-1}}$.

S2.5 Thermal Transport

The fully coupled heat equation for the local time evolution of temperature follows (*48*)

$$\rho_0 c_v \frac{\partial T}{\partial t} = k\nabla^2 T - T\frac{\partial p}{\partial T}C^{-1}:\dot{E}^e + \beta^{av}S^{av}:\dot{E}^e + \beta^p S:\dot{E}^p + \sum_i Q_i Y_i \tag{S.16}$$

where the right-hand side terms represent, in order, thermal diffusion, thermomechanical heating due to change in volume, heating due to the artificial viscosity localized at the shock front, heating due to plastic dissipation, and energy release due to the thermal decomposition of each species. $\boldsymbol{S} = J\boldsymbol{F}^{-1}\boldsymbol{\sigma}\boldsymbol{F}^{-T}$ is the Second Piola-Kirchhoff stress tensor, and $\dot{\boldsymbol{E}} = \frac{1}{2}\left(\dot{\boldsymbol{F}}^{T}\boldsymbol{F} + \boldsymbol{F}^{T}\dot{\boldsymbol{F}}\right)$ is the Total Lagrangian strain rate tensor.

The heating due to compression accounts for the mixture equilibrium between reactants and products, and takes the form

$$\frac{\partial p}{\partial T} = Y_1 \frac{\partial p_r}{\partial T} + (1 - Y_1)\frac{\partial p_p}{\partial T} \tag{S.17}$$

which interpolates the temperature dependent contributions of the pressure response of the material through the equation of state for both phases (reactants and products), using the mass fraction of the condensed phase $Y_1$. We assume that the intermediate products exist for a short enough period during the transition $Y_1 \rightarrow Y_2 \rightarrow Y_3$ that their contribution to pressure and compressive heating can be lumped into the contribution of the gas phase.

S2.6 Takeover Stage

At the FE level, we use the local MISTnetX predictions of post-shock and homogenous temperatures and the associated timescales to inform two subsequent heat sources. The development of shock temperature is assigned a timescale $\tau_{shock} = h/u_s$ where $h$ is the finite element size of 1.9 µm. The characteristic time assigned to the temperature homogenization, $\tau_{hom}$, is a function of the input velocity $u_p$ and whether the system quenches and deflagrates. As described in Section MISTnetX, we predicted the temperature field following shock loading of a nano-PBX systems using MISTnet2 and simulated the thermochemical evolution of the temperature fields to compute the average time required to achieve a homogeneous temperature and the associated standard deviation for $u_p$ between 1.5 km/s to 4.9 km/s in 0.1 km/s increments.

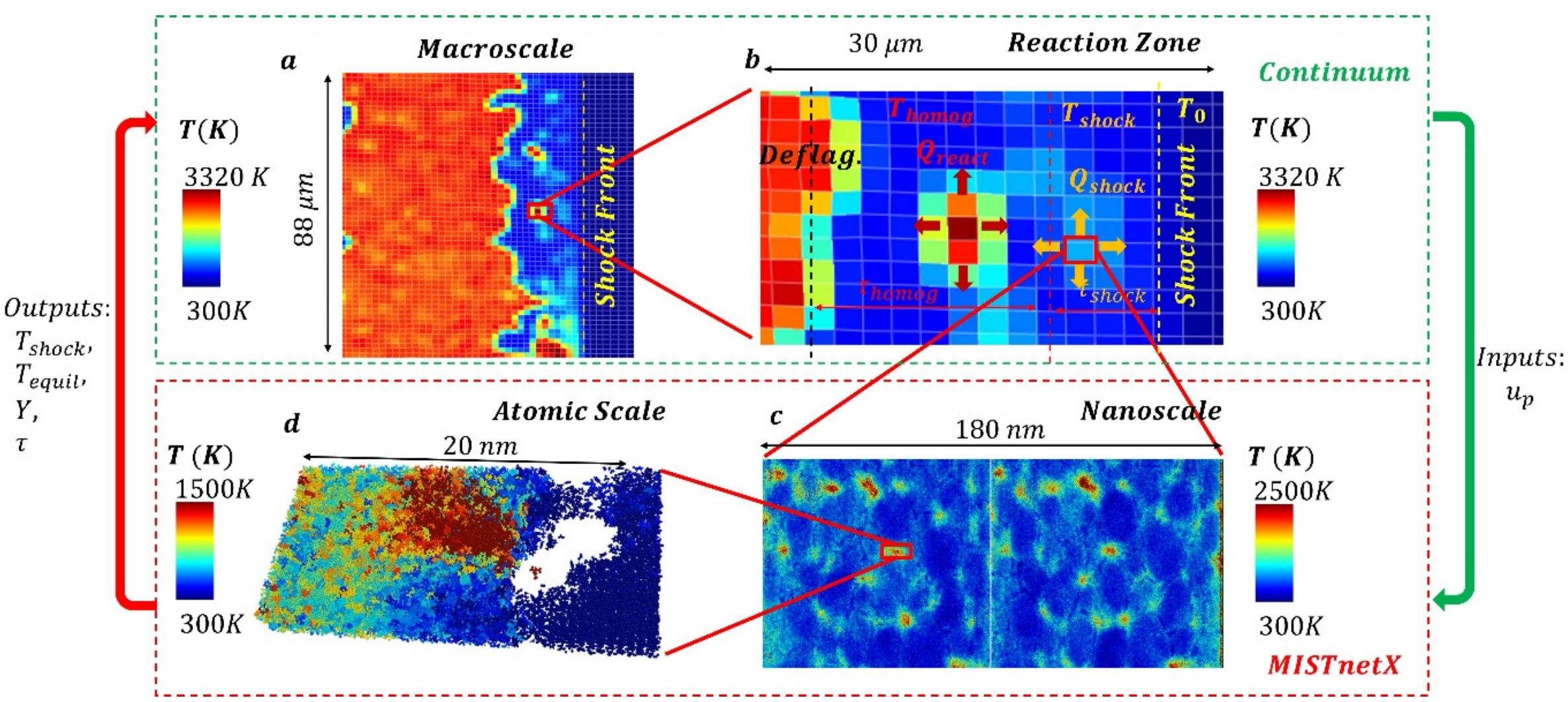

**Fig. S6. Multiscale connection between continuum model and atomistic data.** The connection shows how the continuum level FE mesh communicates with the atomic-scale model to retrieve the post-shock and homogeneous temperatures.

During the takeover stage, surrogate heat sources replace the mechanical and chemical contributions from the fully continuum model in equation (S14). These heat sources are informed by MISTnetX post-shock and equilibrium states for each voxel. During the takeover stage, the heat equation is effectively

$$\rho_0 c_v \frac{\partial T}{\partial t} = k\nabla^2 T + Q_{shock} + Q_{hom} \tag{S.18}$$

with

$$Q_{shock} = \left(\frac{1}{\tau_{shock}}\right) \rho_0 c_v (T_{shock} - T_0) \tag{S.19}$$

$$Q_{hom} = \left(\frac{1}{\tau_{hom}}\right) \rho_0 c_v (T_{hom} - T_{shock}) \tag{S.20}$$

where $\tau_{shock} = h/u_s$ is the time for the establishment of the post-shock state and is equal to the time it takes the shock front to travel an element size distance of $h$ at a shock front velocity $u_s$, and $\tau_{hom}$ is the time it takes for the temperature inside each microstructure to stabilize both spatially and over time. The homogenization time $\tau_{hom}$ is shown for different MISTnetX predictions as a function of $u_p$ for several deflagrated and quenched microstructures in Figure S7.

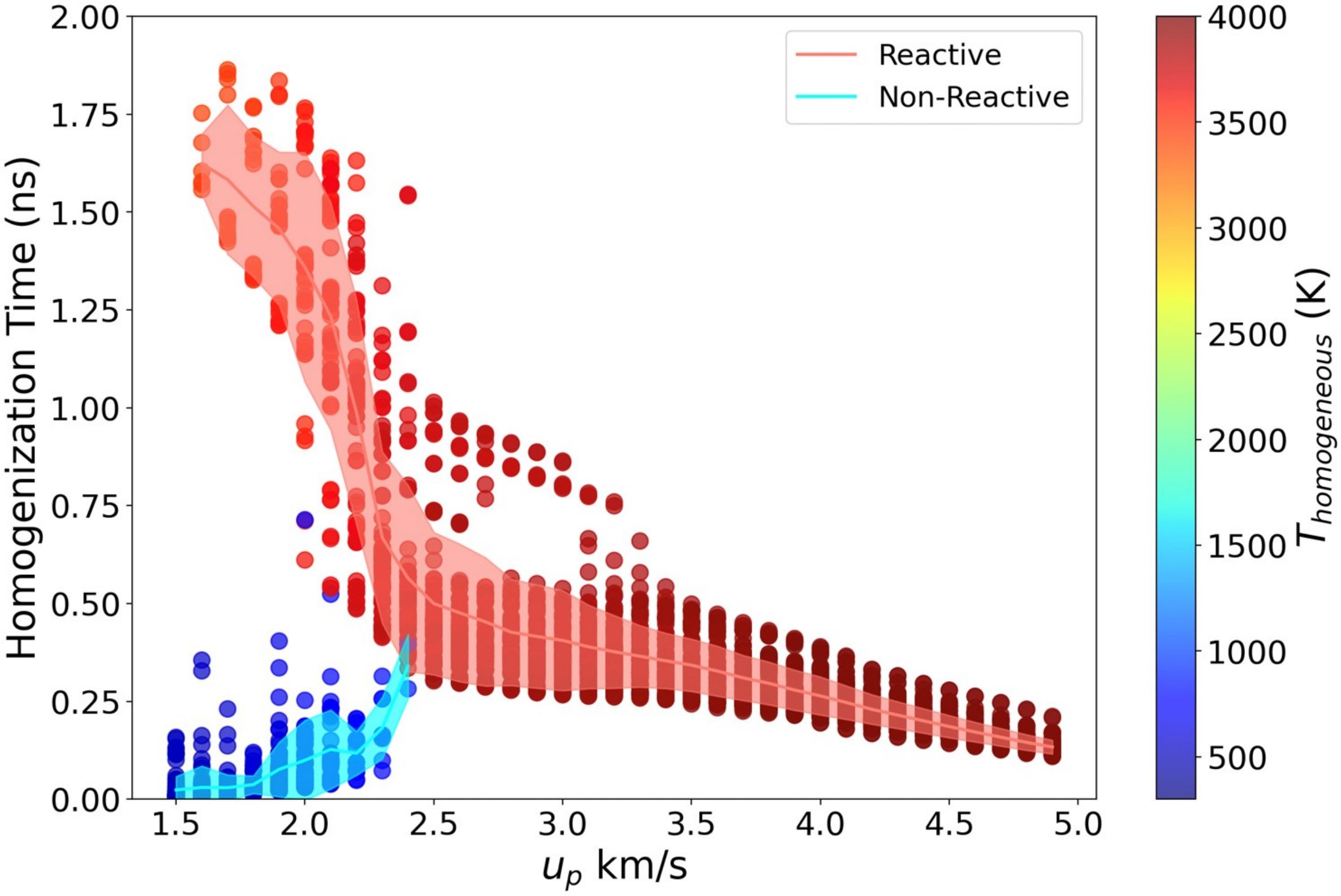

**Fig. S7. Homogenization time as a function of the particle velocity.** The solid lines represent the mean value of the distribution, and the filling represents the first standard deviation region.

Before the takeover, the chemical composition of each element is that of unreacted PBX ($Y_1 = Y_{RDX}$, $Y_2 = Y_3 = 0$), with $Y_{RDX}$ being the initial mass fraction of RDX present on voxel. During the takeover, the reaction kinetics is replaced by the surrogate rates shown in Table S5. Their final states after the takeover can only be either a fully deflagrated state or a quenched state, as shown in Table S5.

| Case | Surrogate Rates | Final State |
|---|---|---|
| Deflagration: $T_{react} \gg T_{shock}$ | $\dot{Y}_1 = -Y_3/\tau_{hom}$<br>$\dot{Y}_2 = 0$<br>$\dot{Y}_3 = Y_3/\tau_{hom}$ | $Y_1 = 0$<br>$Y_2 = 0$<br>$Y_3 = 1$ |
| Quench: $T_{react} \approx T_{shock}$ | $\dot{Y}_1 = 0$<br>$\dot{Y}_2 = 0$<br>$\dot{Y}_3 = 0$ | $Y_1 = 1$<br>$Y_2 = 0$<br>$Y_3 = 0$ |

**Table S5. Predicted final states after takeover stage.** The final states follow a binary distribution based on the possible outcomes predicted by MISTnetX.

S2.7 Determination of the shock and particle velocities

To evaluate MISTnetX the stable value of $u_p$ for each element at the shock front is needed. To estimate the local $u_p$, the shock smearing from artificial viscosity is used as a reference to estimate the stabilization time for the shock wave from the unshocked state ($u_p = 0$) to the local shocked state. Equation (6) introduces a second order artificial viscosity pressure of the form proposed by (*49*). This equation smears the shock front across a characteristic length, which is calibrated and controlled by the numerical parameters $C_0$ and $C_1$. The shock front width is parametrically determined from these parameters, and it remains constant at a value of $w \approx 3h$. Based on the constant shock width, we obtain an estimation for the time that it takes for the pressure profile to stabilize based on the time it takes for the shock front to travel a distance of $w = 3h$. This estimation is then used as a local measure of shock development to obtain the local $u_p$ after the shock has reached each element.

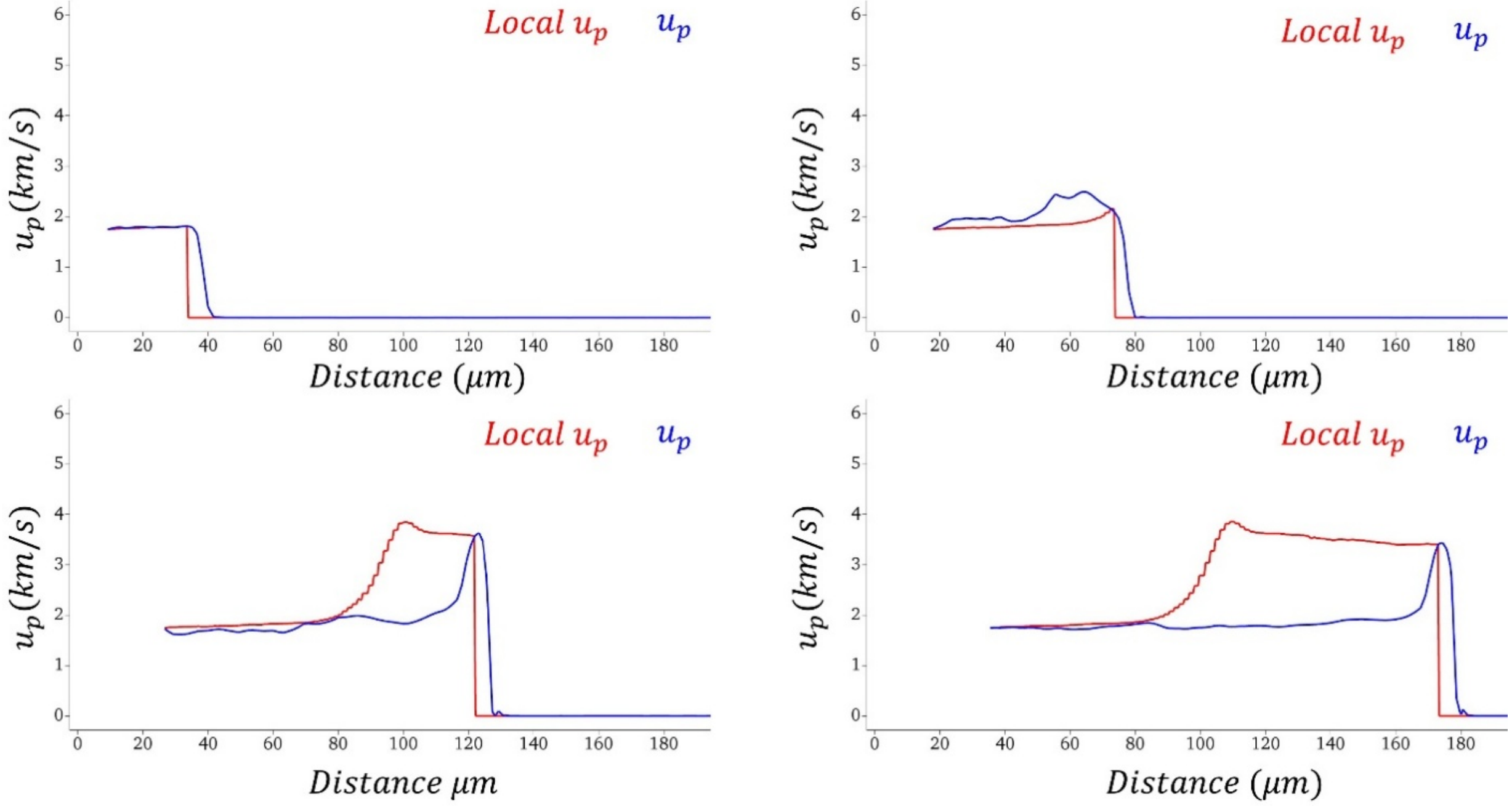

**Fig. S8. Local and actual particle velocity profiles.** Profiles at 5, 10, 15, and 20 ns after initial impact show the close agreement between the local particle velocity and the computed stable particle velocity used for MISTnetX.

To track the shock front, we define a local time metric $\tau_{track}$, which represents the time that has passed locally at each element after the shock front has reached it. From a Lagrangian perspective, the local $u_p$ is reached after a time corresponding to a travel distance $w = 3h$ has passed. Therefore, we define a tracking metric that evolves through the ODE

$$d\delta_{track} = u_s/3hd\tau \tag{S.21}$$

$$\delta_{track} = \int_0^{t_{call}} \frac{u_s}{3h} d\tau \tag{S.22}$$

Equation (S.22) is implicitly solved for $t_{call}$, which represents the time after the shock front arrival at which the local $u_p$ has stabilized. Once the stabilized local $u_p$ has been established, it is employed by equations (S.18), (S.19), and (S.20) to perform the takeover update. Once the takeover stage ends, the continuum constitutive equations (S2), (S17), and (S18) regain full control and continue to evolve the temperature, chemical composition, and coupled mechanical response.

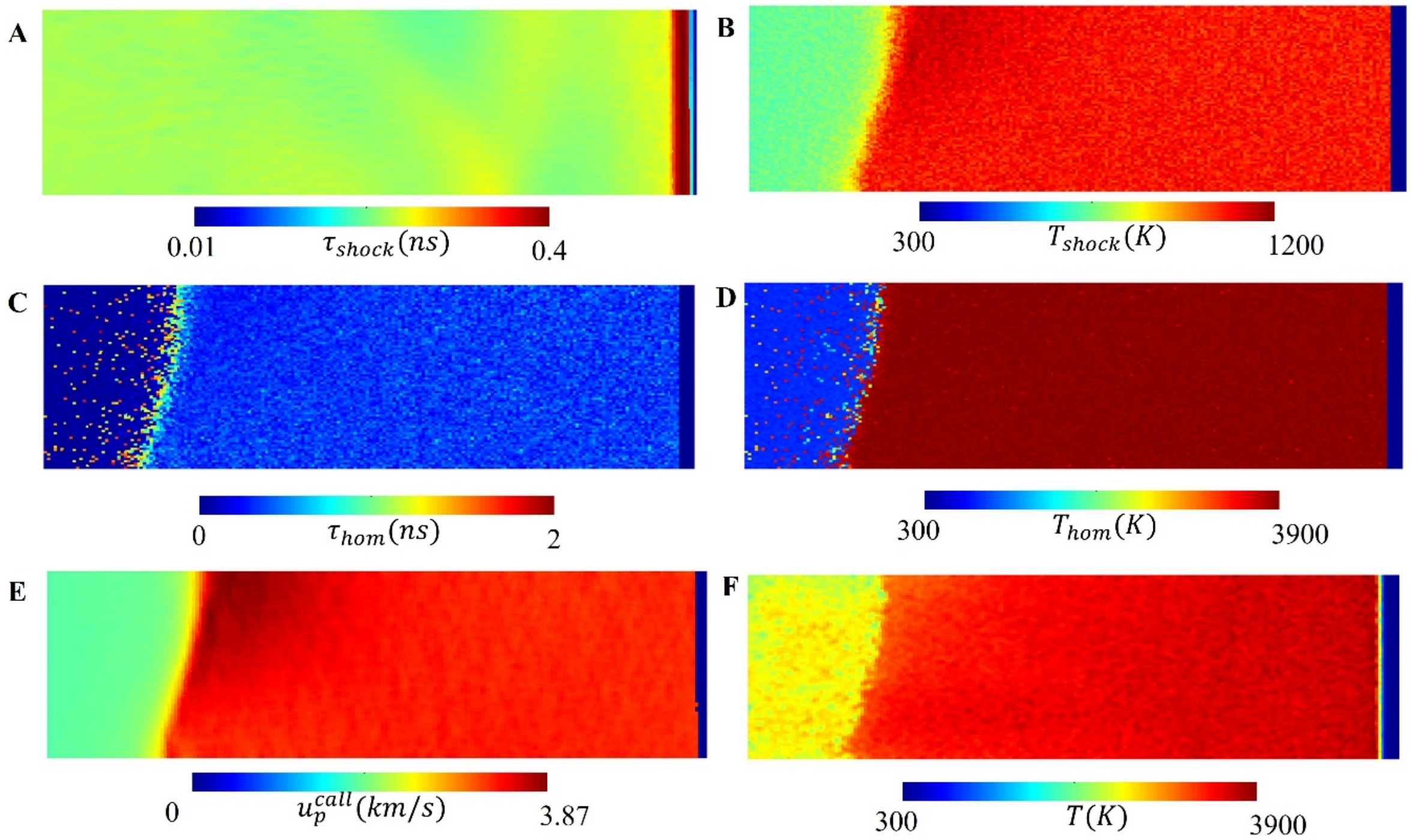

**Fig. S9. Snapshots of MISTnetX predictions during STDT.** (**A**) The shock heating time at later stages. (**B**) The accumulated history of post shock temperatures predicted during STDT. (**C**) The accumulated history of homogenization times predicted during STDT. (**D**) The accumulated history of predicted homogenization temperatures during STDT. (**E**) The accumulated history of call particle velocities used to retrieve the MISTnetX predictions. (**F**) The evolved continuum temperature field during detonation.

The post-shock and homogeneous temperatures increase for all microstructures as the local call $u_p$ increases. The number of deflagrating microstructures increases as well, and at a threshold particle velocity value conditioned by the underlying particular distribution of microstructures, all microstructures become critical. The effect of this strong threshold on the predicted temperatures as a function of $u_p$ is seen in the actual temperature profiles, where there is a sharp transition between the deflagration regime to the detonation regime at DDT.

S2.8 Determination of the Run to Detonation Distance

In order to determine the run to detonation distance, a position-time plot is constructed using data for the shock front location over time, and two separate lines are constructed using two points before and after the DDT transient. The intersection point between these two lines is determined using a bisection method, which represents the average location of the DDT point. Due to this method depending on the choice of points to form the lines before and after DDT, we study the distance obtained as a function of all possible data pairs for velocity before and after DDT. This variability is considered to estimate the uncertainty and error introduced by the datapoint selection. Shock velocity can deviate from constant behavior due to plastic dissipation and acceleration from deflagration interactions.

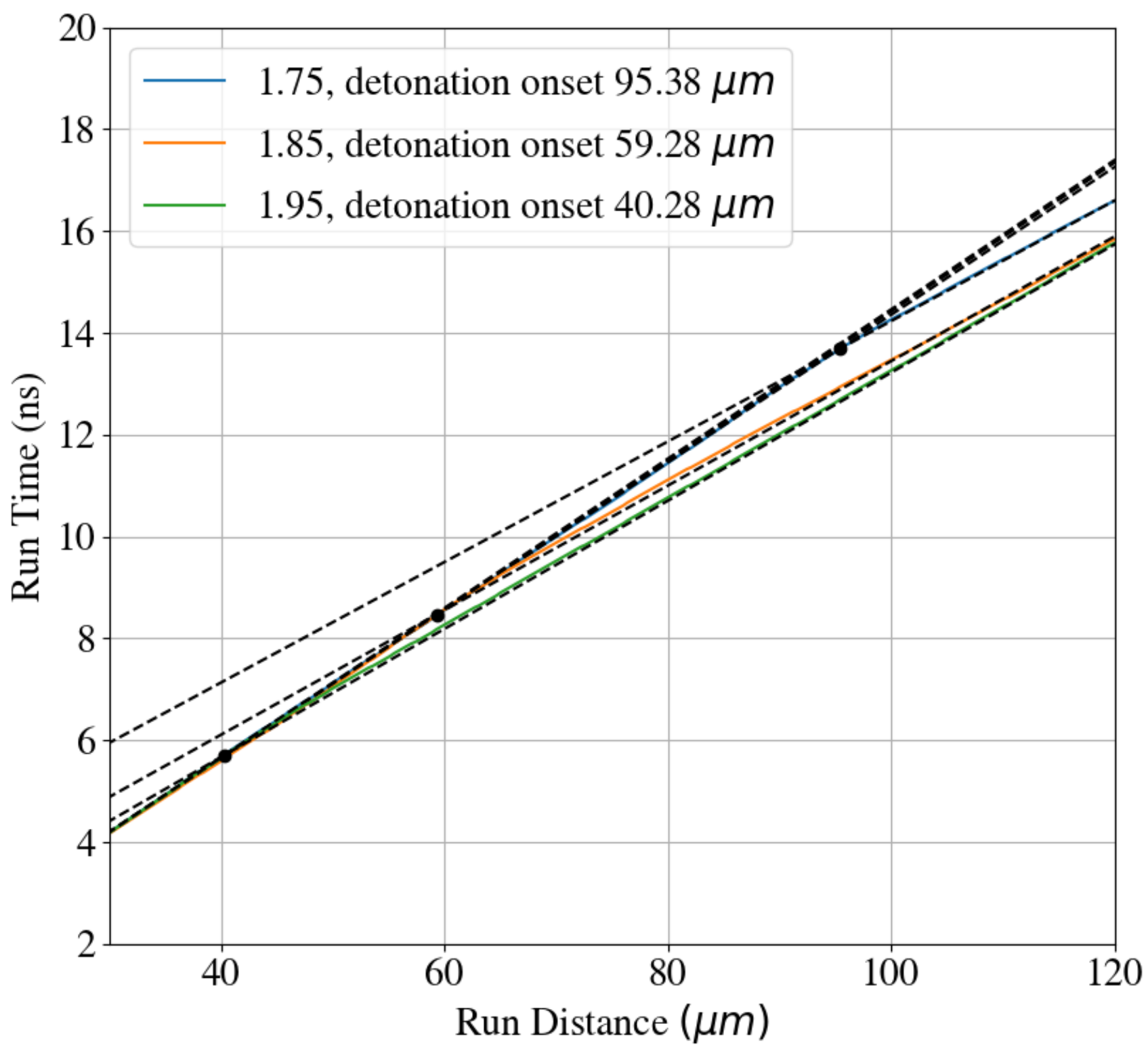

**Fig. S10. Position-time plot for the shock front location at different initial impact velocities.** The dashed lines represent the extension of the fitted first degree polynomial for the portions of each history before and after DDT.

## S3. Shock to Detonation Transition

Figure S11 shows snapshots of the key events in STDT in our nanocomposite system shocked with an impact velocity $u_p = 2$ km/s; pressure is shown as a transparent field and temperature as solid surface for volumes where final products exceed 95% of completion

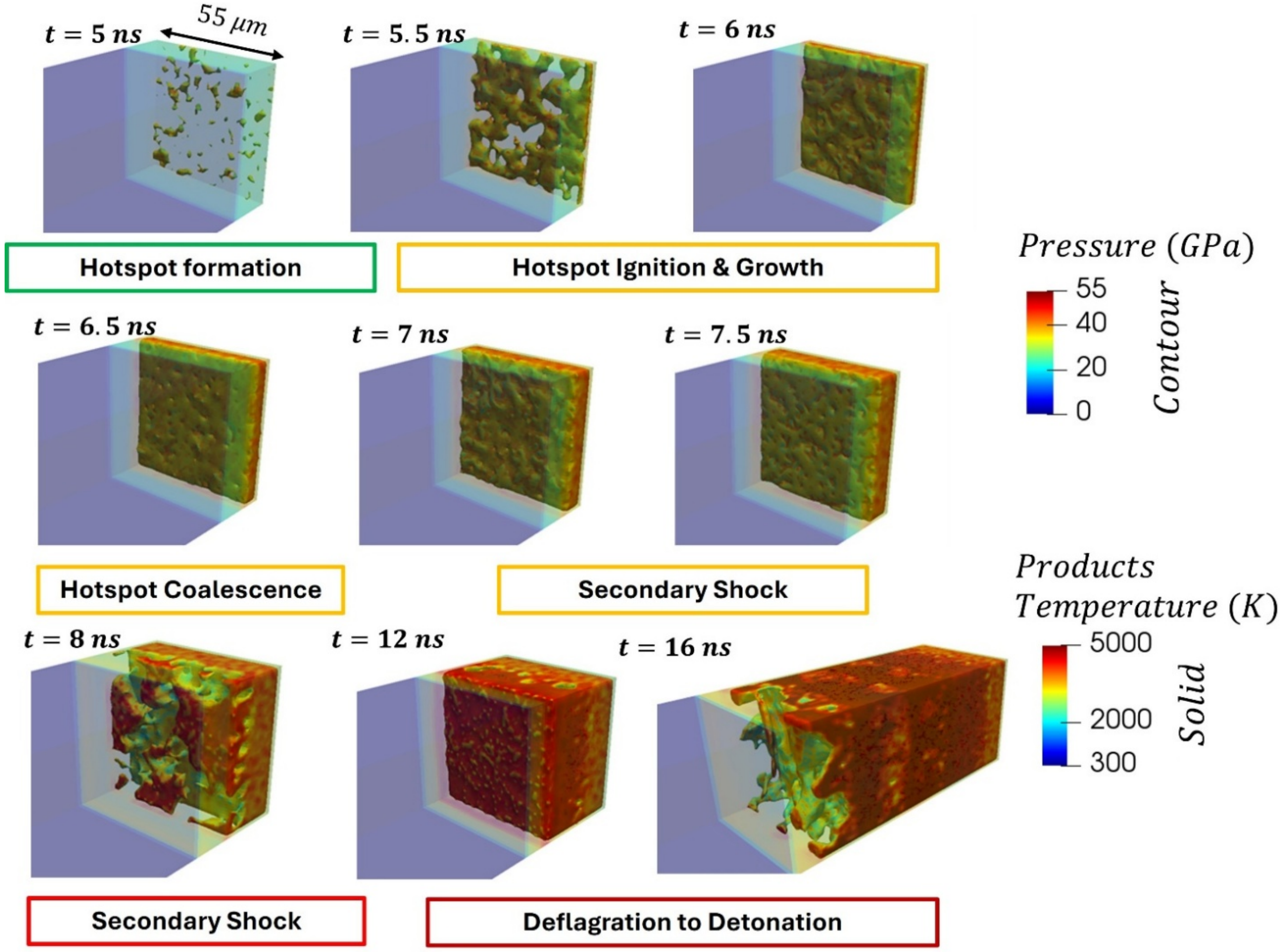

**Fig. S2. STDT for an initial impact of 2 km/s.** The same processes and succession of events are seen as the ones for the lower impact velocity cases.

## APPENDIX: MISTnetX Dataset

There are **7 PBX systems** shocked at a particle velocity of 1.0, 1.5, and 2.0 km/s using atomistic simulation. Four systems are used for training. Two systems are used for validation. One system is used for testing. Each system contains a top and bottom set.

System 1:

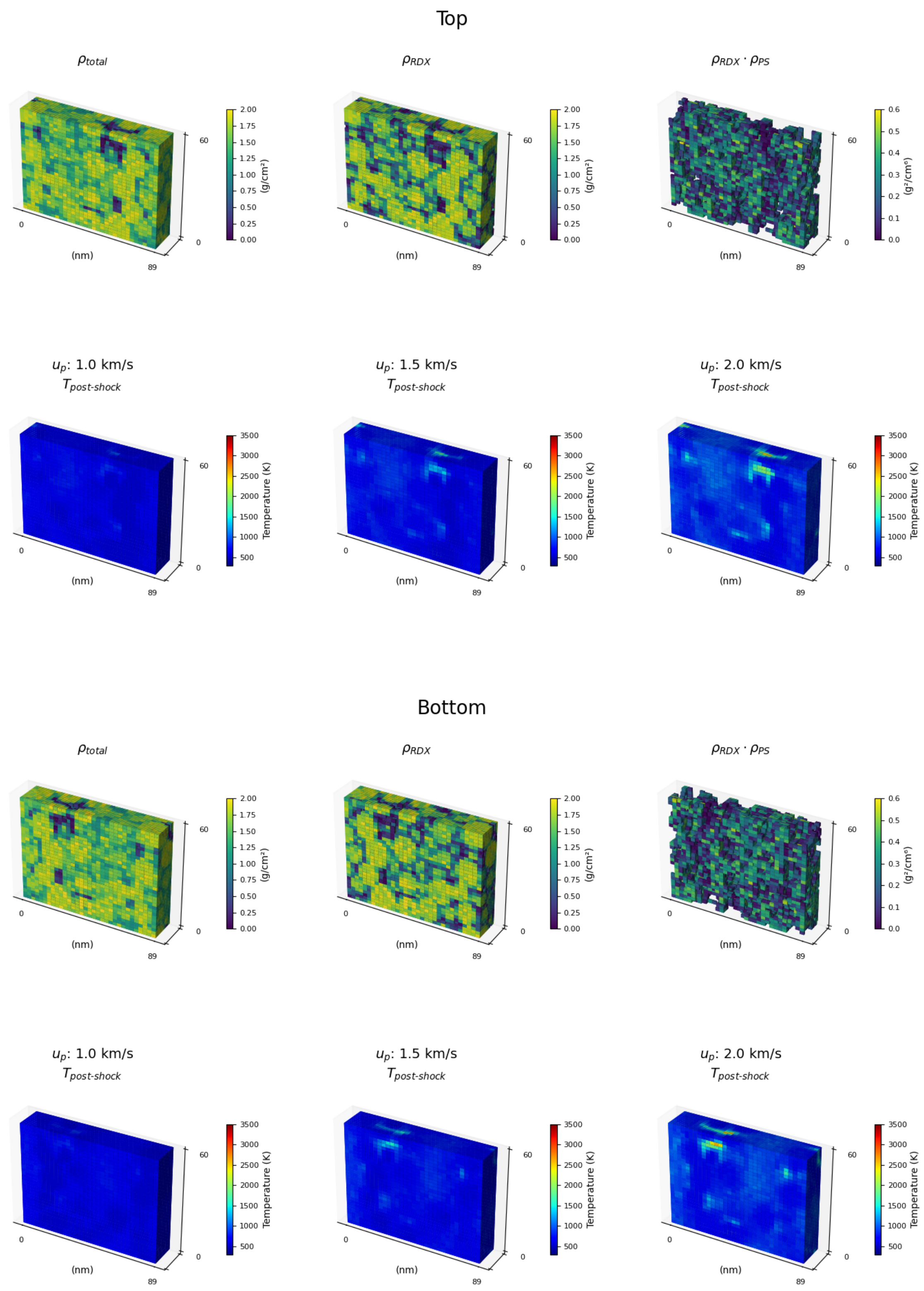

Top
$\rho_{total}$
$\rho_{RDX}$
$\rho_{RDX} \cdot \rho_{PS}$
(nm)
(g/cm³)
(g²/cm⁶)
$u_p$: 1.0 km/s
$T_{post\text{-}shock}$
$u_p$: 1.5 km/s
$u_p$: 2.0 km/s
Temperature (K)
Bottom

System 2:

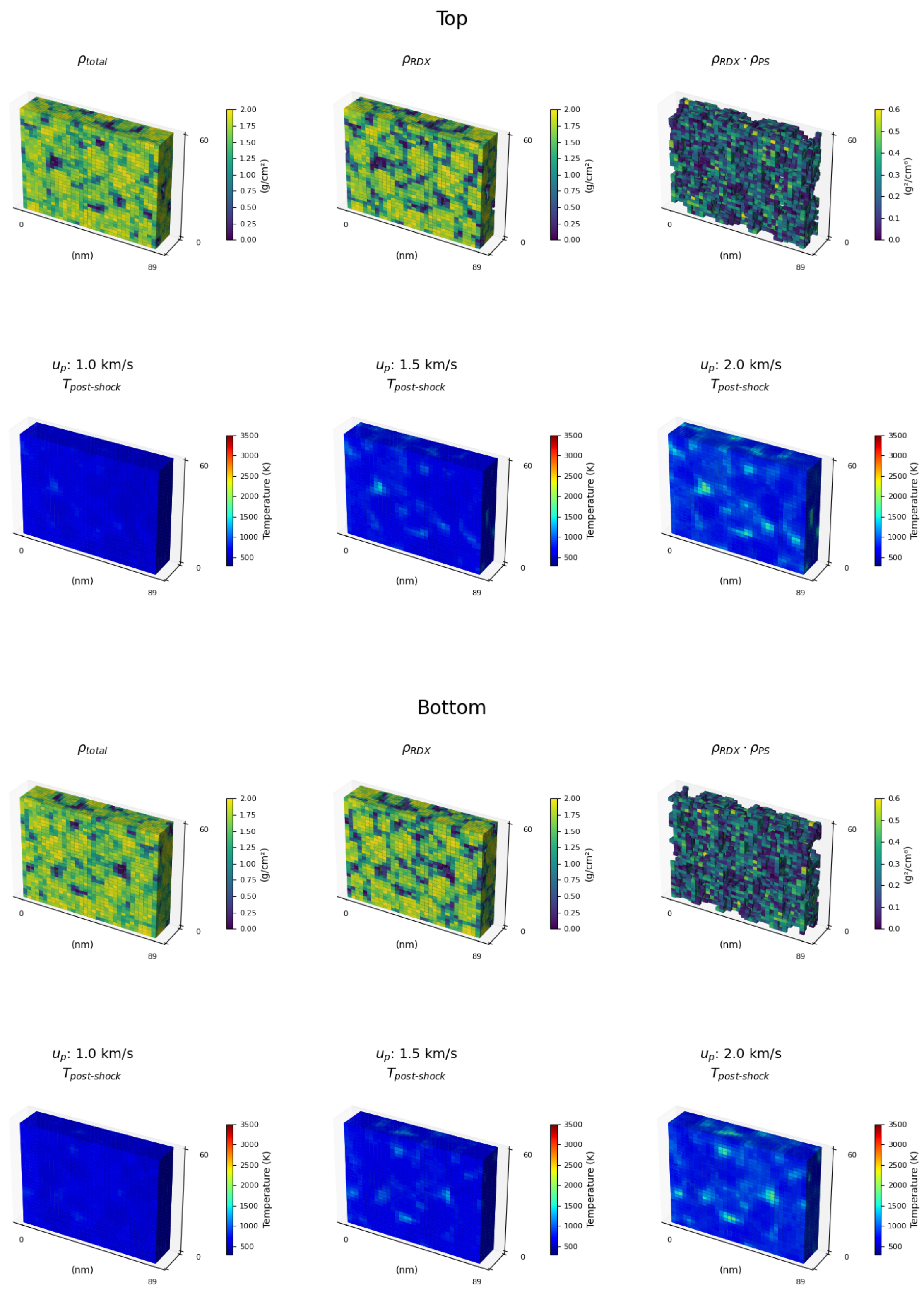
Top
$\rho_{total}$
$\rho_{RDX}$
$\rho_{RDX} \cdot \rho_{PS}$
$u_p$: 1.0 km/s
$T_{post\text{-}shock}$
$u_p$: 1.5 km/s
$T_{post\text{-}shock}$
$u_p$: 2.0 km/s
$T_{post\text{-}shock}$
Bottom
$\rho_{total}$
$\rho_{RDX}$
$\rho_{RDX} \cdot \rho_{PS}$
$u_p$: 1.0 km/s
$T_{post\text{-}shock}$
$u_p$: 1.5 km/s
$T_{post\text{-}shock}$
$u_p$: 2.0 km/s
$T_{post\text{-}shock}$
(nm)
Temperature (K)

System 3:

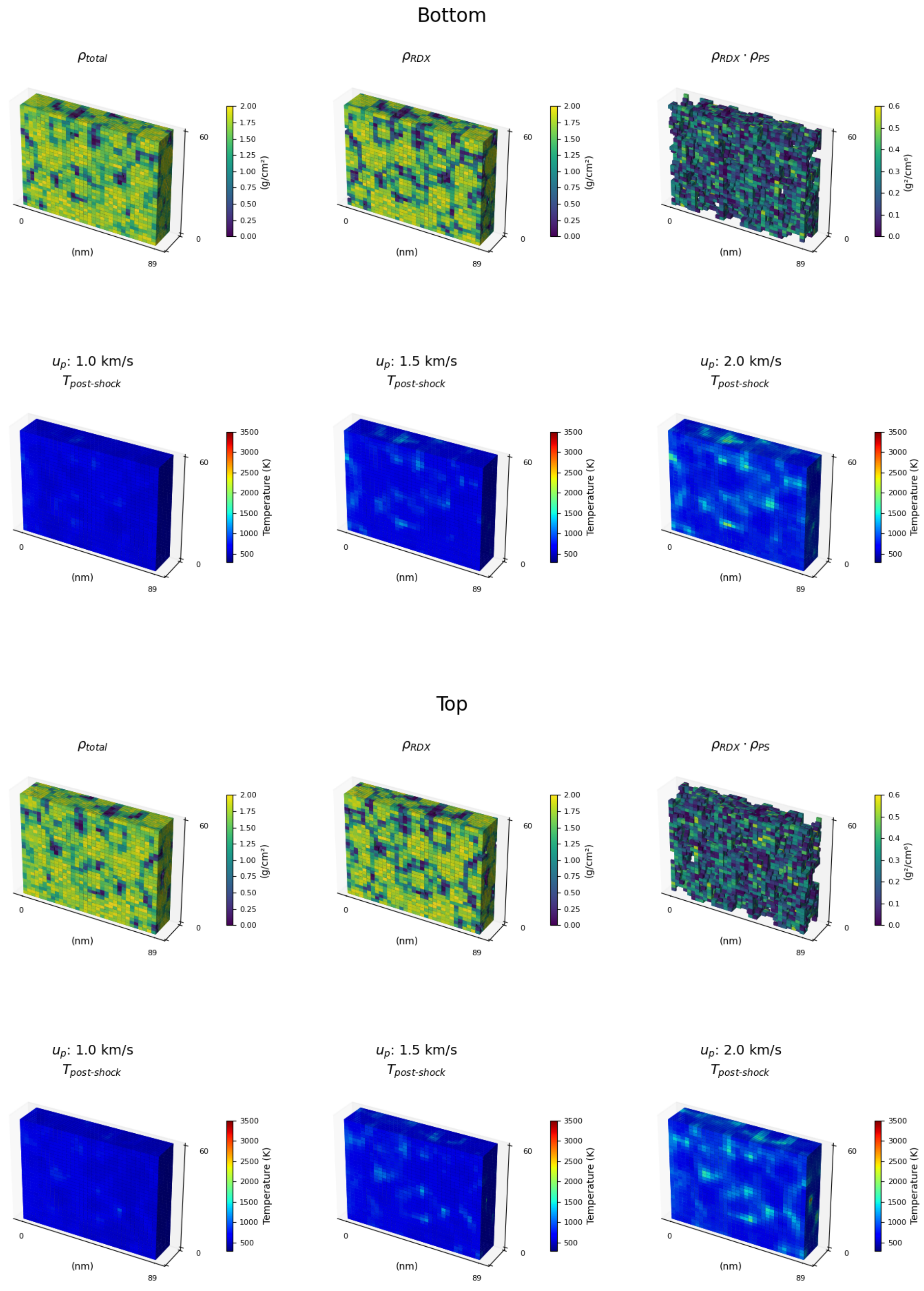

Bottom
ρtotal
ρRDX
ρRDX · ρPS
(g/cm³)
(g²/cm⁶)
(nm)
0
89
60
2.00
1.75
1.50
1.25
1.00
0.75
0.50
0.25
0.00
0.6
0.5
0.4
0.3
0.2
0.1
0.0
up: 1.0 km/s
up: 1.5 km/s
up: 2.0 km/s
Tpost-shock
Temperature (K)
3500
3000
2500
2000
1500
1000
500
Top

System 4:

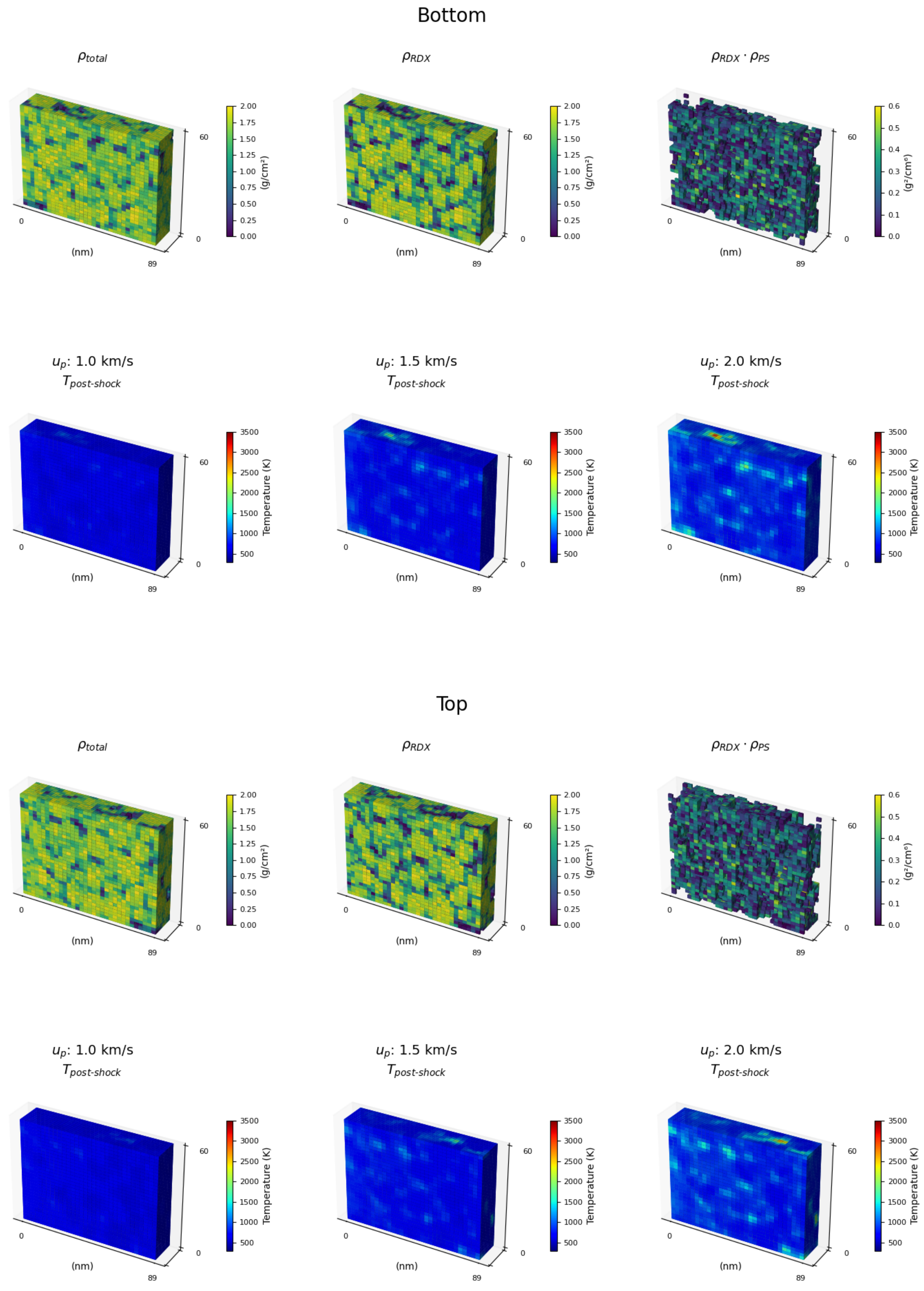

Bottom
$\rho_{total}$
$\rho_{RDX}$
$\rho_{RDX} \cdot \rho_{PS}$
(g/cm³)
(g²/cm⁶)
(nm)
$u_p$: 1.0 km/s
$u_p$: 1.5 km/s
$u_p$: 2.0 km/s
$T_{post\text{-}shock}$
Temperature (K)
Top

System 5:

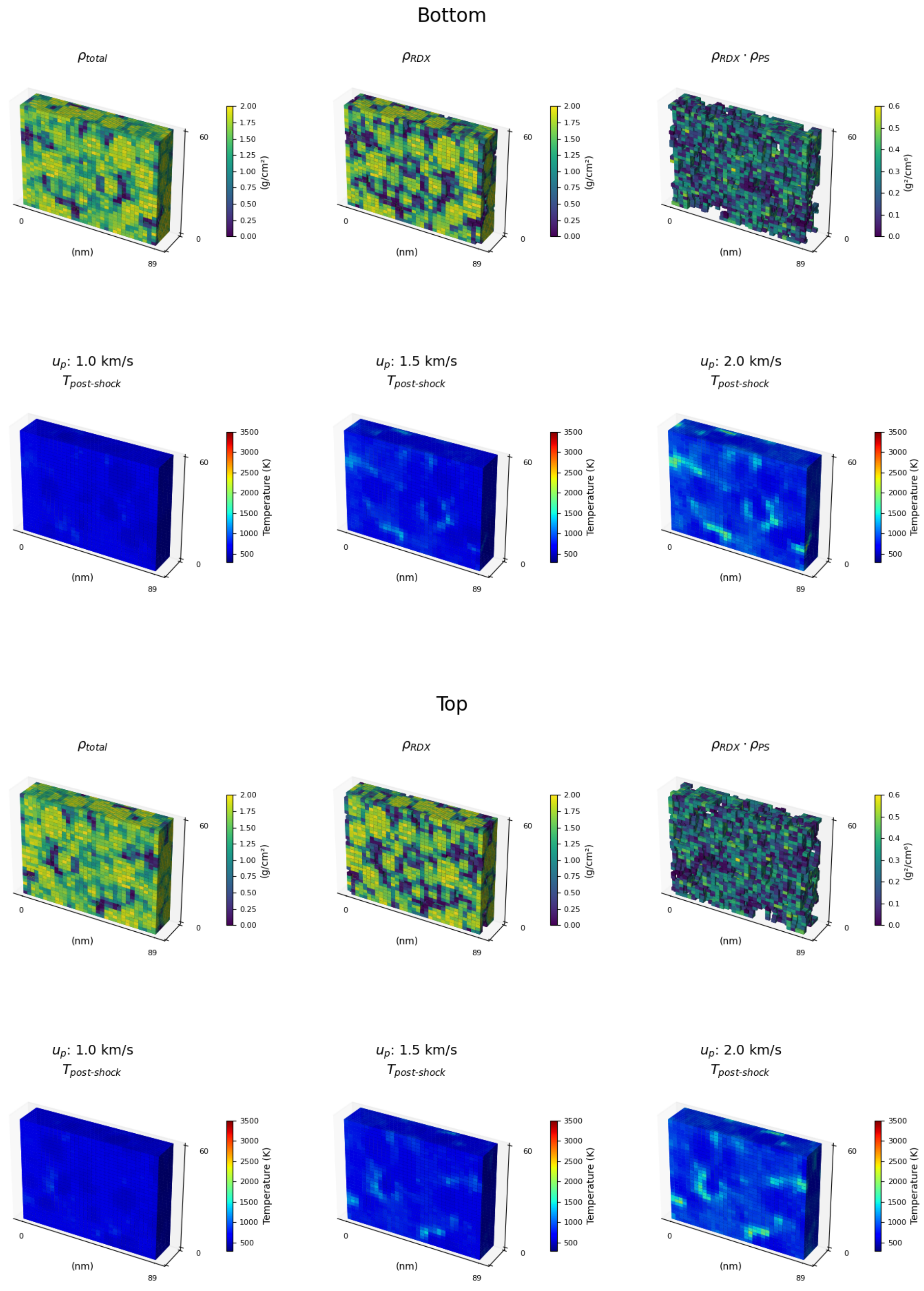

Bottom
$\rho_{total}$
$\rho_{RDX}$
$\rho_{RDX} \cdot \rho_{PS}$
(nm)
(g/cm³)
(g²/cm⁶)
$u_p$: 1.0 km/s
$u_p$: 1.5 km/s
$u_p$: 2.0 km/s
$T_{post\text{-}shock}$
Temperature (K)
Top

System 6:

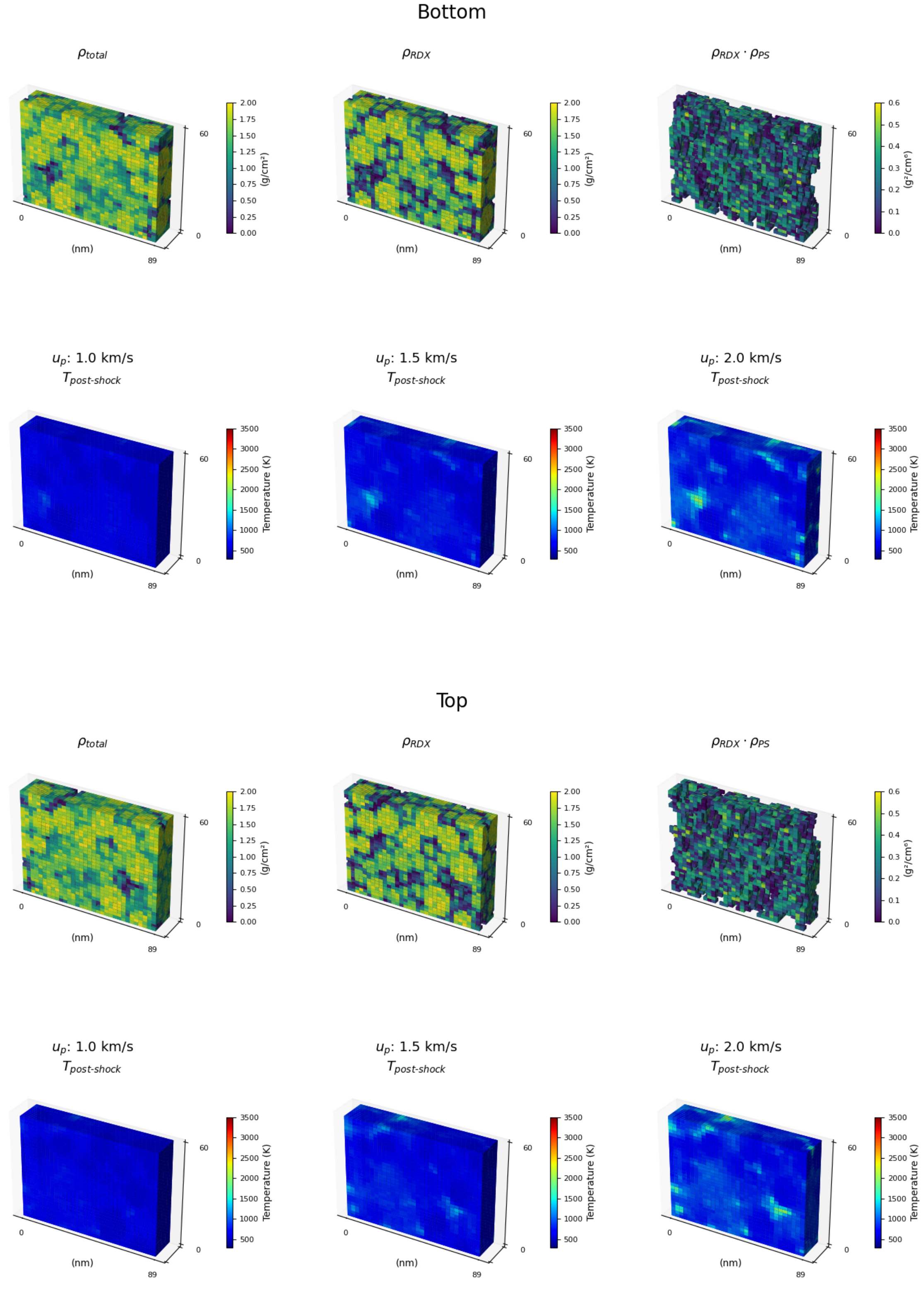
Bottom
$\rho_{total}$
$\rho_{RDX}$
$\rho_{RDX} \cdot \rho_{PS}$
(g/cm³)
(g²/cm⁶)
(nm)
$u_p$: 1.0 km/s
$u_p$: 1.5 km/s
$u_p$: 2.0 km/s
$T_{post\text{-}shock}$
Temperature (K)
Top

System 7:

In addition to the seven PBX systems listed above, there are four larger PBX systems. The seven small and four larger PBX systems are shocked at a particle velocity of 2.5 km/s. Of the four larger systems, two are used for training, one is used for validation, and one is used for testing. The smaller systems are depicted below, followed by the larger systems.

System 1:

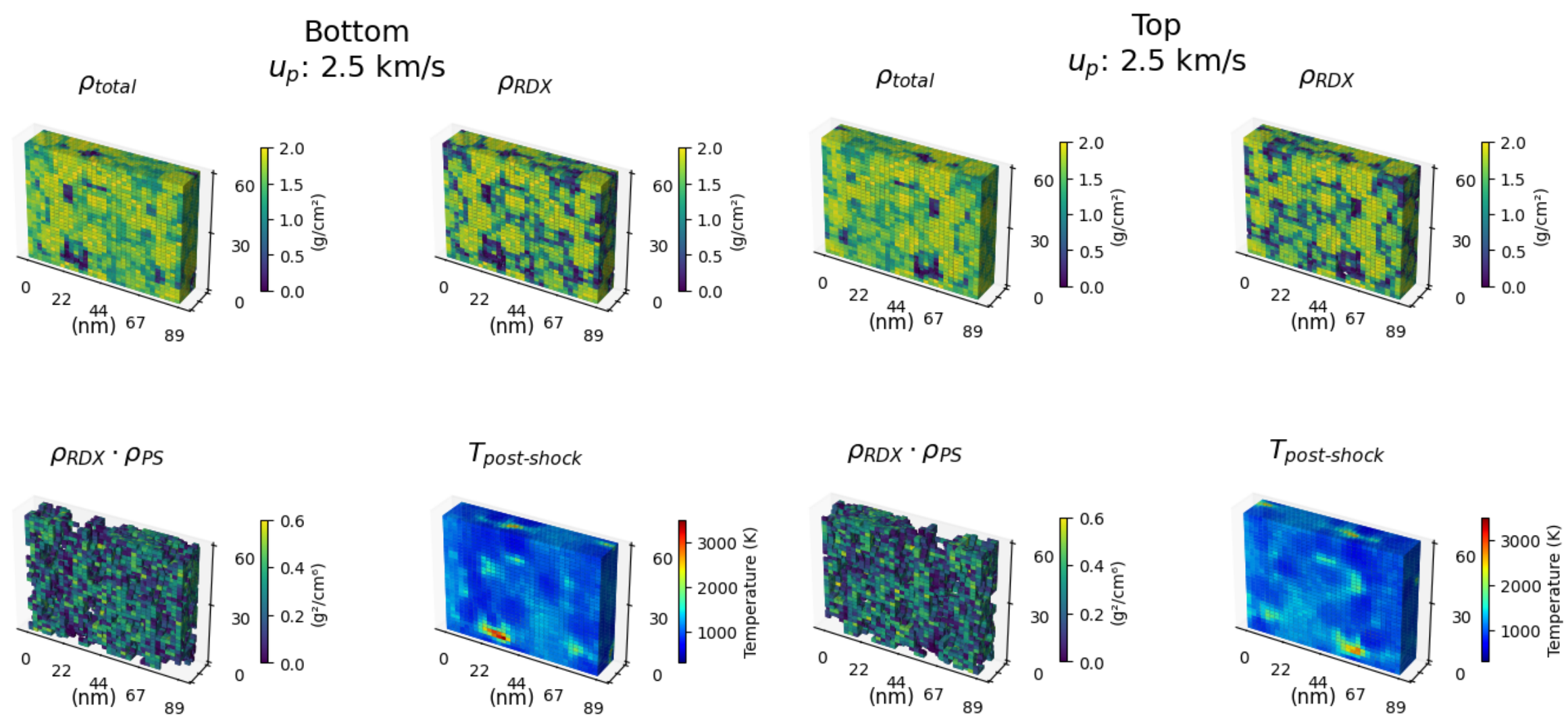

System 2:

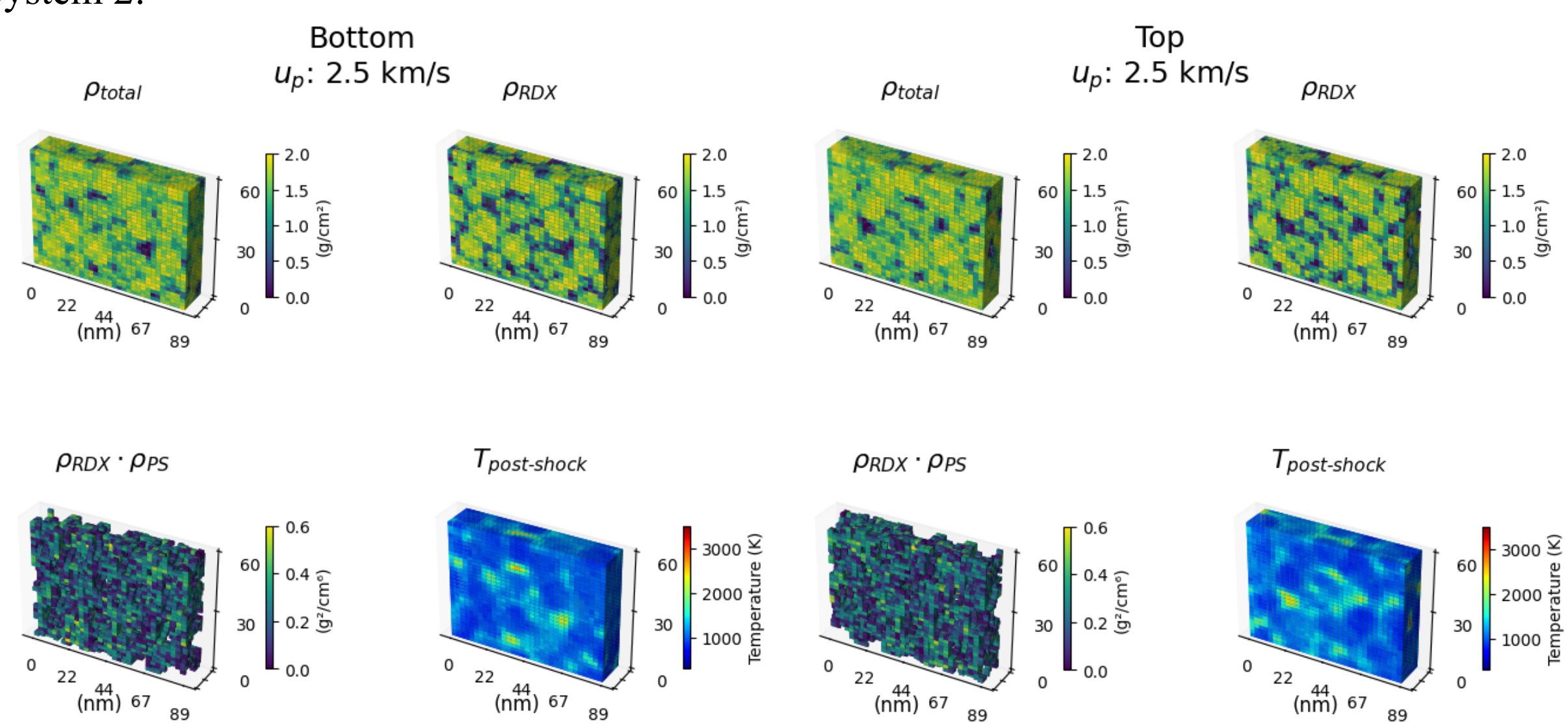

System 3:

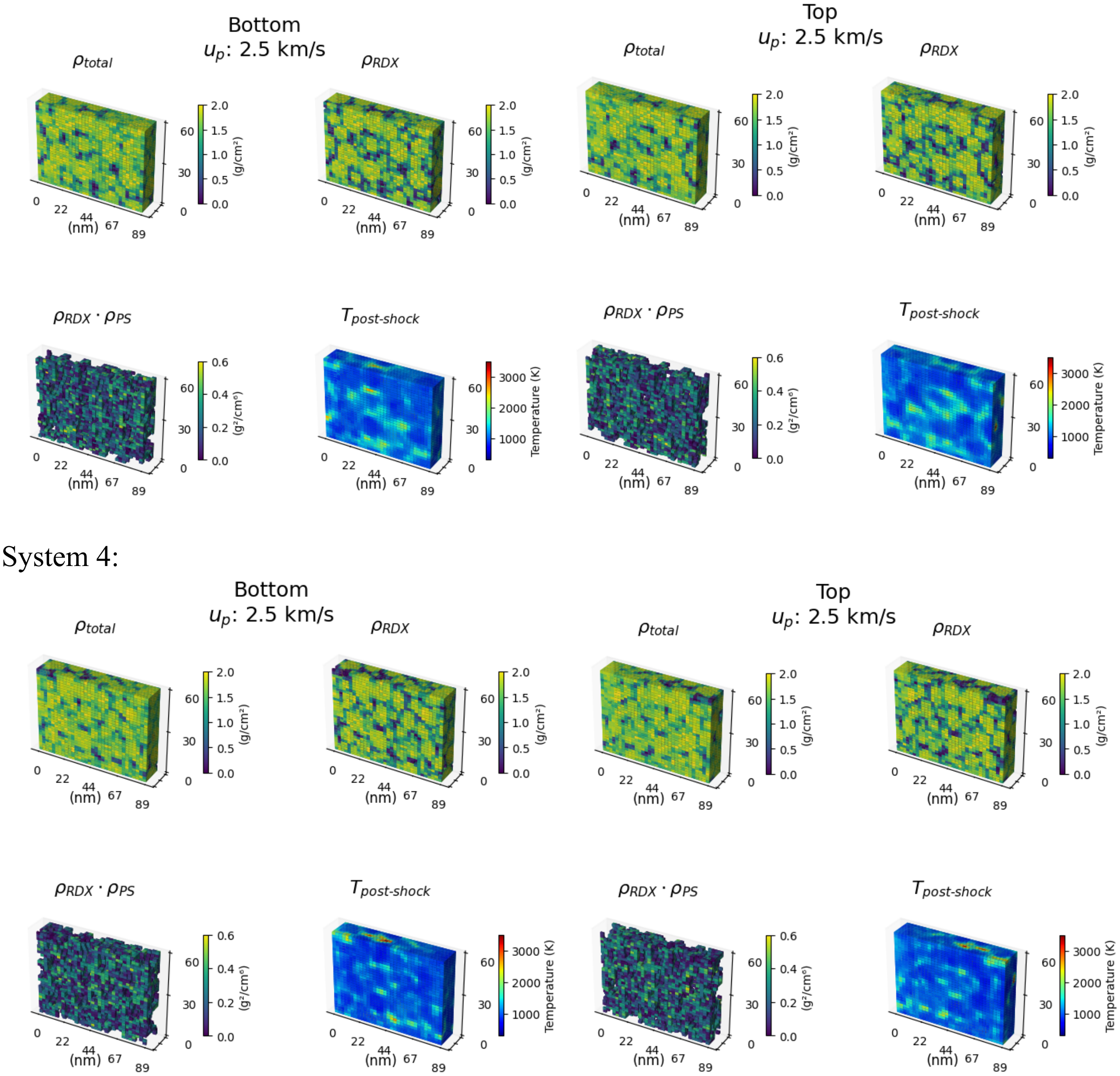

System 4:

System 5:

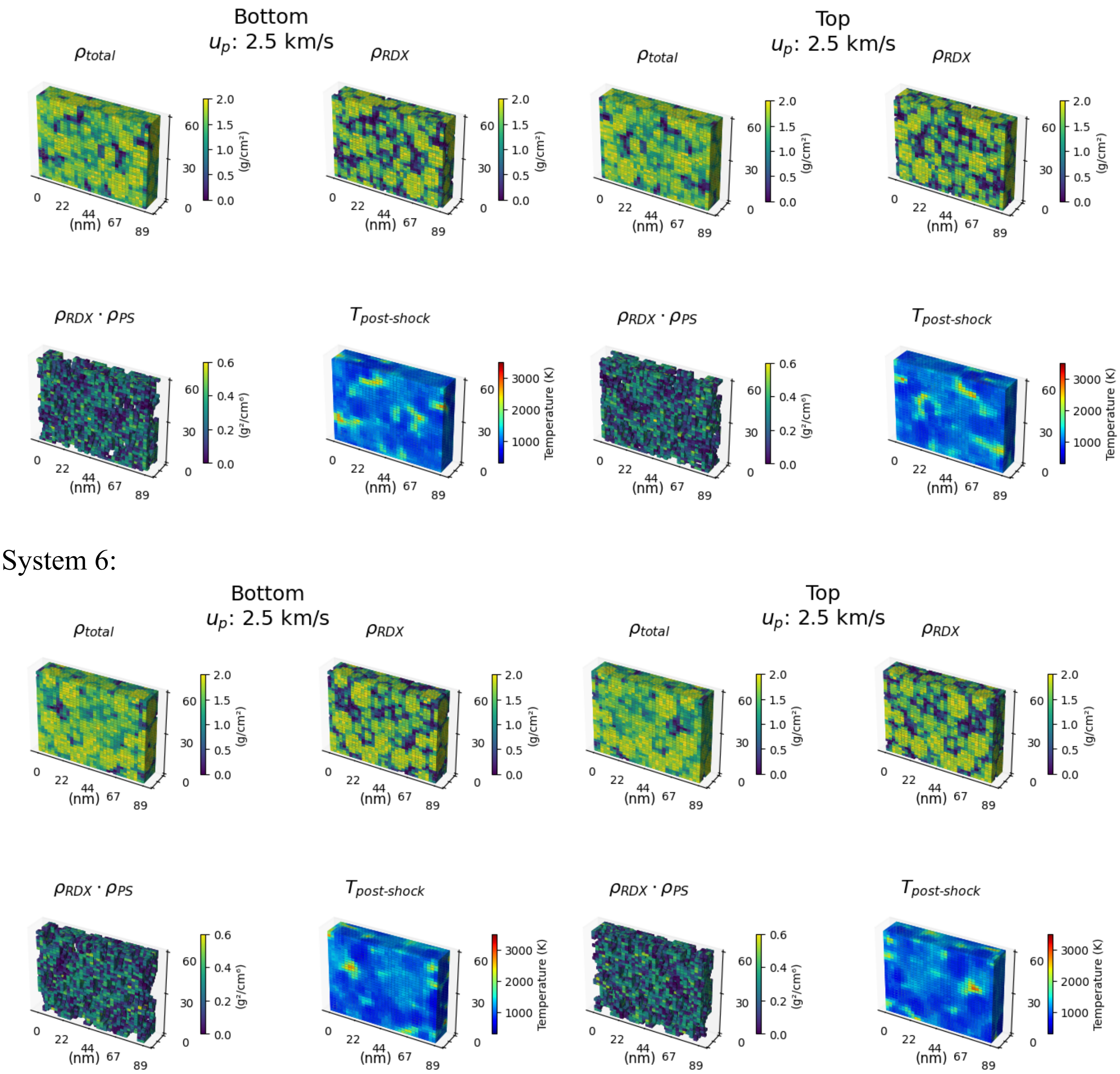

System 6:

System 7:

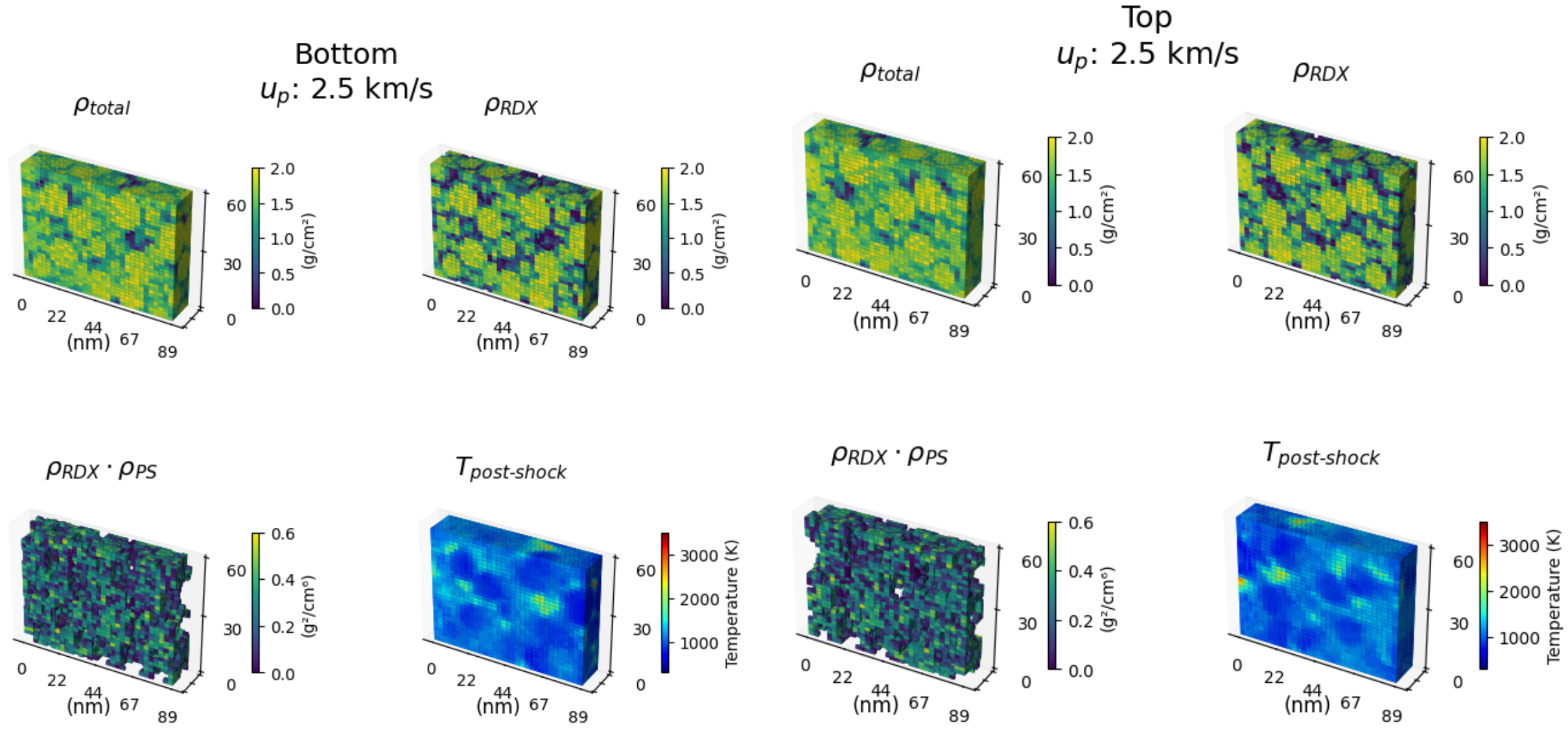

**Larger PBX systems:** These are binned in quadrants across the top and bottom, with a few nanometers of overlap. Each system creates 8 unique microstructures (bottom and top, each with a top left, top right, lower left, and lower right configuration):

System 1:

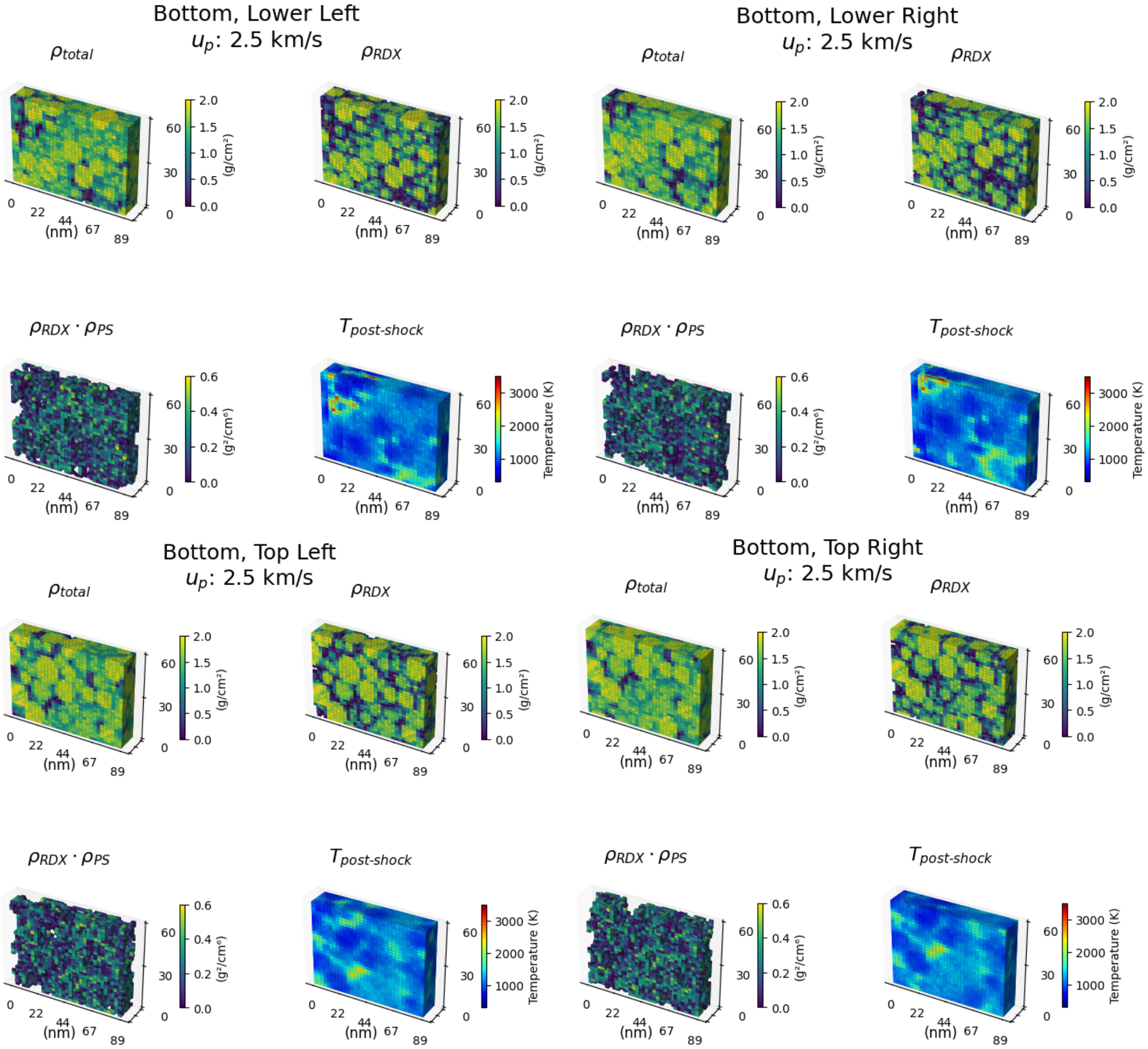

Bottom, Lower Left
$u_p$: 2.5 km/s
$\rho_{total}$
$\rho_{RDX}$
$\rho_{RDX} \cdot \rho_{PS}$
$T_{post\text{-}shock}$
Bottom, Lower Right
$u_p$: 2.5 km/s
$\rho_{total}$
$\rho_{RDX}$
$\rho_{RDX} \cdot \rho_{PS}$
$T_{post\text{-}shock}$
Bottom, Top Left
$u_p$: 2.5 km/s
$\rho_{total}$
$\rho_{RDX}$
$\rho_{RDX} \cdot \rho_{PS}$
$T_{post\text{-}shock}$
Bottom, Top Right
$u_p$: 2.5 km/s
$\rho_{total}$
$\rho_{RDX}$
$\rho_{RDX} \cdot \rho_{PS}$
$T_{post\text{-}shock}$
(nm)
(g/cm²)
(g²/cm⁶)
Temperature (K)

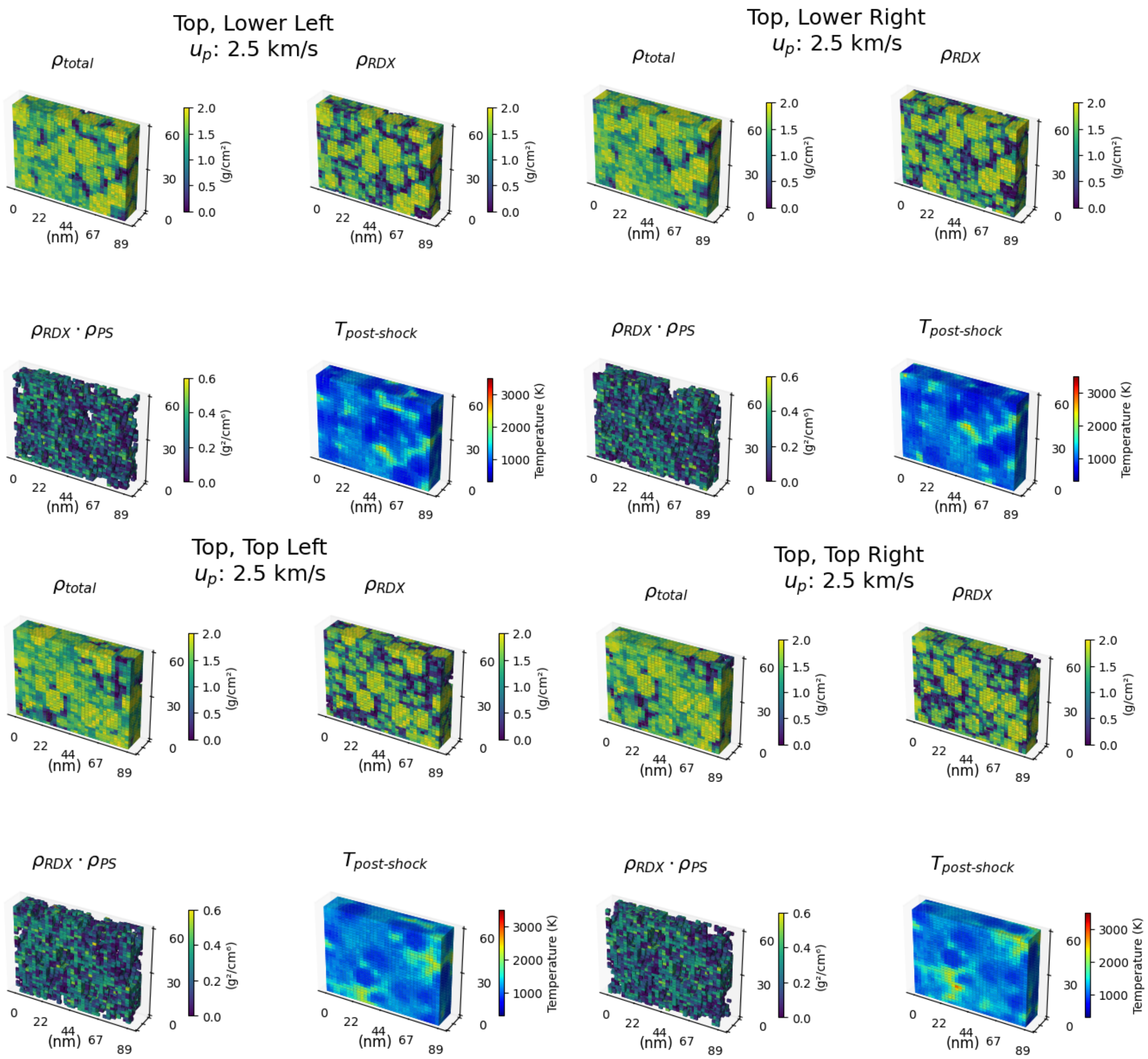

Top, Lower Left
$u_p$: 2.5 km/s
$\rho_{total}$
$\rho_{RDX}$
$\rho_{RDX} \cdot \rho_{PS}$
$T_{post\text{-}shock}$
Top, Lower Right
$u_p$: 2.5 km/s
Top, Top Left
$u_p$: 2.5 km/s
Top, Top Right
$u_p$: 2.5 km/s
(nm)
(g/cm³)
(g²/cm⁶)
Temperature (K)

System 2:

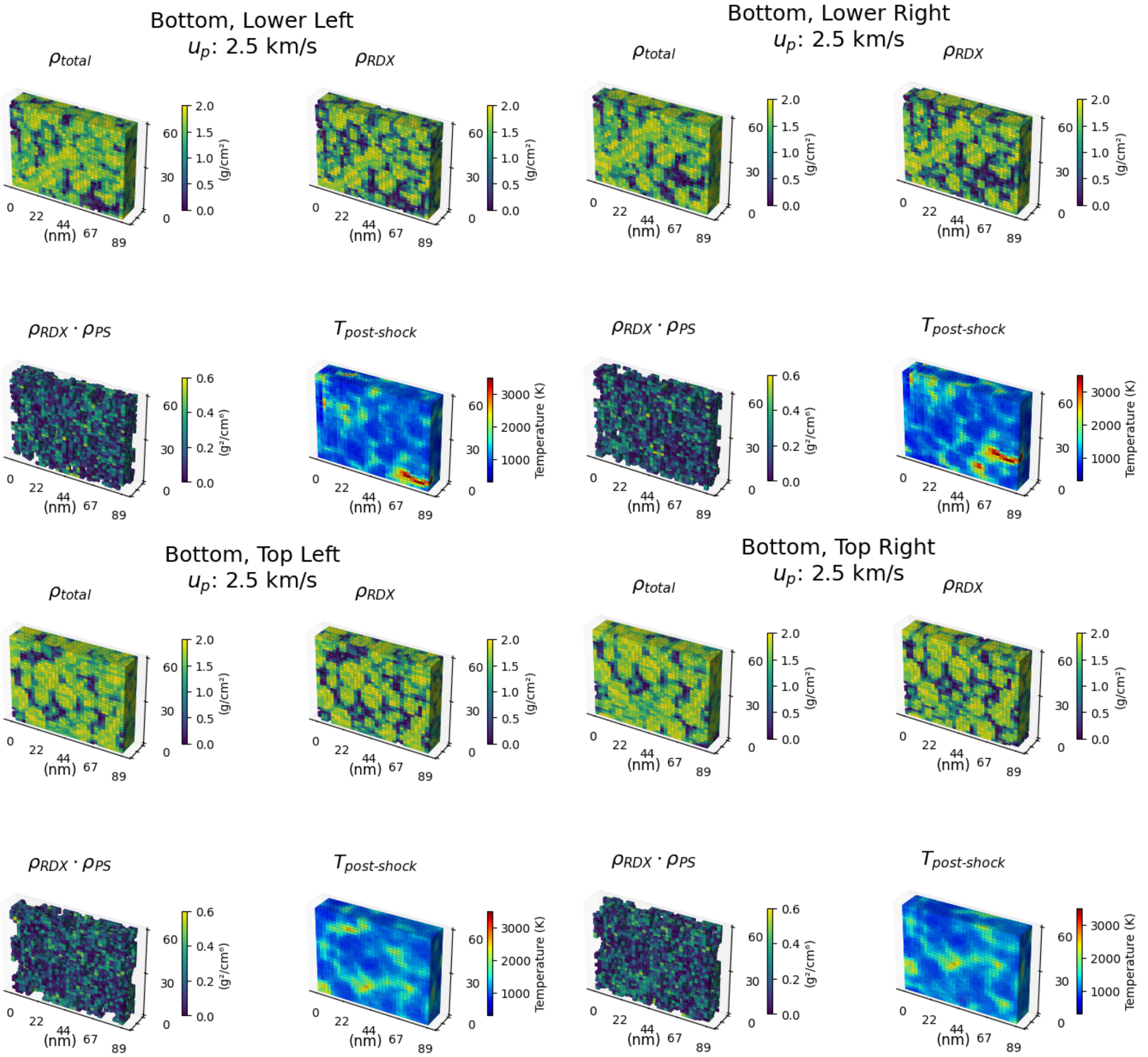

Bottom, Lower Left
$u_p$: 2.5 km/s
$\rho_{total}$
$\rho_{RDX}$
$\rho_{RDX} \cdot \rho_{PS}$
$T_{post\text{-}shock}$
Bottom, Lower Right
$u_p$: 2.5 km/s
$\rho_{total}$
$\rho_{RDX}$
$\rho_{RDX} \cdot \rho_{PS}$
$T_{post\text{-}shock}$
Bottom, Top Left
$u_p$: 2.5 km/s
$\rho_{total}$
$\rho_{RDX}$
$\rho_{RDX} \cdot \rho_{PS}$
$T_{post\text{-}shock}$
Bottom, Top Right
$u_p$: 2.5 km/s
$\rho_{total}$
$\rho_{RDX}$
$\rho_{RDX} \cdot \rho_{PS}$
$T_{post\text{-}shock}$
(nm)
(g/cm³)
(g²/cm⁶)
Temperature (K)

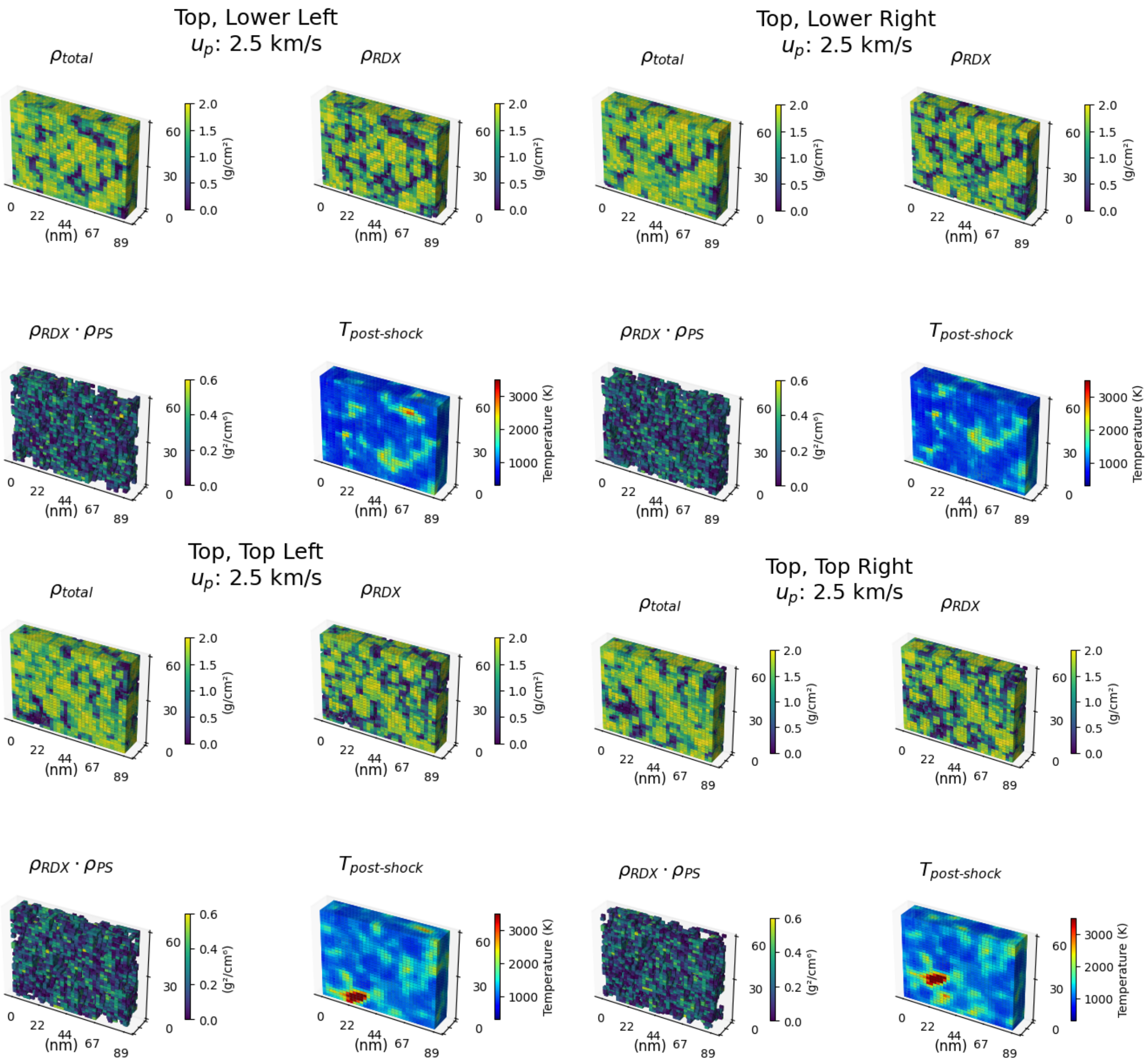
Top, Lower Left
$u_p$: 2.5 km/s
$\rho_{total}$
$\rho_{RDX}$
$\rho_{RDX} \cdot \rho_{PS}$
$T_{post\text{-}shock}$
(nm)
(g/cm³)
(g²/cm⁶)
Temperature (K)
Top, Lower Right
$u_p$: 2.5 km/s
Top, Top Left
$u_p$: 2.5 km/s
Top, Top Right
$u_p$: 2.5 km/s

System 3:

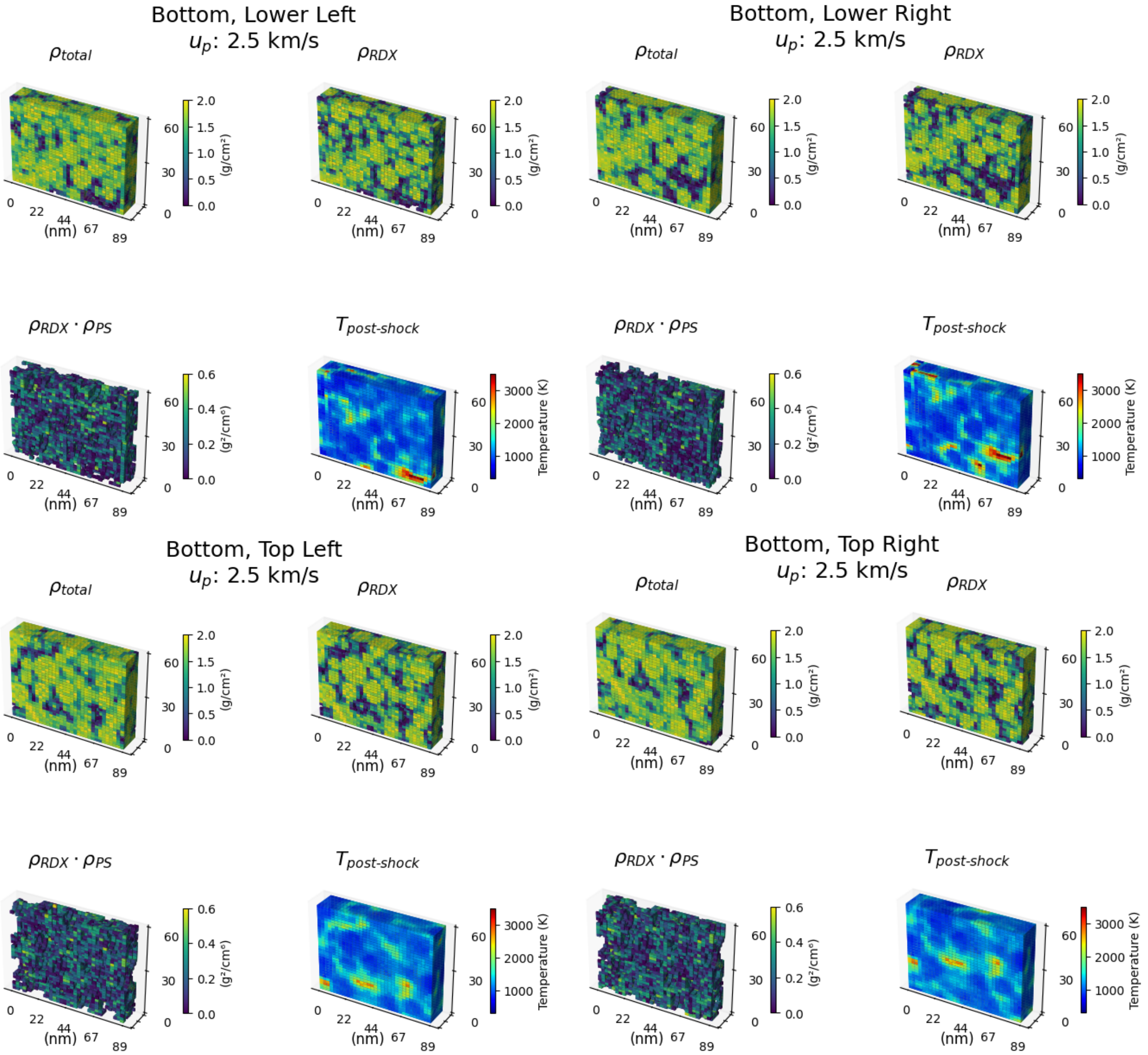

Bottom, Lower Left
$u_p$: 2.5 km/s
$\rho_{total}$
$\rho_{RDX}$
$\rho_{RDX} \cdot \rho_{PS}$
$T_{post\text{-}shock}$
Bottom, Lower Right
$u_p$: 2.5 km/s
Bottom, Top Left
$u_p$: 2.5 km/s
Bottom, Top Right
$u_p$: 2.5 km/s
(nm)
(g/cm²)
(g²/cm⁶)
Temperature (K)

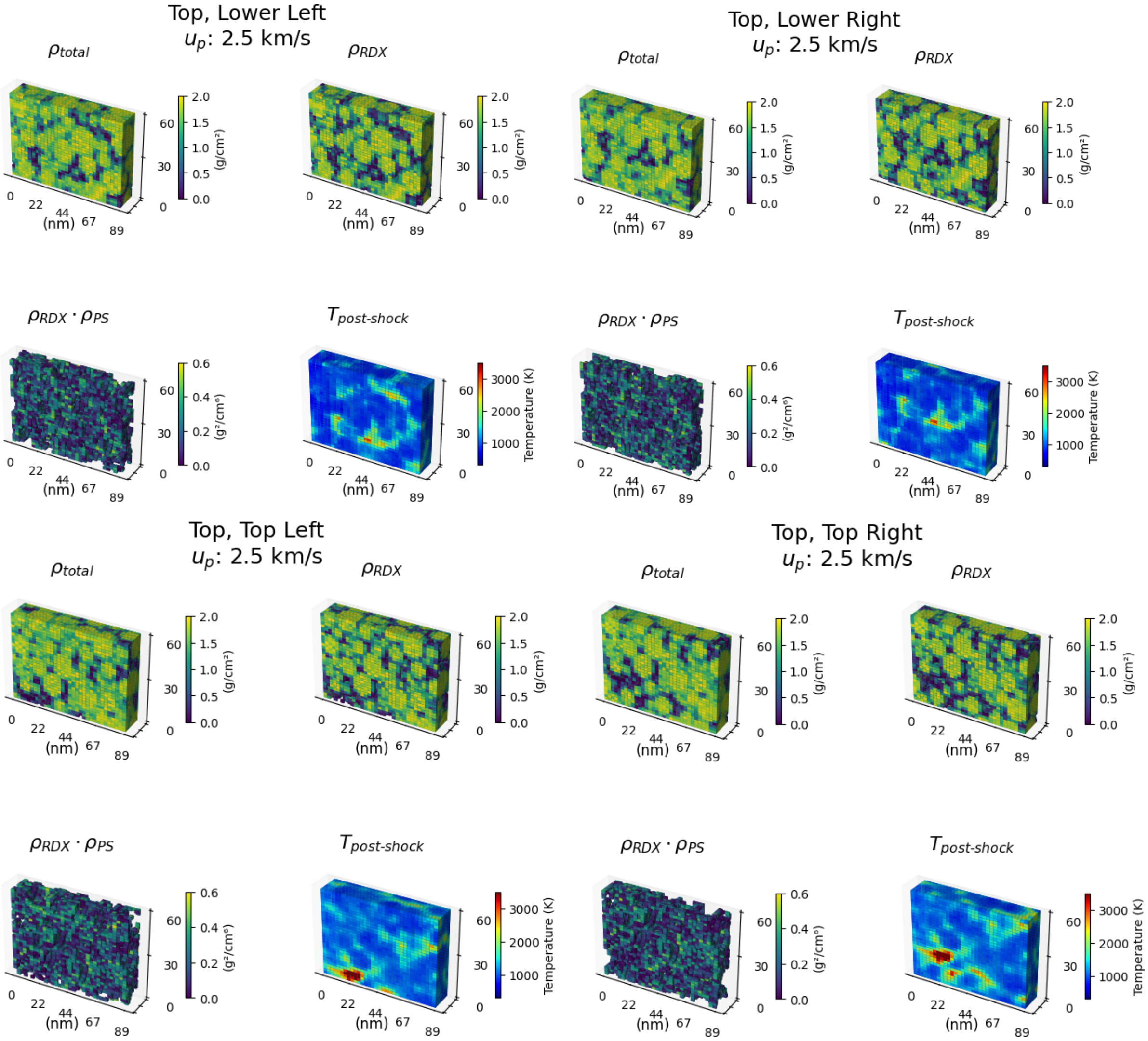

Top, Lower Left
$u_p$: 2.5 km/s
$\rho_{total}$
$\rho_{RDX}$
$\rho_{RDX} \cdot \rho_{PS}$
$T_{post\text{-}shock}$
Top, Lower Right
$u_p$: 2.5 km/s
Top, Top Left
$u_p$: 2.5 km/s
Top, Top Right
$u_p$: 2.5 km/s
(nm)
(g/cm²)
Temperature (K)

System 4:

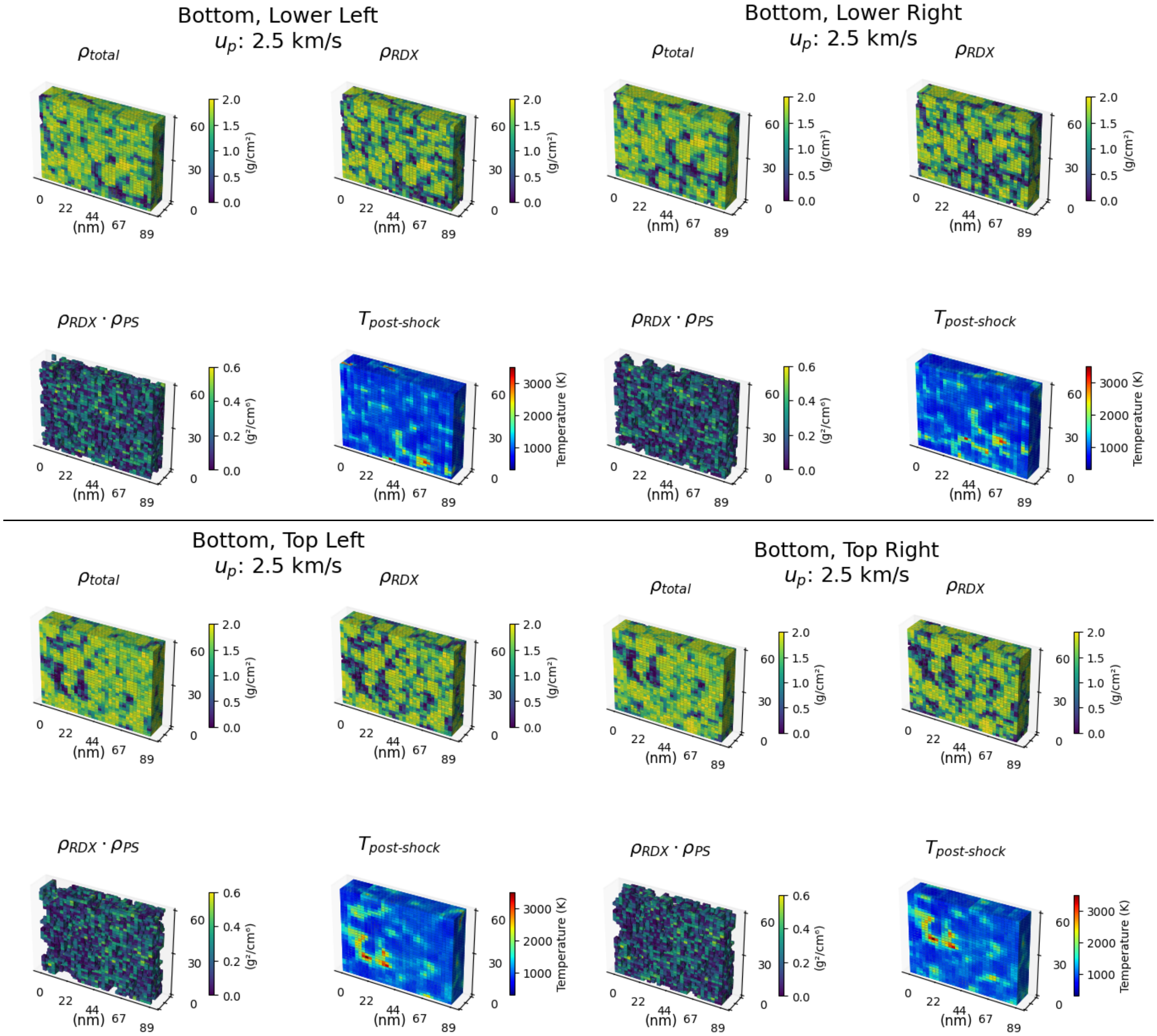

Bottom, Lower Left
$u_p$: 2.5 km/s
Bottom, Lower Right
$u_p$: 2.5 km/s
Bottom, Top Left
$u_p$: 2.5 km/s
Bottom, Top Right
$u_p$: 2.5 km/s
$\rho_{total}$
$\rho_{RDX}$
$\rho_{RDX} \cdot \rho_{PS}$
$T_{post\text{-}shock}$
(nm)
(g/cm³)
(g²/cm⁶)
Temperature (K)

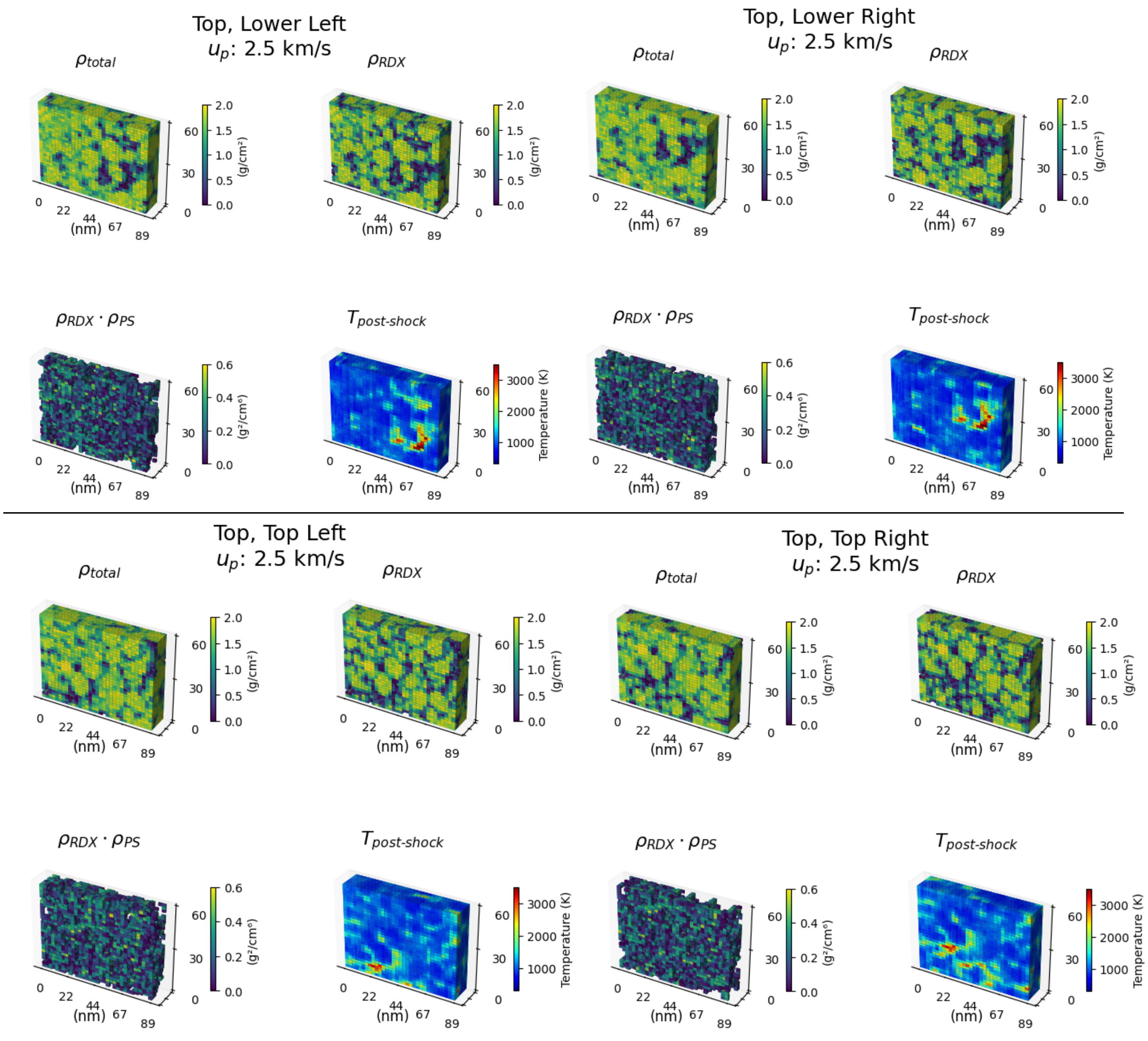

There are **10 multi-pore systems** shocked at a particle velocity of 1.0, 1.5, and 2.0 km/s using dissipative particle dynamic simulation. Six systems are used for training. Two systems are used for validation. Two systems are used for testing.

System 1:

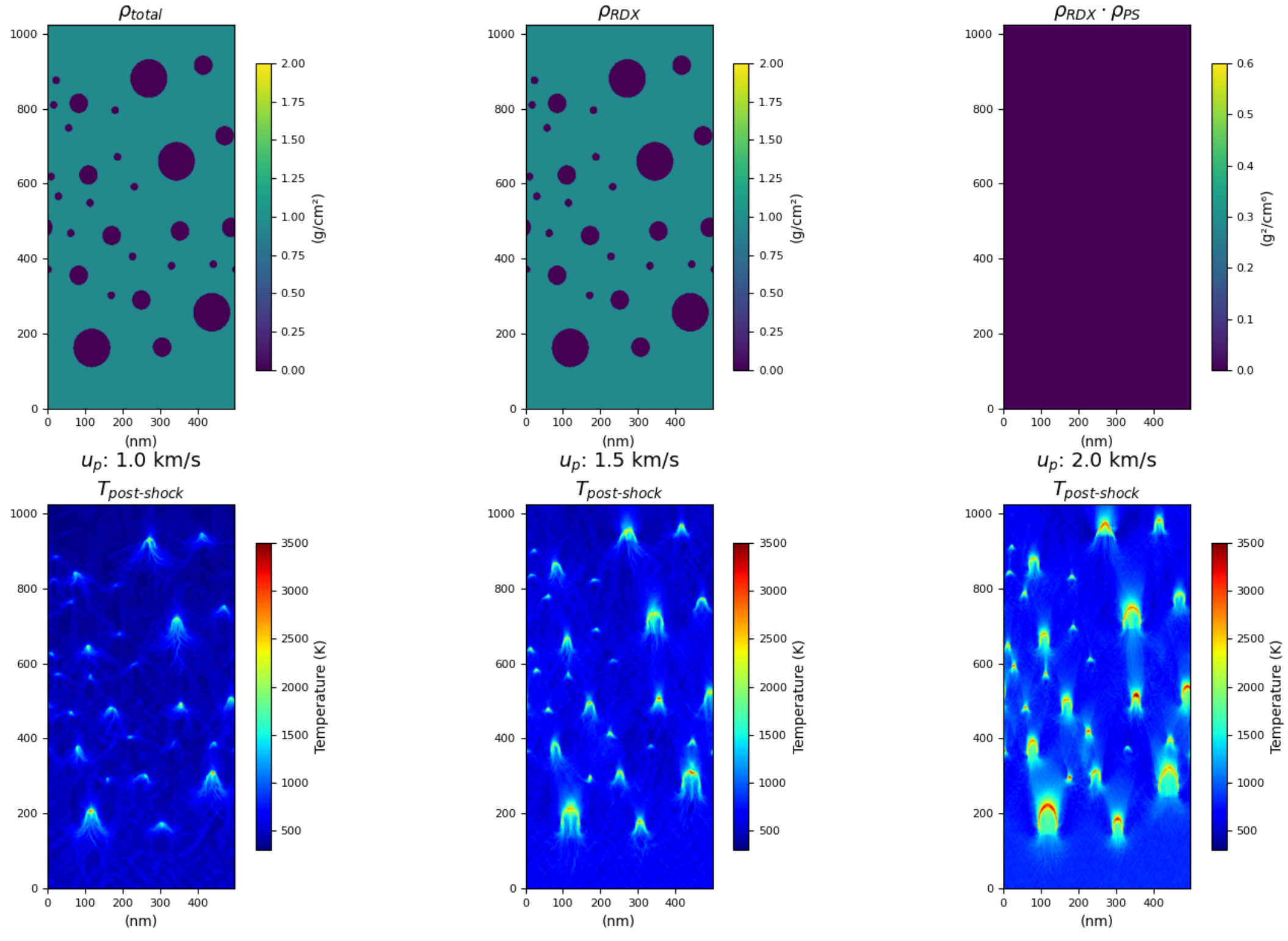
$\rho_{total}$
$\rho_{RDX}$
$\rho_{RDX} \cdot \rho_{PS}$
(g/cm³)
(g²/cm⁶)
(nm)
$u_p$: 1.0 km/s
$u_p$: 1.5 km/s
$u_p$: 2.0 km/s
$T_{post\text{-}shock}$
Temperature (K)

System 2:

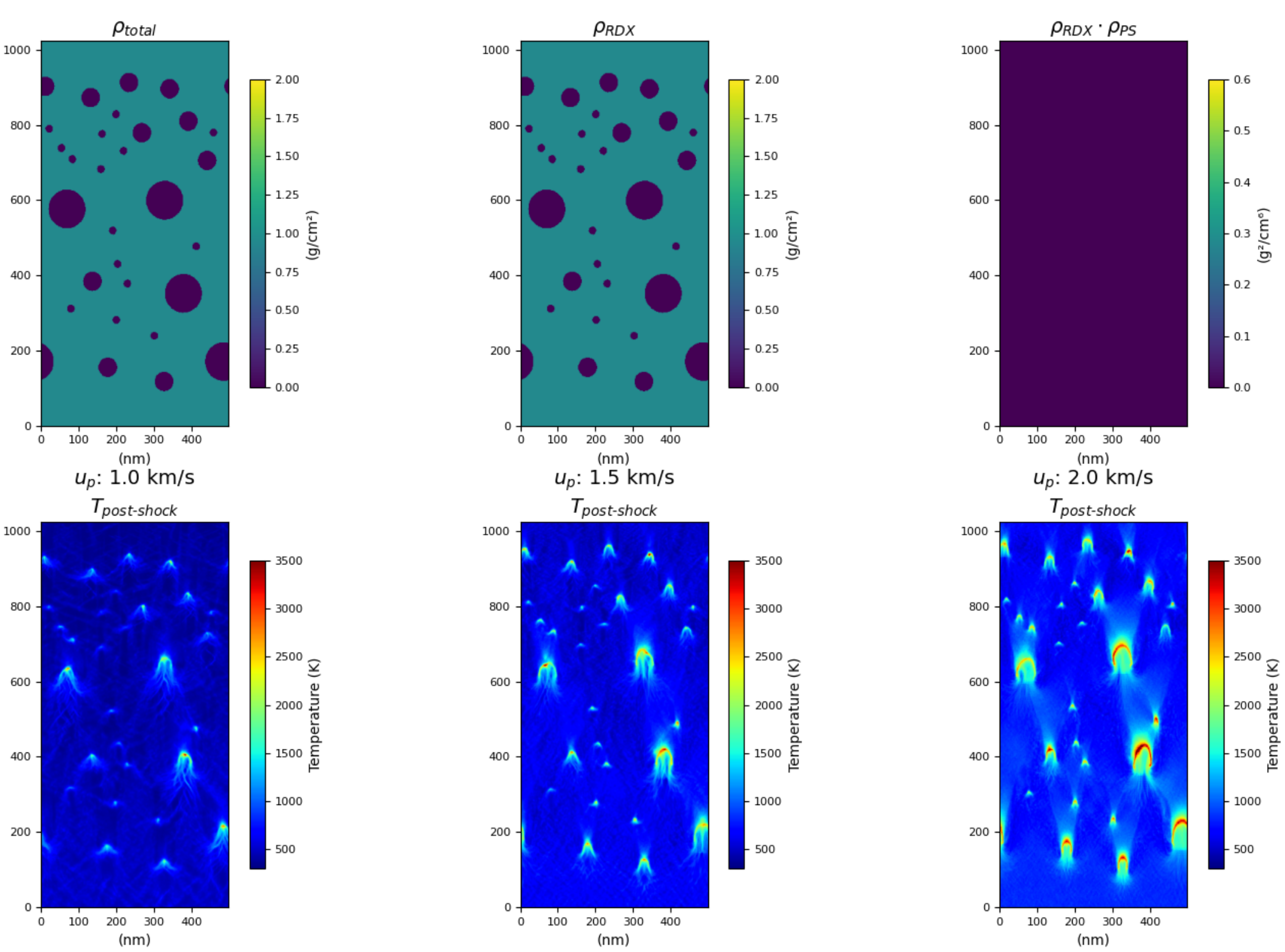
$\rho_{total}$
$\rho_{RDX}$
$\rho_{RDX} \cdot \rho_{PS}$
(g/cm³)
(g²/cm⁶)
(nm)
$u_p$: 1.0 km/s
$u_p$: 1.5 km/s
$u_p$: 2.0 km/s
$T_{post\text{-}shock}$
Temperature (K)

System 3:

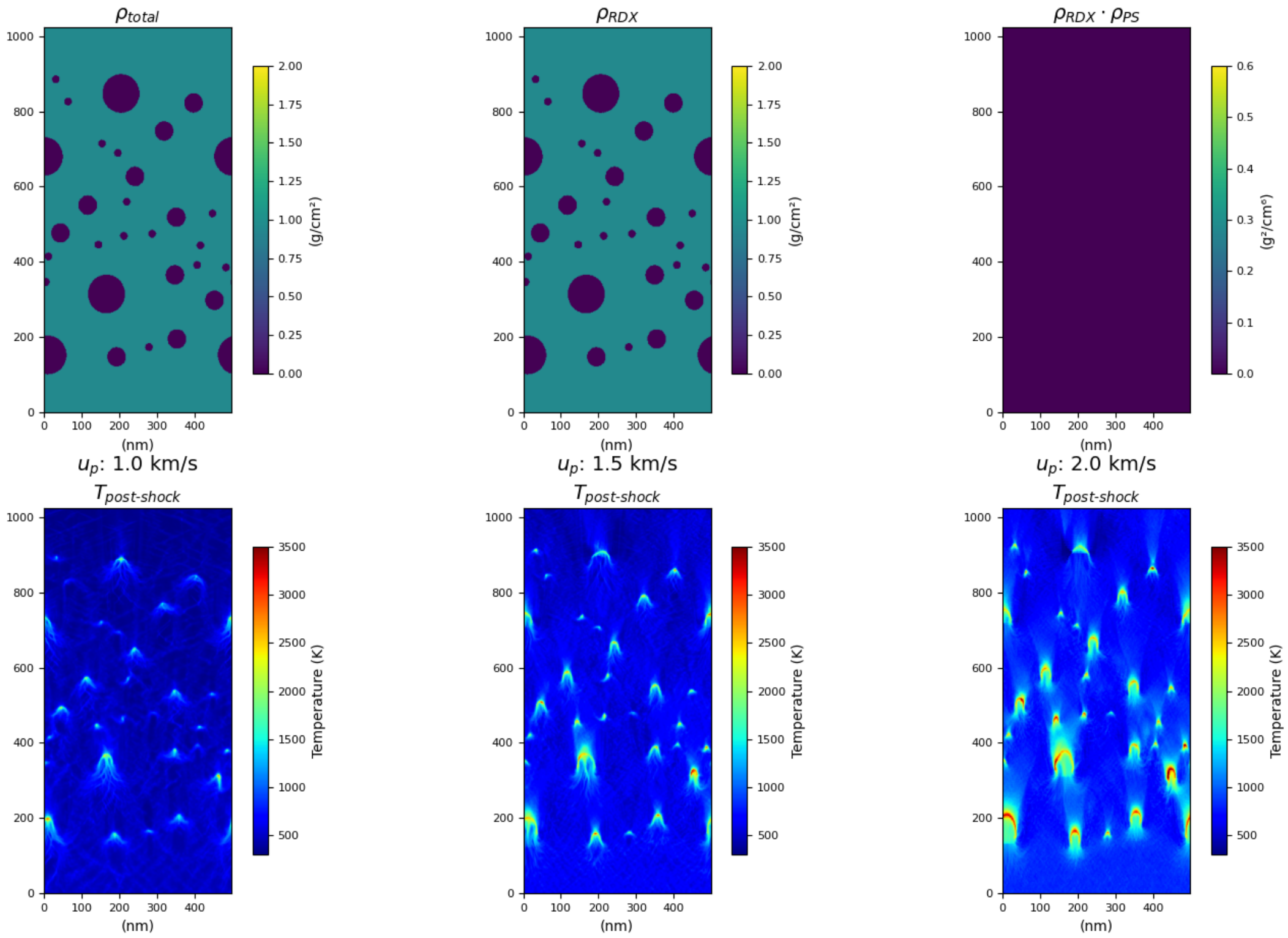

System 4:

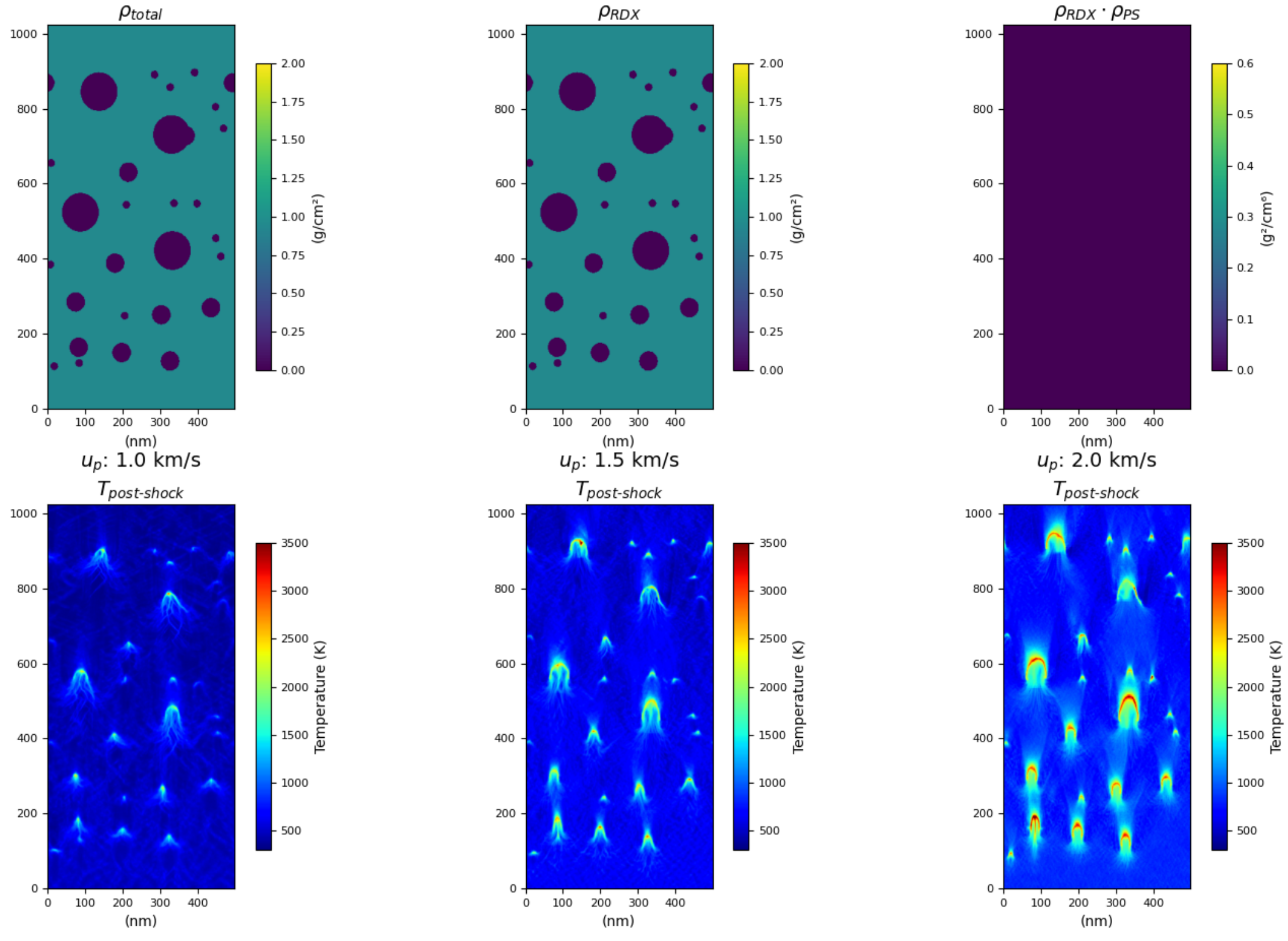
$\rho_{total}$
$\rho_{RDX}$
$\rho_{RDX} \cdot \rho_{PS}$
(g/cm³)
(g²/cm⁶)
(nm)
$u_p$: 1.0 km/s
$u_p$: 1.5 km/s
$u_p$: 2.0 km/s
$T_{post\text{-}shock}$
Temperature (K)

System 5:

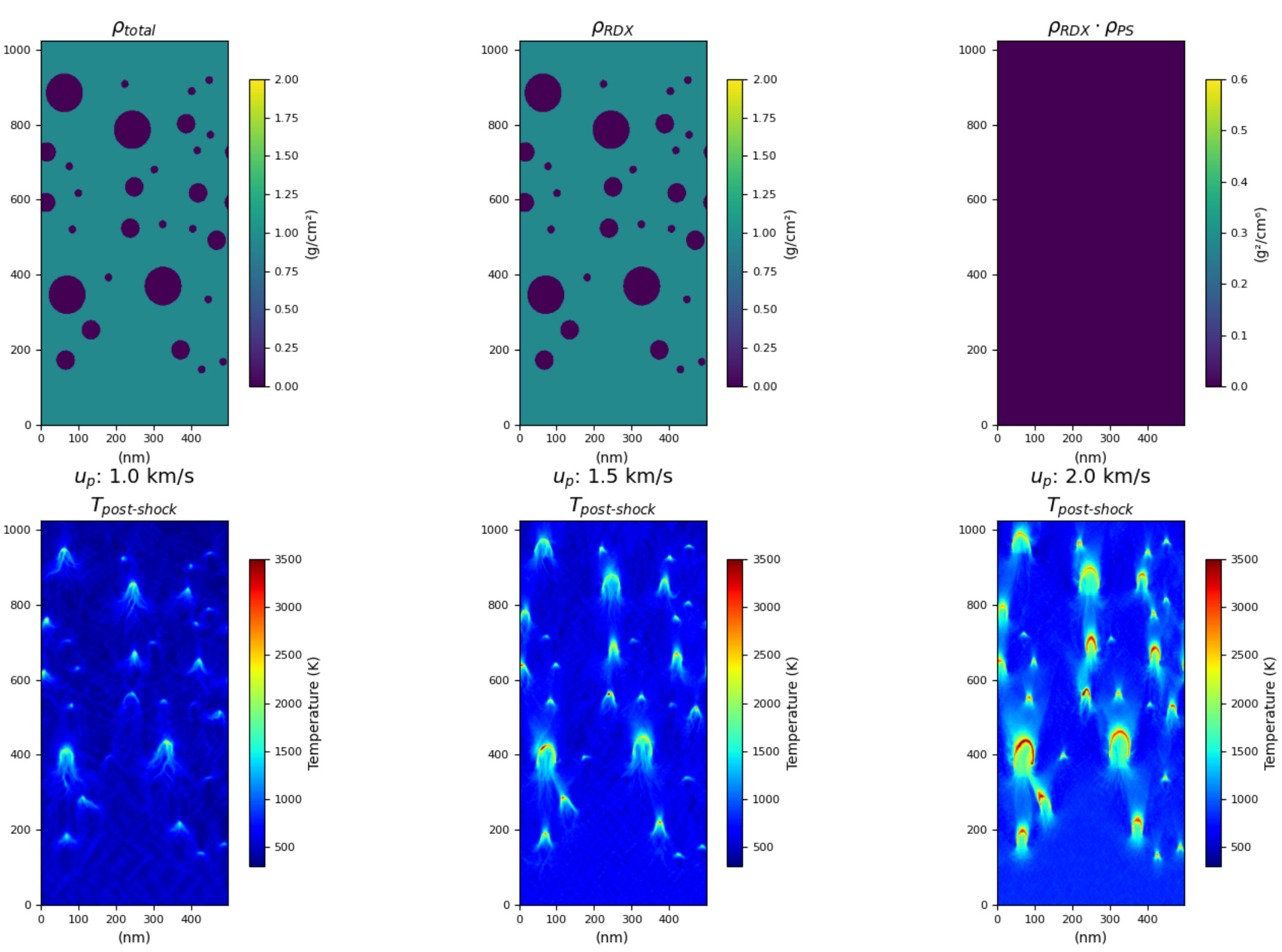
$\rho_{total}$
$\rho_{RDX}$
$\rho_{RDX} \cdot \rho_{PS}$
(g/cm³)
(g²/cm⁶)
(nm)
$u_p$: 1.0 km/s
$u_p$: 1.5 km/s
$u_p$: 2.0 km/s
$T_{post\text{-}shock}$
Temperature (K)

System 6:

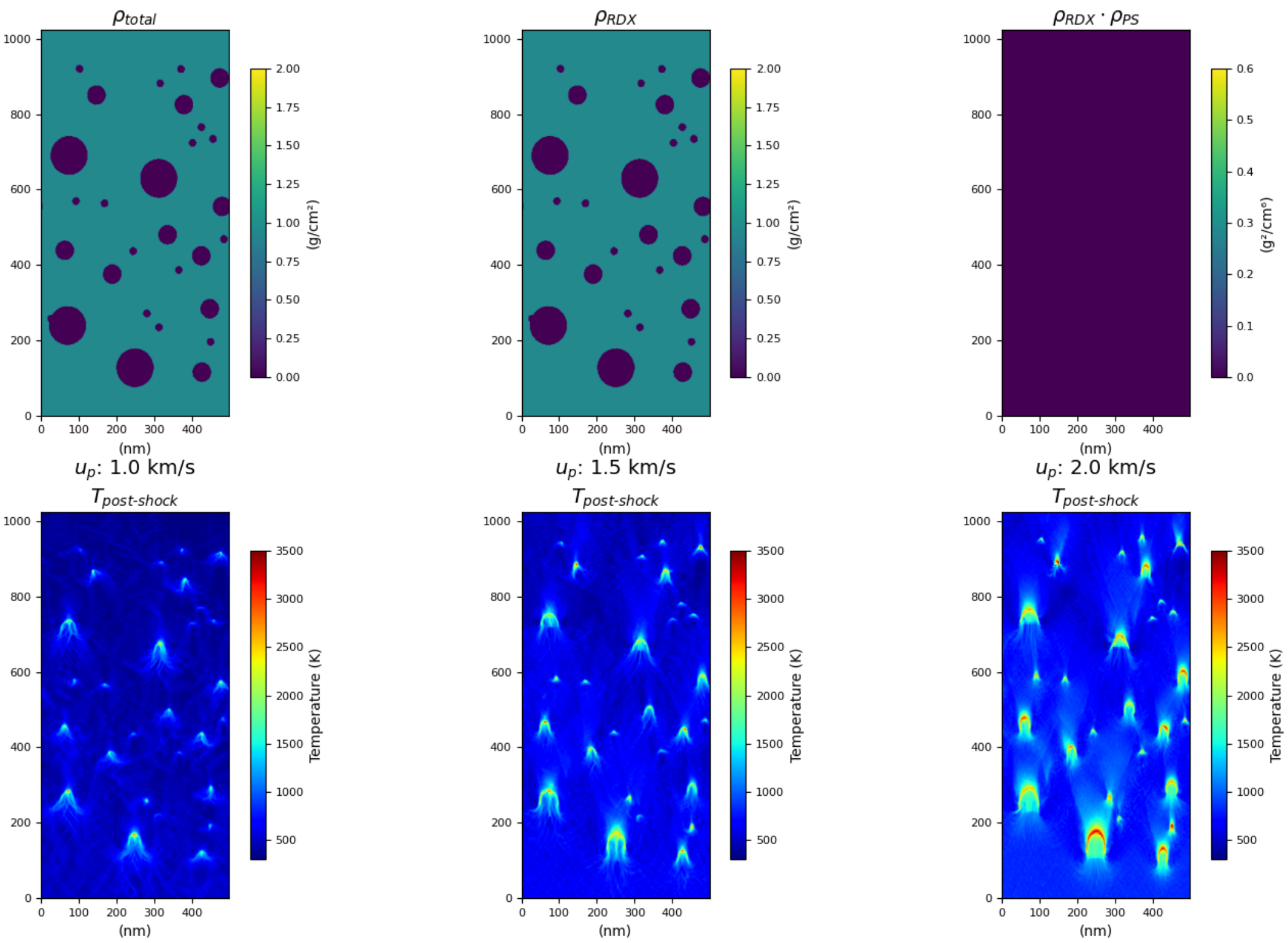

System 7:

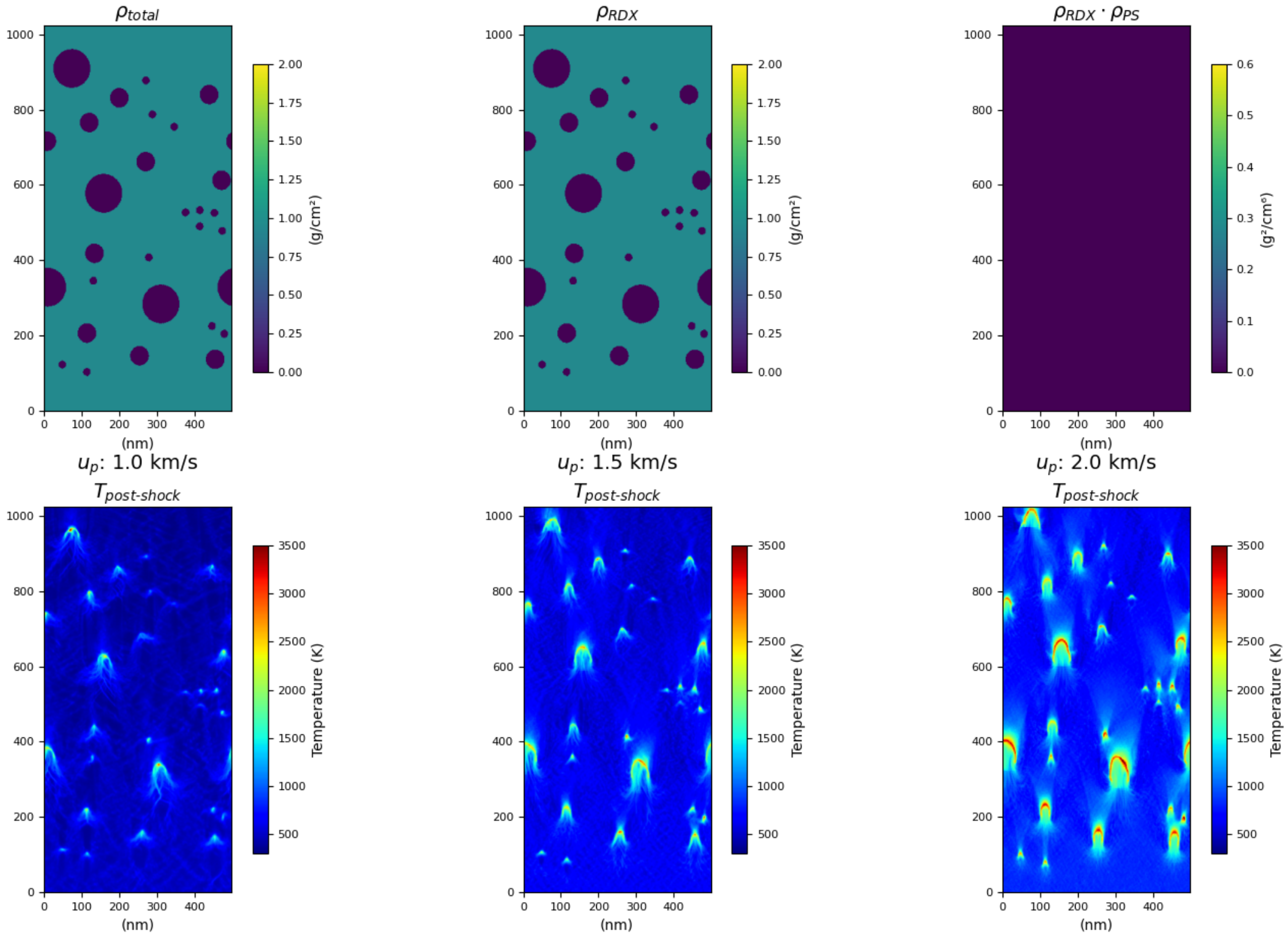

$\rho_{total}$
$\rho_{RDX}$
$\rho_{RDX} \cdot \rho_{PS}$
(g/cm³)
(g²/cm⁶)
(nm)
$u_p$: 1.0 km/s
$u_p$: 1.5 km/s
$u_p$: 2.0 km/s
$T_{post\text{-}shock}$
Temperature (K)

System 8:

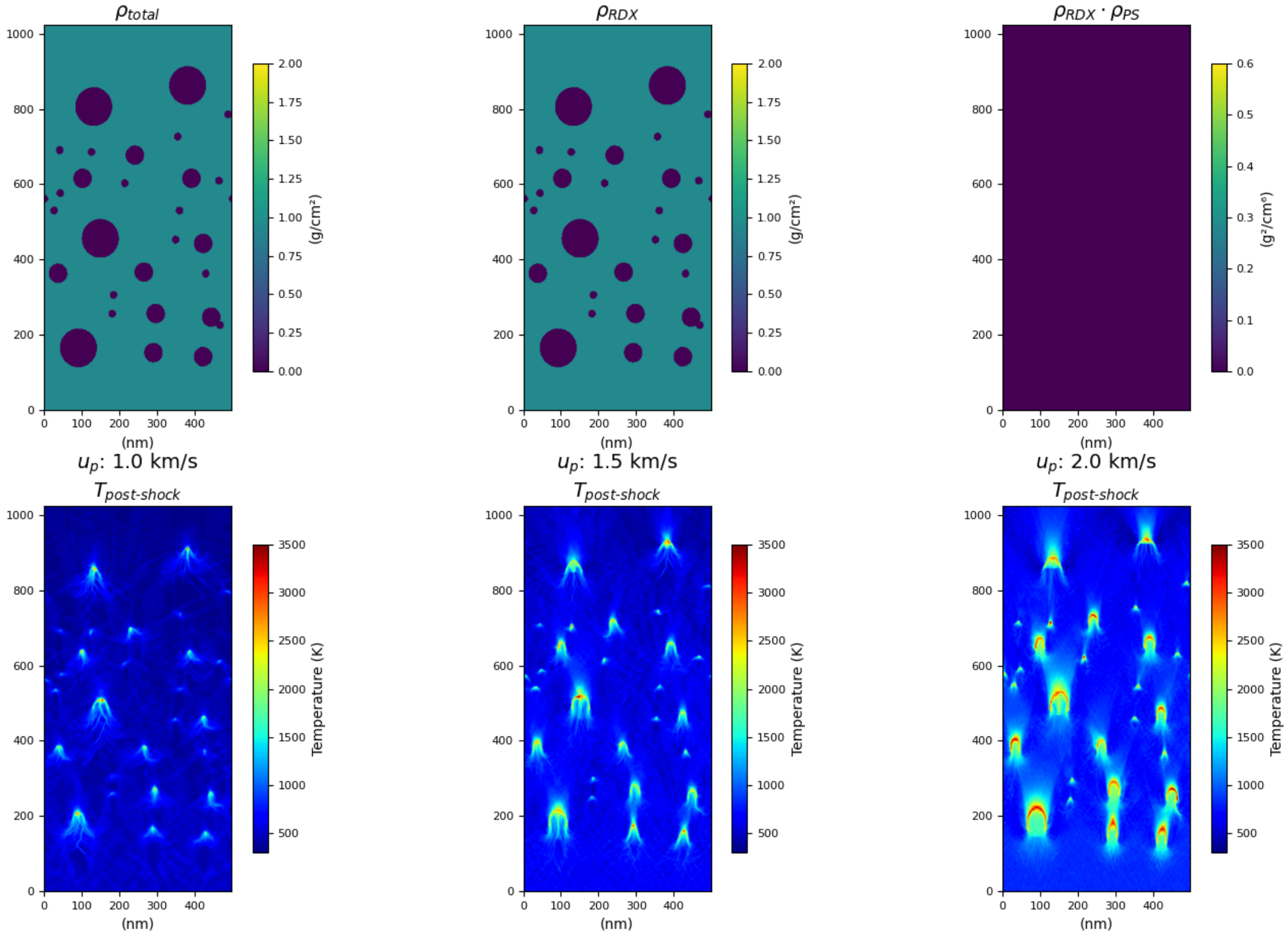
$\rho_{total}$
$\rho_{RDX}$
$\rho_{RDX} \cdot \rho_{PS}$
(g/cm³)
(g²/cm⁶)
(nm)
$u_p$: 1.0 km/s
$u_p$: 1.5 km/s
$u_p$: 2.0 km/s
$T_{post\text{-}shock}$
Temperature (K)

System 9:

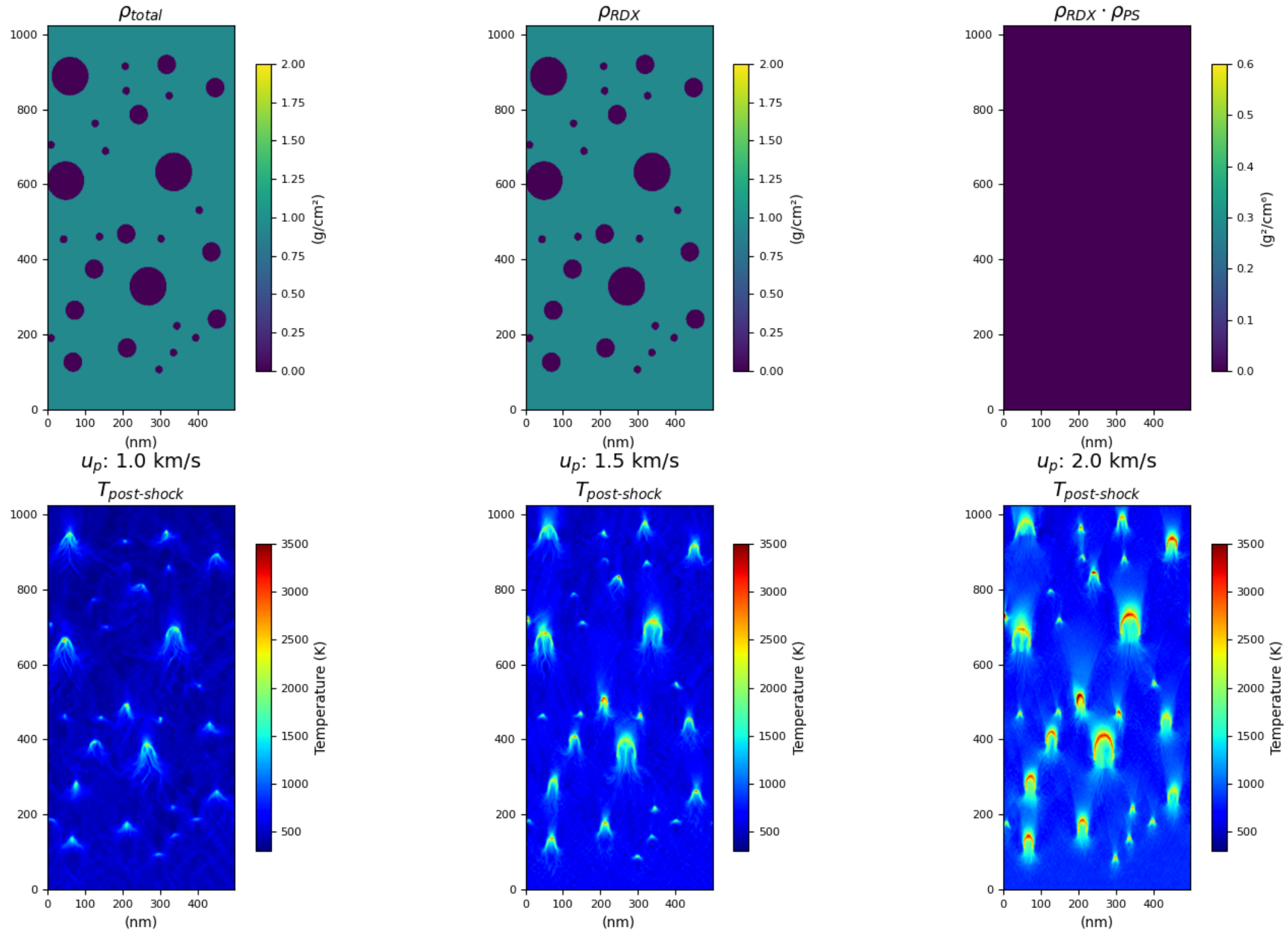
$\rho_{total}$
$\rho_{RDX}$
$\rho_{RDX} \cdot \rho_{PS}$
(g/cm³)
(g²/cm⁶)
(nm)
$u_p$: 1.0 km/s
$u_p$: 1.5 km/s
$u_p$: 2.0 km/s
$T_{post\text{-}shock}$
Temperature (K)

System 10:

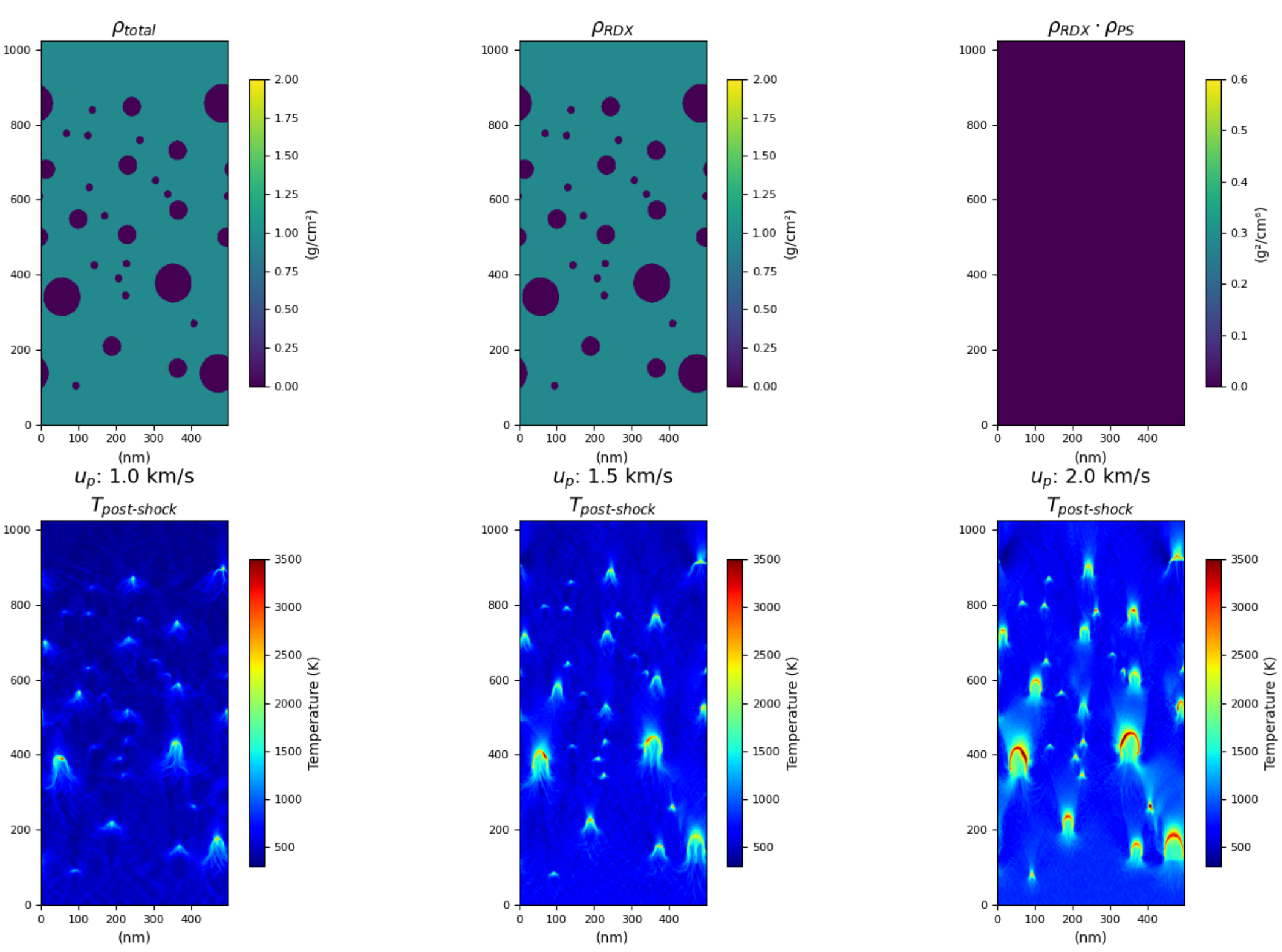
$\rho_{total}$
$\rho_{RDX}$
$\rho_{RDX} \cdot \rho_{PS}$
(g/cm³)
(g²/cm⁶)
(nm)
$u_p$: 1.0 km/s
$u_p$: 1.5 km/s
$u_p$: 2.0 km/s
$T_{post\text{-}shock}$
Temperature (K)

There are **3 single-pore systems** shocked at a particle velocity of 1.0, 1.5, and 2.0 km/s using dissipative particle dynamic simulation. All are used for training:

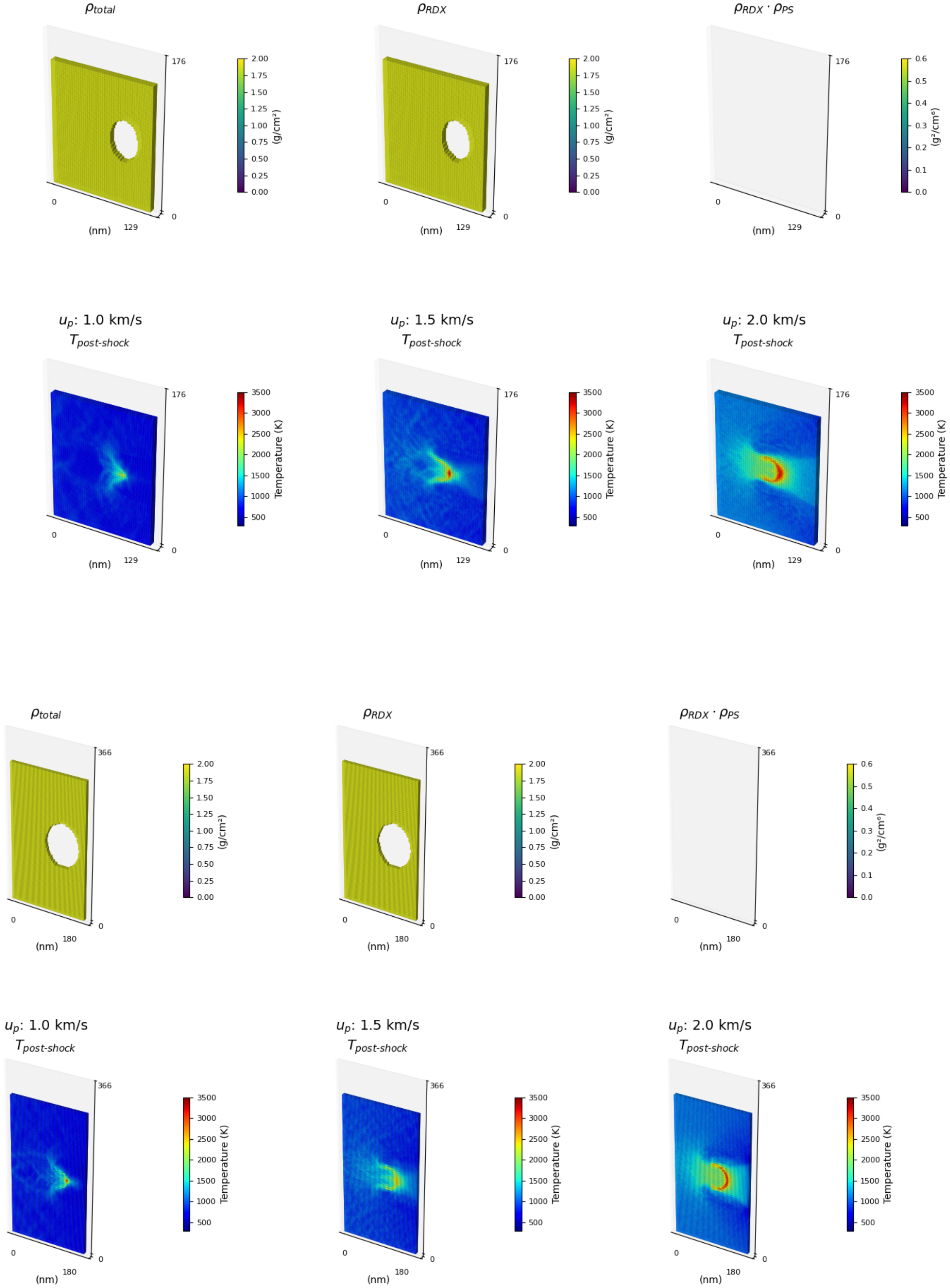

$\rho_{total}$
$\rho_{RDX}$
$\rho_{RDX} \cdot \rho_{PS}$
(g/cm³)
(g²/cm⁶)
(nm)
$u_p$: 1.0 km/s
$u_p$: 1.5 km/s
$u_p$: 2.0 km/s
$T_{post\text{-}shock}$
Temperature (K)

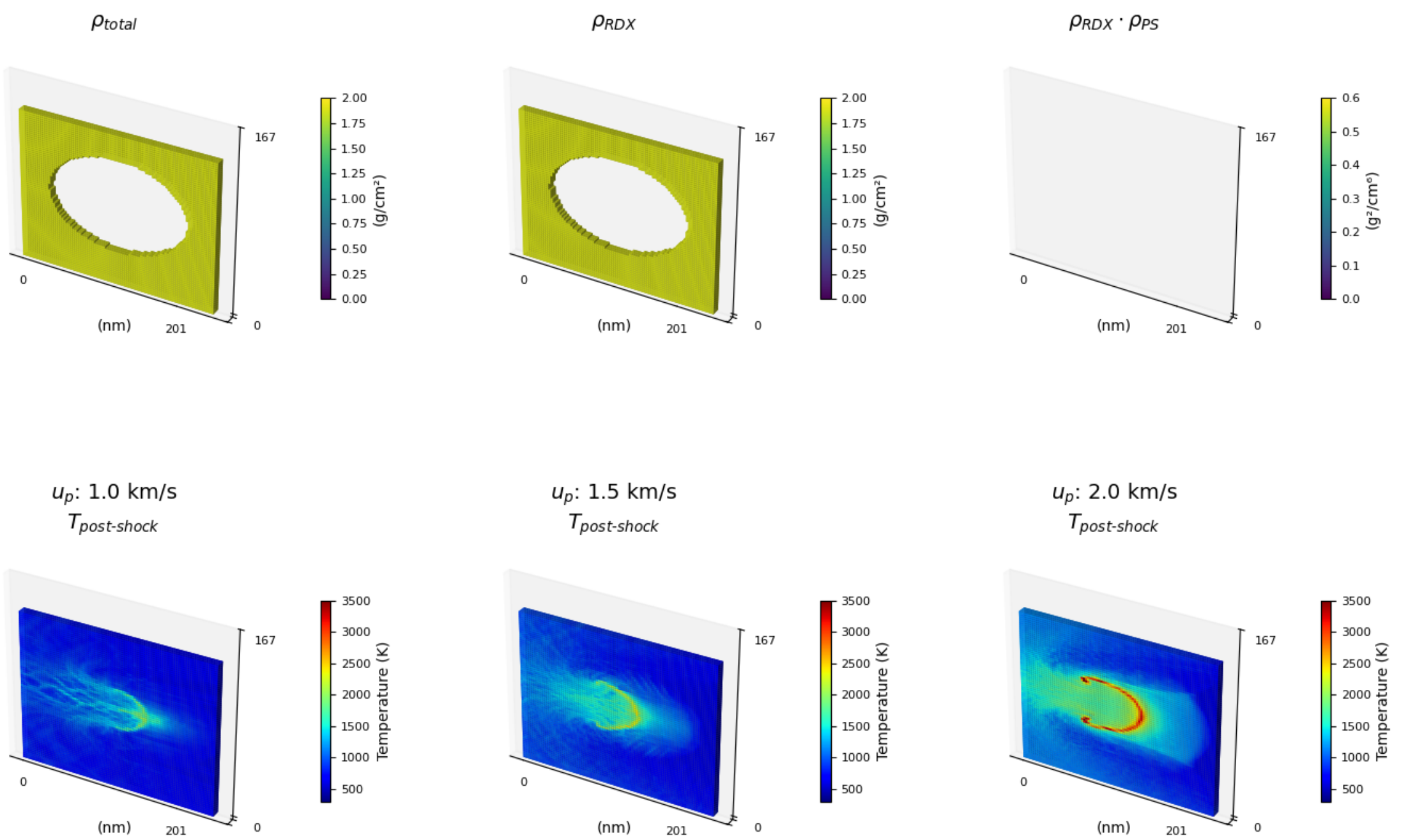

There are **3 single-pore systems** shocked at a particle velocity of 1.0 and 2.0 km/s using atomistic simulation. All are used for training:

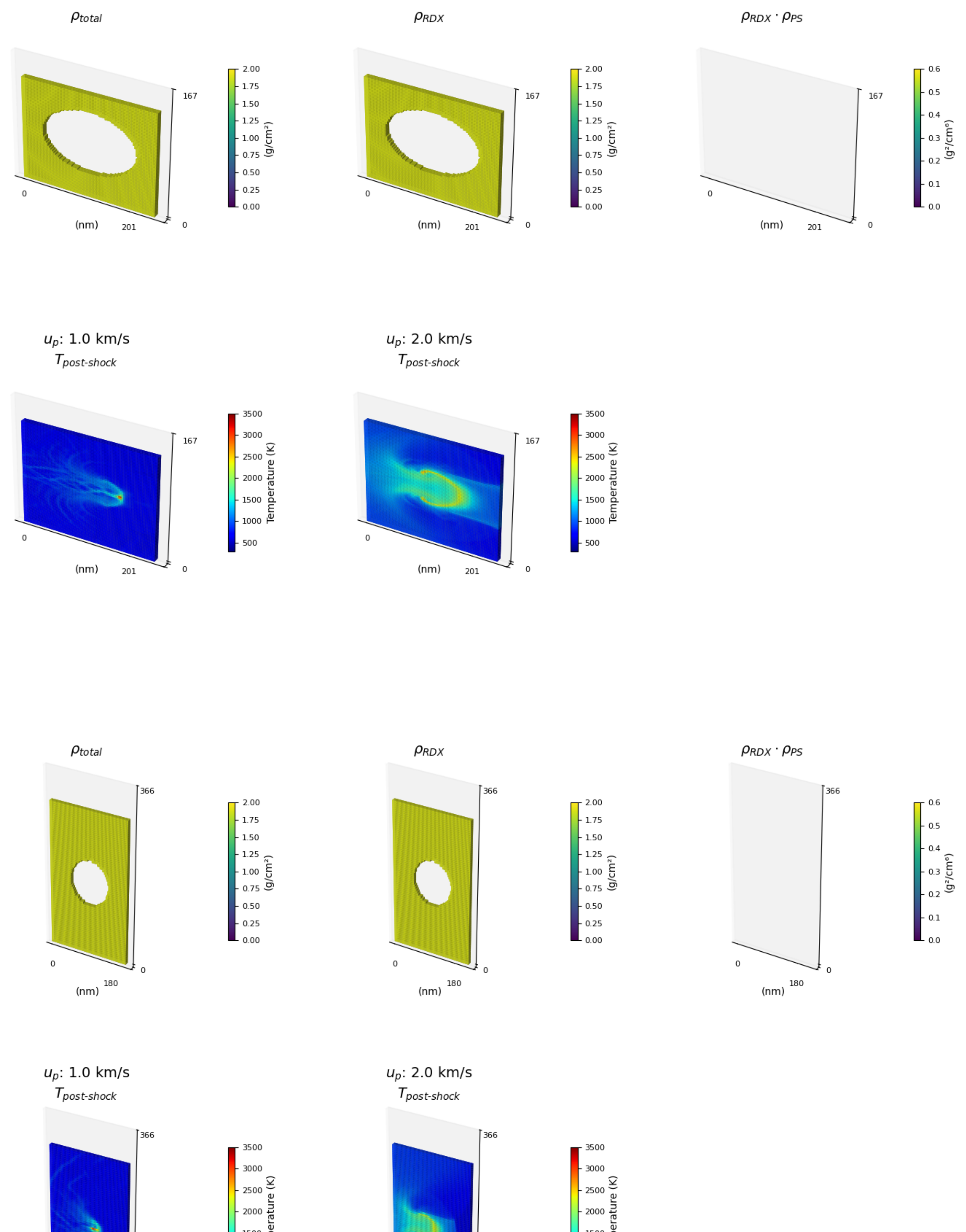
$\rho_{total}$
$\rho_{RDX}$
$\rho_{RDX} \cdot \rho_{PS}$
(g/cm³)
(g²/cm⁶)
(nm)
$u_p$: 1.0 km/s
$T_{post\text{-}shock}$
$u_p$: 2.0 km/s
Temperature (K)

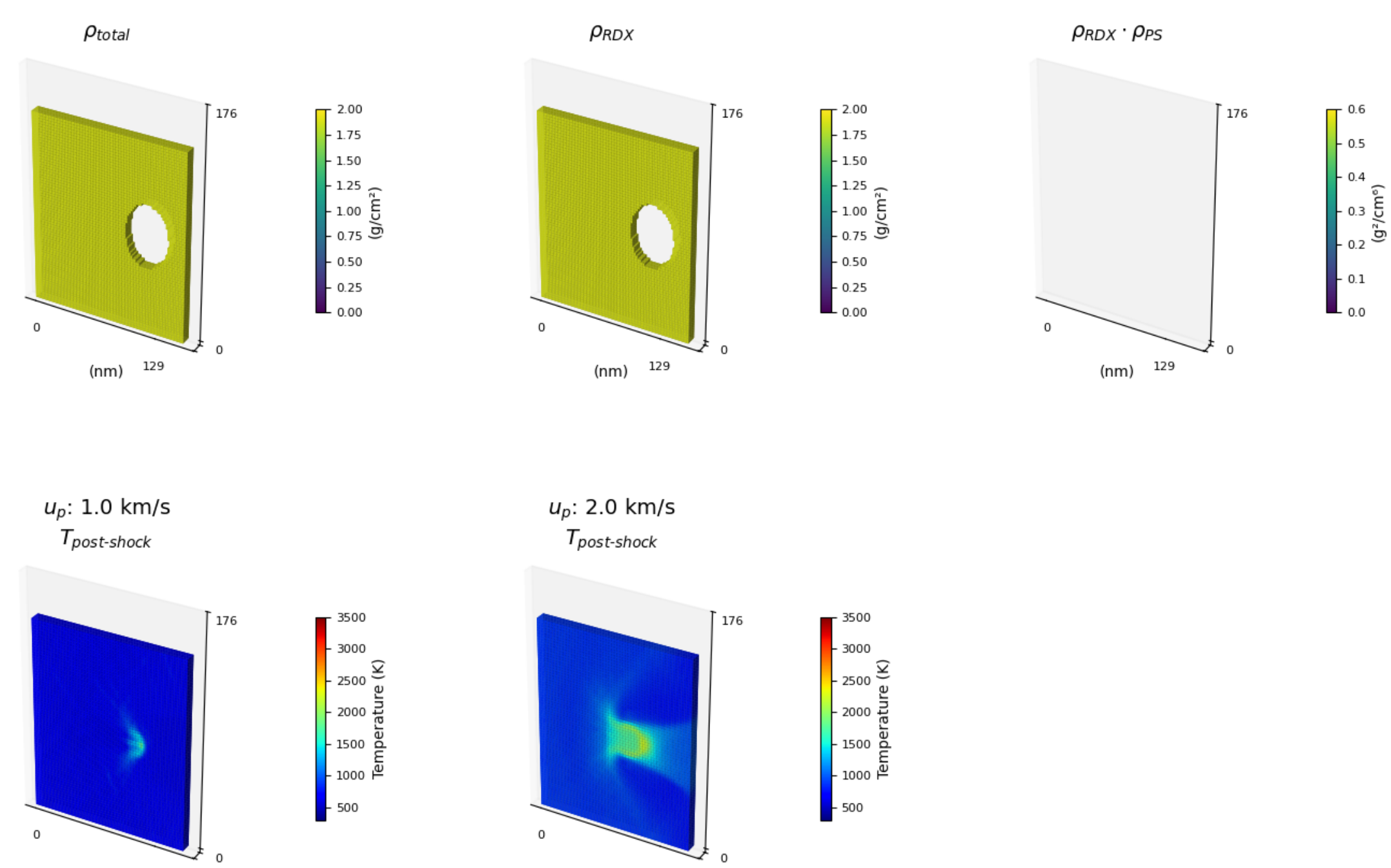

$\rho_{total}$
$\rho_{RDX}$
$\rho_{RDX} \cdot \rho_{PS}$
176
0
(nm)
129
(g/cm³)
(g²/cm⁶)
2.00
1.75
1.50
1.25
1.00
0.75
0.50
0.25
0.00
0.6
0.5
0.4
0.3
0.2
0.1
0.0
$u_p$: 1.0 km/s
$T_{post\text{-}shock}$
$u_p$: 2.0 km/s
$T_{post\text{-}shock}$
Temperature (K)
3500
3000
2500
2000
1500
1000
500